\documentclass{aa}  

\newcommand{\Ni}{\rm $^{56}$Ni}
\newcommand{\Co}{\rm $^{56}$Co}
\newcommand{\Cos}{\rm $^{57}$Co}

\newcommand{\Nis}{\rm$^{57}$Ni}
\newcommand{\Ti}{\rm$^{44}$Ti}

\newcommand{\Ca}{\rm$^{44}$Ca}
\newcommand{\Ms}{${\rm M_{\odot}}$}

\newcommand{\env}{$\sim$}
\newcommand{\cmcub}{cm$^{-3}$}
\newcommand{\grays}{{\rm $\gamma$-rays}}
\newcommand{\gray}{{\rm $\gamma$-ray}}
\newcommand{\arp}{Ar$^+$}
\newcommand{\hep}{$\rm He^+$}
\newcommand{\nep}{$\rm Ne^+$}

\newcommand{\fors}{Mg$_2$SiO$_4$}
\newcommand{\ens}{MgSiO$_3$}

\newcommand{\sili}{SiO$_2$}
\newcommand{\alu}{Al$_2$O$_3$}

\newcommand{\mic}{$\mu$m}

\newcommand{\cmc}{cm$^{-3}$}

\usepackage{graphicx}
\usepackage{mathtools}
\usepackage{txfonts}
\usepackage{amsmath}
\usepackage{subcaption}
\usepackage{lipsum}
\usepackage{placeins}
\begin{document}

   \title{Revisiting the formation of molecules and dust in core collapse supernovae.}

%   \subtitle{Subtitle}

   \author{I. Cherchneff\inst{1}
        \and D. Talbi\inst{1}
        \and J. Cernicharo\inst{2}
        %\fnmsep\thanks{Shows the usage of elements in the author field}
        }

   \institute{Université de Montpellier, CNRS, Montpellier, France \\
             \email{isabelle.cherchneff@umontpellier.fr}
 %            \thanks{Shows the usage of elements in the author field}
             \and Consejo Superior de Investigaciones Cientificas, Instituto de Fisica Fundamental, C/ Serrano 121, 28006 Madrid, Spain \\}
   \date{Received September 30, 20XX}

% \abstract{}{}{}{}{}
% 5 {} token are mandatory
 
  \abstract
 %\context heading (optional)
  {Core-collapse Supernovae of Type II contribute the chemical enrichment of galaxies through explosion. Their role as dust producers in the high-redshift Universe may be of paramount importance. However, the type and amount of dust they synthesise after outburst is still a matter of debate and the formation processes remain unclear.}
  % aims heading (mandatory)
   {We aim to identify and understand the chemical processes at play in the dust formation scenario, and derive mass yields for molecules and dust clusters at late post-explosion time.}
   % methods heading (mandatory)
   {We revisit existing models by improving on the physics and chemistry of the supernova ejecta. We identify and consider new chemical species and pathways underpinning the formation of dust clusters, and apply a unique exhaustive chemical network to the entire ejecta of a Supernova with a 15~\Ms\ progenitor. We test this new chemistry for various gas conditions in the ejecta, and derive mass yields for molecules and dust clusters.}
  % results heading (mandatory)
   {We obtain the molecular component of the ejecta up to 11 years after explosion. The most abundant species are, in order of decreasing masses, O$_2,$ CO, SiS, SiO, CO$_2$, SO$_2$, CaS, N$_2$, and CS. We identify molecules that are tracers of high-density clumps. As for dust clusters, we find the composition is dominated by silicates and silica, along with carbon dust, but with modest amounts of alumina. Pure metal clusters and metal sulphide and oxide clusters have negligible masses. High-density gas favours the formation of carbon clusters in the outer ejecta region whereas low temperatures hamper the formation of silicates in the oxygen core. The results are in good agreement with existing astronomical data and recent observations with the James Webb Space Telescope. They highlight the importance of chemistry for the derivation of dust budget from Supernovae.}
  % conclusions heading (optional), leave it empty if necessary
   {}

   \keywords{Supernovae: general --
                Astrochemistry --
                Molecular processes
               }

   \maketitle

%%%%%%%%%%%%%%%%%%%%%%%%%%%%%%%%%%%%%%%%%%%%%%%%%%%%%%%%%%%%%%
\section{Introduction}
Supernovae (SNe) hold a special place among  the dust contributors to galaxies. Indeed, the explosion of massive stars with initial mass on the Zero Age Main Sequence (ZAMS) comprised between 8~\Ms~and 30~\Ms, leads to dust formation in the ejected gas (hererafter, the ejecta) but the exact dust amount formed is still a matter of debate. Determining the SN dust yields is of paramount importance for assessing the dust budgets of local and high-redshift galaxies and the competition as dust providers between SNe and evolved, low-mass stars on the Asymptotic Giant Branch. According to several studies, SNe may represent the main dust factories in primeval galaxies at high redshift  \citep{dwek11,schnei24}, but the contribution of AGB stars may be more important than initially estimated \citep{boy25}. Despite the huge energy released by the explosion ($\sim 1$~B $= 1\times 10^{51}$~ergs) and the harsh physical conditions experienced by the ejecta, the presence of molecules and dust have been inferred in many SNe a few hundred days after explosion. 

The explosion of the blue supergiant Sanduleak$-$69 202 in the Large Magellanic Cloud more than 35 years ago leading to SN1987A provided the first evidence for molecule and dust synthesis in SN ejecta. The fundamental and overtone transitions of a few molecules, specifically carbon monoxide, CO, and silicon monoxide, SiO, were detected as early as 120 days post-explosion (hereafter, day~120) \citep{catch88,spyro88,meik89}, while an asymmetry of optical emission lines combined with a decrease in luminosity were interpreted as evidence of dust production at day 530 \citep{lu89,dan91}. 
Since then, emission lines of CO, SiO, SO, ${\rm SO_2}$, $\rm HCO^+$ and possibly SiS, were detected in the young remnant of SN 1987A with the Atacama Large Millimetre/sub-millimetre Array, ALMA \citep{kam13,mat17,cig19}. Other Type~II~SNe were investigated with the detection of CO and/or SiO molecules and evidence for dust formation a few hundred days after outburst \citep[e.g.][]{kot05,kot06,kot09,rho18,park25,med25}. 
%Warm dust has been detected in several SNe (e.g., Elmhamdi et al. 2003b; Kotak et al. 2005, 2006, 2009; Sugerman et al. 2006; Inserra et al. 2011b; Gallagher et al. 2012). An excess in the mid-IR, combined with a decrease of several magnitudes in the optical light curve and blue-shifted emission lines, are the usual indicators of the synthesis of dust in the ejecta. Therefore, astronomical studies provide evidence of the presence of molecules and dust in Type~II~SN ejecta. 

A direct proof of the existence of dust grains produced in SNe is delivered by studies of primitive Solar System materials (e.g. meteorites and interplanetary dust). The extracted pre-solar grains show anomalies in their isotopic composition characteristics of the nucleosynthesis of their parent stars. Pre-solar grains with SN origin include mainly silicates, with most grains having compositions consistent with olivine (e.g. forsterite Mg$_2$SiO$_4$) or pyroxene (e.g. enstatite MgSiO$_3$) or intermediate between these two, graphite, and to a lesser extent, silicon carbide and metal oxides such as alumina (Al$_2$O$_3$) and spinel (MgAl$_2$O$_4$). Rare pre-solar grains with SN origin such as silicon nitride (Si$_3$N$_4$) and silica (SiO$_2$) have also been identified \citep{nit95,haen13}.

The first exhaustive physico-chemical models of molecule and dust formation in SN ejecta show that specific molecules such as {\rm O$_2$}, CO, SiO, and SO form in large quantities some hundred days after explosion in local and high-redshift SNe \citep{cher08,cher09,cher10,sar13}. In the case of SN1987A, the predicted large CO mass of \env $0.1$ \Ms~is confirmed by ALMA observations \citep{kam13}. The molecular component of the ejecta represents between $10-40$ \% of the ejected matter, depending mainly on the progenitor mass and \Ni~content \citep{cher09,sar13}. Along with molecules, these models predict the synthesis of large quantities of dust equivalent to \env 8 \% and \env 2 \% of the ejecta mass for high-z Pair Instability and local Type~II SNe, respectively \citep{cher10,sar15}, with dust compositions including silicates, alumina, silica, and carbon. Further studies of SN1987A based on a comparable approach \citep{slu18} and using a similar chemical network \citep{sar22} corroborate these findings for local SNe. However, there are several drawbacks and simplifications in the existing chemical models. For example, different chemical schemes are applied to the various ejecta regions while most of the nucleation pathways and reaction rates for dust clusters are greatly optimised. 

The present study is the first in a series of new investigations on molecule and dust formation in local and high-z SNe. Here we revisit the chemistry of a non-interacting Type~II SN of progenitor mass equals to 15~\Ms. We aim at shedding light on the processes at the origin of the formation and evolution of molecules and the nucleation of dust in the nebular phase. In doing so, we consider a large number of new chemical species and processes not included in previous studies and define one unique chemical scheme that we apply to the entire ejecta. We re-investigate dust nucleation routes in light of recent theoretical and experimental studies for several dust types, thereby greatly improving on previous chemical descriptions of dust synthesis. This new model confirms the presence of already detected key molecules and predicts new species widely present in the ejecta. We assess dust masses and point to specific ejecta conditions necessary to foster the production of silicate and carbon dust. In the era of the James Webb Space Telescope (JWST), we believe this new model provides useful information to be used in the interpretation of current and future astronomical data. The improved physical and chemical models are described in \S~\ref{phy} and \S~\ref{chem}, while results are presented and discussed in \S~\ref{res}, and compared to observations and exiting studies in \S~\ref{obs} and \S~\ref{dis}. Finally, conclusions are outlined in \S~\ref{con}. 

%Since then, warm dust has been detected in several SNe (e.g., Elmhamdi et al. 2003b; Kotak et al. 2005, 2006, 2009; Sugerman et al. 2006; Inserra et al. 2011b; Gallagher et al. 2012). An excess in the mid-IR, combined with a decrease of several magnitudes in the optical light curve and blueshifted emission lines, are the usual indicators of the synthesis of dust in the ejecta. 
%%%%%%%%%%%%%%%%%%%%%%%%%%%%%%%%%%%%%%%%%%%%%%%%%%%%%%%%%%%%%%
\section{The physical model}
\label{phy}
%_____________________________________________________________
%                    FIG1- elemental composition
%-------------------------------------------------------------
 \begin{figure*}
 \centering
        \includegraphics[width=0.85\linewidth]{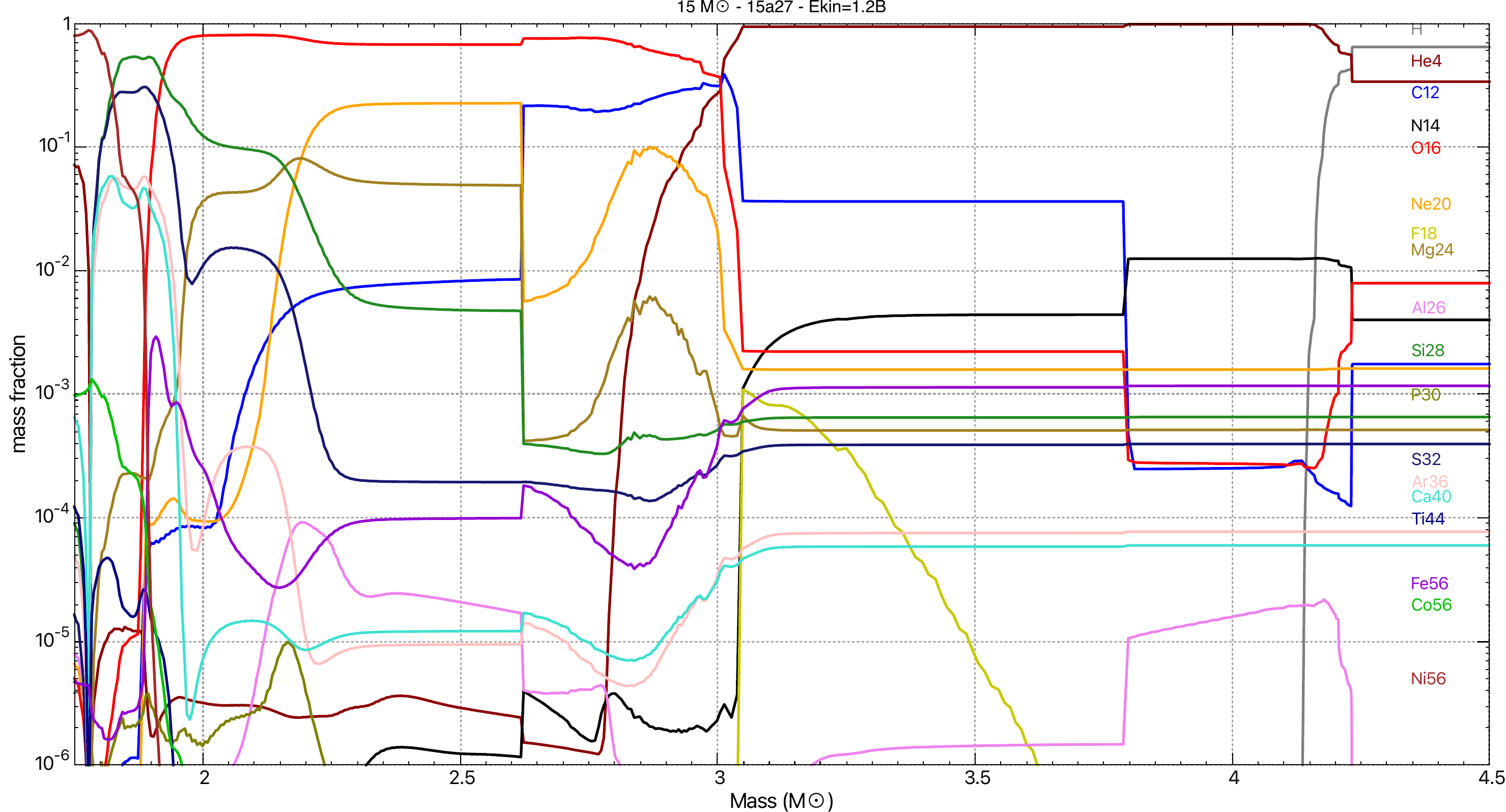}
      \caption{Elemental composition as a function of enclosed mass for a Type II-P SN with 15~\Ms\ progenitor from \citet{rau02}.}
%        \ContinuedFloat %      \usepackage{subcaption}
%        \includegraphics[angle=90]{figure.eps}
%     \includegraphics{figure.eps}
%       \includegraphics[angle=90]{figure.eps}
%    \caption{continued.}
       \label{fig01}
 \end{figure*}
%-------------------------------------
%
We choose a 15~\Ms~stellar progenitor as a template in this study. This mass is in line with the progenitor masses proposed for several Type IIP SNe recently observed and monitor by JWST, e.g. SN 2017eaw \citep{kil18,sza19} and SN 2023ixf \citep{vand23}. Larger and smaller progenitor masses will be investigated in a future study. Radioactive elements such as \Ni, \Co, and \Ti, create a flux of \gray~photons that pervades the ejecta. The degrading of \grays~to X-rays and ultraviolet (UV) photons occurs by Compton scattering and creates a population of fast Compton electrons in the ejecta. These fast electrons directly destroy molecular species and ionise the gas to produce noble gas ions such as \arp, \nep, and \hep,~that are key species to the ejecta chemistry. 
%
%----------- parameters radioactivity deposition -------
\begin{table}
\caption{Parameters for the 15~\Ms\ ejecta model by \citet{rau02}. }                 % title of Table
\label{tab1}    % is used to refer this table in the text
\centering                        % used for centering table
\resizebox{\columnwidth}{!}{\begin{tabular}{c c c c c c}      % centered columns (4 columns)
\hline            % inserts double horizontal lines
t$_{\rm 0}$ & E$_{\rm ex}$ & M$_{\mathrm{He}}$ & v$_{\rm He}$ &v$_{\rm c}$ & $\rho_\mathrm{gas}(t_\mathrm{0})$ \\
\hline\hline
100 & 1.2  & 4.23& 2693 \tablefootmark{a} &3522 \tablefootmark{a}& $1.35\times 10^{-13}$ \\
(days)&(B)&( \Ms)&(km s$^{-1}$)&(km s$^{-1}$)  &(g\ cm$^{-3}$) \\
\hline\hline
Isotope $i$ & $1 / {\lambda_{\rm i}}$ & M$_{\rm i}$ \tablefootmark{b}& E$_{\rm i}^{\gamma}$  \tablefootmark{c} & $\kappa_{\rm i} $  \tablefootmark{d}& $ \tau_{\mathrm{i,0}}$  \\
 &(days)&(\Ms) &(MeV) & (cm$^{-2}$g$^{-1}$) &$-$ \\
 \hline
${\mathrm {^{56}Co}}$ & 111.4& 0.1261& 3.606& 0.033&12.24 \\
${\mathrm {^{57}Co}}$ & 392.0& $4.211\times10^{-3}$& 0.1216&0.0792 &29.39 \\
${\mathrm {^{44}Ti}}$ &31036.6 &$3.421\times10^{-5}$& 2.275& 0.04&14.84 \\
\hline
\end{tabular}}
\tablefoot{
\tablefoottext{a} {See Equations \ref{vcore} and \ref{vel} below,}
\tablefoottext{b} {\citet{rau02}}
\tablefoottext{c} {\citet{seit14}}
\tablefoottext{d} {\citet{woo89}}
}\
\end{table}
% ----------------------------------------
%
In order to estimate the ionisation rates of Compton electrons in \S~\ref{chem}, we first need to assess the radioactive energy deposition rate in the ejecta. The SN light curve is powered by the radioactivity of several decay chains, among which the most important for our study are listed below with the corresponding half-lives $t_{\rm 1/2}$ \citep{seit14}
%
% ---------------------- radioactive processes 
\begin{equation}
\label{nicafe}
\begin{aligned}
{^{56}Ni}\ \xrightarrow{\text{$t_{\rm 1/2}=5$ days}}\ {^{56}Co} \xrightarrow{\text{$t_{1/2}=77.2$ days}} {^{56}Fe}, 
\end{aligned}
\end{equation}
\begin{equation}
\begin{aligned}
\label{nicafe2}
{^{57}Ni}\ \xrightarrow{\text{$t_{1/2}=1.5$ days}} {^{57}Co}\ \xrightarrow{\text{$t_{1/2}=271.8$ days}}\ {^{57}Fe},
\end{aligned}
\end{equation}
\begin{equation}
\begin{aligned}
\label{tiscca}
 {^{44}Ti}\ \xrightarrow{\text{$t_{1/2}=9453.5$ days}} \ {^{44}Sc} \xrightarrow{\text{$t_{1/2}=0.17$ days}}\ {^{44}Ca}.
 \end{aligned}
 \end{equation}
%--------------------------
%
For an isotope $i$, the energy deposition rate through radioactive decay  and emission of  \grays\ at time $t$ is given by

% energy depo rate -----------
\begin{equation}
\label{decay}
L_{\rm i}(t)={\lambda_{\rm i}}\ {N_{\rm i}(0)}\ {e^{-\lambda_{\rm i} t}}\ (1-e^{\tau_{i,0}(t/t_0)^{-2}})\ E_{\rm i}^{\gamma}
\end{equation}
% -------------------
where $\lambda_{\rm i} = ln(2)\ / t_{\rm 1/2,i}$ is the decay constant, $N_{\rm i}(0)$ is the total number of isotope $i$ at time $t=0$, ${\tau_{i,0}}$ is the optical depth from the centre at the reference time $t_{\rm 0}$, and $E_{\rm i}^\gamma$ is the energy emitted per decay in \grays. 
For each isotope, ${\tau_{i,0}}$ is defined as \citep{woo89,cher09}

% optical depth
\begin{equation}
\label{taui}
\tau_{\mathrm{i,0}}={\kappa_i} \frac{3}{4\pi} \frac{M_{\mathrm{He}}}{r(t_{\rm0})^2}
\end{equation}

where $M_{\mathrm{He}}$ is the mass of the helium core, ${\kappa_i}$ is the average \gray~mass absorption coefficient and $r(t_{\rm0})$ is the radius of the ejecta at time $t_{\rm0}$. 

Owing to the rapid decay of \Ni\ and \Nis\ compared to the ejecta dynamical time, we assume M(\Co)$=$ M(\Ni) and M(\Cos)$=$ M(\Nis). Similarly, we take M(\Ca)$=$ M(\Ti). The relevant parameters for the decay chains we consider and the calculation of the energy deposition rate as given by Equation \ref{decay} are summarised in Table \ref{tab1}. We choose $t_{\rm 0}=100$ days post-explosion as a reference time to start our calculations, leading to the following energy deposition rates as a function of time $t$, in ergs\ ${\rm s^{-1}}$
% deposition rates ------
%\begin{subequations}
\begin{equation}
\label{dep56}
%\begin{align}
L_{\mathrm {^{56}Co}}(t) = {1.61868\times 10^{42}} \times {e^{(-t/111.42)} }\times {[1-e^{-12.24(t/100)^{-2}}]},
\end{equation}
\begin{equation}
\label{dep57}
L_{\mathrm {^{57}Co}}(t) = {5.09014\times 10^{38}} \times {e^{(-t/392.00)} }\times {[1-e^{-29.39(t/100)^{-2}}]},
\end{equation}
\begin{equation}
\label{dep44}
L_{\mathrm {^{44}Ti}}(t) = {1.26584\times 10^{36}} \times {e^{(-t/31039.6)} }\times {[1-e^{-14.84(t/100)^{-2}}]}.
%\end{align}
\end{equation}
%\end{subequations}
%----------------------------------------

The total \gray-energy deposition rate to be considered in the calculation of the ionisation rates by Compton electrons is then
%
%.---- total depo rate ----
\begin{equation}
\label{deptot}
L_{\mathrm {tot}}(t) = L_{\mathrm {^{56}Co}}(t) + L_{\mathrm {^{57}Co}}(t)+L_{\mathrm {^{44}Ti}}(t).
%{1.26584\times 10^{36}} \times {e^{(-t/31039.6)} }\times {[1-e^{-14.84(t/100)^{-2}}]}.
%\end{align}
\end{equation}
%-----------------------------
The elemental composition after explosion is that derived by \citet{rau02} and is represented in Fig.~\ref{fig01}. In Rauscher's model, the SN ejecta core has a mass of 12.61 \Ms\ whereas the helium-rich (hereafter He-rich) ejecta mass is 4.23 \Ms. For the purpose of this study, which focuses on the synthesis of molecules and dust clusters, we define the ejecta as the matter comprised between the mass zones 1.78~\Ms\, and 4.14~\Ms, that is, 2.36~\Ms\, of hydrogen-free (hereafter H-free) ejected gas. We assume a spherically symmetric ejecta which elemental composition remains stratified after the star has exploded, despite the existence of Rayleigh-Taylor instabilities and macroscopic mixing within the helium core triggered by the reverse shock at the base of the hydrogen envelope. 

In Rauscher's explosion model, this H-free ejecta is represented by 332 zones, each zone having a specific velocity and being of variable mass in the range $5\times 10^{-3}- 1.35\times10^{-2}$~\Ms. To facilitate comparison with molecular lines and dust observations, we assign a velocity for each mass zone following the description of \citet{tru99}, that was later used by \citet{sar22} for his model of SN1987A. According to \citet{tru99}, the ejecta follows homologous expansion ($v_\mathrm{ej} \propto r$, where $r$ is the radius)  and is made of a homogeneous core and an envelope with a density profile $\rho_\mathrm{en} \propto r^{-n}$, where $ n \simeq 12$  in case the progenitor is a red supergiant (RSG) \citep{mat99}. The velocity of the SN ejecta core is given by
 %---------
\begin{equation}
\label{vcore}
  v_{\mathrm{c}}=  
     \sqrt{ \frac{10(n-5)}{3(n-3)} \frac{E_{\mathrm{ex}}}{M_{\mathrm{ej}}},
     }\,
\end{equation}
%-------------------------------------------
%   \begin{equation}
%      \tau_{\mathrm{ff}} =
%         \sqrt{ \frac{3 \pi}{32 G} \frac{4\pi r_0^3}{3 M_{\mathrm{r}}}
%}\,,
 %  \end{equation}
%
where $ E_\mathrm{ex}$ is the explosion energy and $ M_\mathrm{ej}$ the mass of the ejecta core. According to \citet{tru99}, the density of the ejecta core is given by 
\begin{equation}
\label{rhoc}
  \rho_{\mathrm{c}}=  
   \frac{3}{4\pi} \frac{n-3}{n} \frac{M_{\mathrm{ej}}}{(v_{\mathrm{c}} t)^3}.\
\end{equation}
The enclosed mass $M_{\mathrm{enc}}$, that is the mass comprised under a certain mass zone in the ejecta, is given by 
\begin{equation}
\label{menc}
 M_{\mathrm{enc}}=  
   \frac{4\pi}{3}\rho_{\mathrm{c}} (v  t)^3,\
\end{equation}

where $v$ is the zone velocity. By combining equations \ref{vcore}, \ref{rhoc} and \ref{menc}, we derive the following expression for the velocity as a function of enclosed mass, thereby translating zone position into zone velocity in the ejecta
\begin{equation}
\label{vel}
  v=  
%   \frac{n}{n-3} \biggl[ \frac{M_{\mathrm{enc}}}{M_{\mathrm{ej}}} \biggr]^{1/3}\times  v_{\mathrm{c}} \,
 \frac{n}{n-3} \left[ \frac{M_{\mathrm{enc}}}{M_{\mathrm{ej}}} \right]^{1/3}\times v_{\mathrm{c}}.\
\end{equation}
% 
%_____________________________________________________________
%                                            Zone parameters
%-------------------------------------------------------------
\begin{table*}[!h]
\caption{Parameters of regions in the stratified ejecta with 15~\Ms\ progenitor \citep{rau02}.}                 % title of Table
\label{par}    % is used to refer this table in the text
\centering                        % used for centering table
\begin{tabular}{c c c c c c}      % centered columns (4 columns)
\hline\hline               % inserts double horizontal lines
Region& Si/S/Ca & O/Si/Mg& O/C/Mg & He/C/N  & Ejecta\\ 
\hline
Major elements &${\rm Si-S-Ca-Ar}$& ${\rm O-Si-Mg-Ne}$&${\rm O-C-Ne-Mg}$&${\rm He-C-N-O}$\\
Boundaries \tablefootmark{a}& $1.783 - 1.894$& $1.894 - 2.623$& $2.623 - 3.013$& $3.013 - 4.141 $\\
Velocitiy range  \tablefootmark{b} &$774-990$ & $991-1630$ & $1631-1830$&$1831-2247$ \\
Number of zones &22 &145 &70&  95 & 332 \\
Mass \tablefootmark{a}   &   $0.111$&  $0.729$& $0.390$& $1.128$ &$2.358$ \\ % table heading
%{\rm $\mu_{gas}(t_0)$} \tablefootmark{c}& & & & & \\
%{\rm $N_{gas}(t_0)$}\tablefootmark{d} & & & & & \\
%{\rm $T_{gas}(t_0)$} \tablefootmark{e}& & & && \\
\hline                                  %inserts single line
\end{tabular}
\tablefoot{
\tablefoottext{a} {in \Ms,}
\tablefoottext{b} {in ${\rm km~s ^{-1}}$.}
%\tablefoottext{b} {\Ms,}
%\tablefoottext{c} {g cm$^{-3}$,}
%\tablefoottext{d}{particle cm$^{-3}$,}
%\tablefoottext{e}{K.}
}
\end{table*}
%------------------------------------

%_____________________________________________________________
%                       FIG2 - Velocities versus zones
%-------------------------------------------------------------
   \begin{figure}
   \centering
\resizebox{\hsize}{!}{\includegraphics{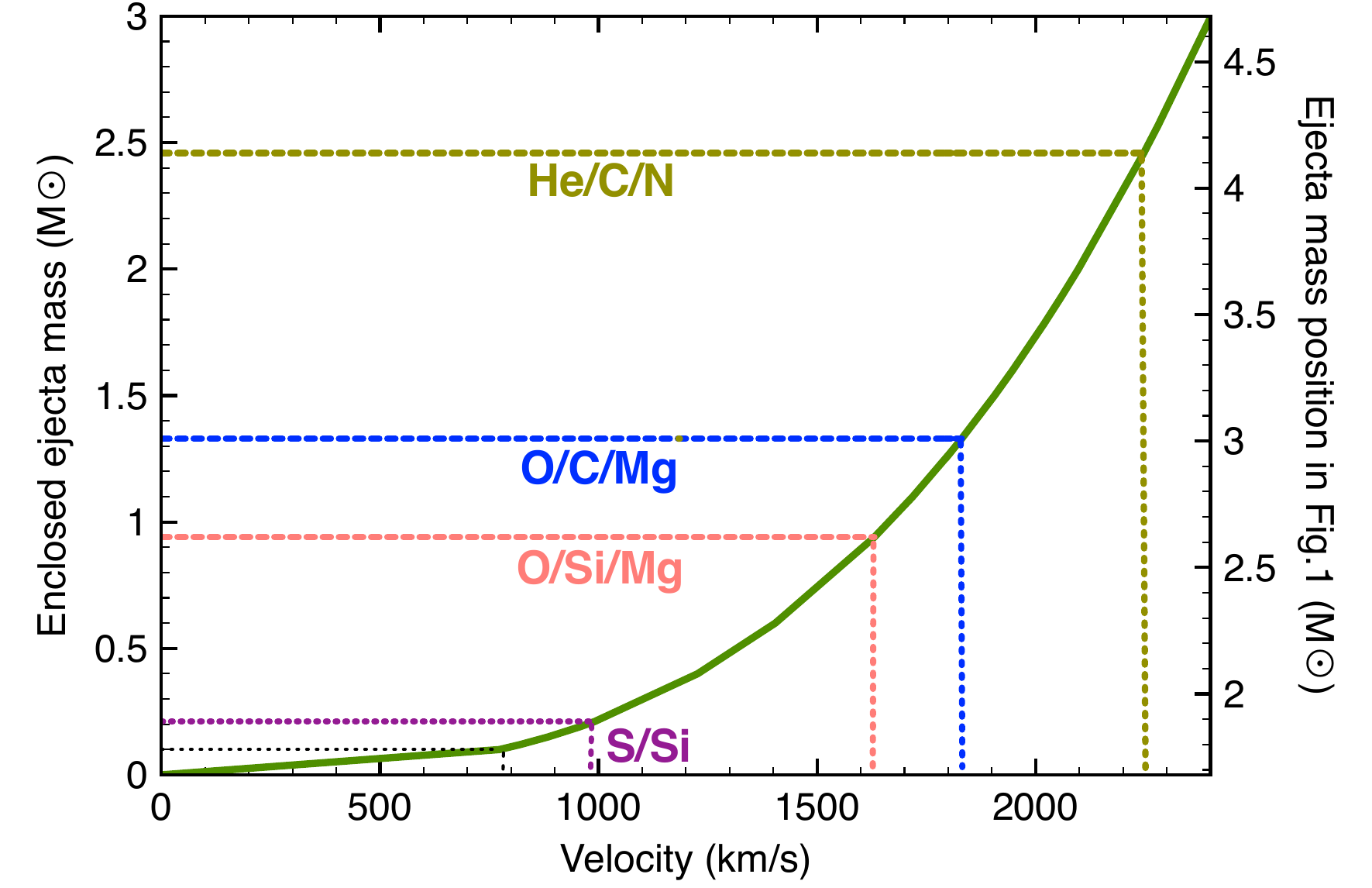}}  
      \caption{Zone velocity as a function of enclosed mass and zone position for the 15~\Ms\ explosion model of \citet{rau02}.}
         \label{fig2}
   \end{figure}
%----------------------------------------

The necessary parameters are listed in Table~\ref{tab1} and the zone position-velocity correspondance is shown in Fig. \ref{fig2}. 

In the perspective of a chemical study of the ejecta, we organise these 332 mass zones in different regions according to the chemical composition and the dominant elemental species present in each region following \citet{koz98}, \citet{cher08}, \citet{cher09}, and \citet{sar13}. However, instead of deriving an averaged elemental composition for each region to study the chemistry \citep[see][]{cher09,sar13}, here we investigate the chemistry of each specific zone included in a region and sum over all zones to derive important quantities like molecule and dust cluster masses. There are four major regions in total that cover the ejecta  and which parameters are listed in Table \ref{par}. 

\subsection{Gas density}
\label{ngas}
We study the time evolution of the entire ejecta by following the time evolution of the chemistry in each zone of the stratified ejecta at a starting time ${ t_0=100}$ days. Each zone expands with time at a certain velocity given by Equation~\ref{vel} and follows homologous expansion so that the density varies with time according to the equation
 %---------
\begin{equation}
\label{den}
n_\mathrm{gas}(M_{r},t) = \rho_\mathrm{gas}(M_r,t_0)/\mu(M_r,t_0)\times (t/t_0)^{-3},
\end{equation}
%---------§
where $\rho_\mathrm{gas}(M_r,t_0)$ and $\mu(M_r, t_0)$  are the gas density and the gas mean molecular weight at ${\rm t_0=}100$~days, respectively, in the mass zone of coordinate $M_r$. 

In their study of type IIP SNe, \citet{utro17} model the light curve of a typical 15~\Ms\, progenitor and derive gas densities for the ejecta out of three-dimensional (3D) simulations of neutrino-driven explosions. Gas density profiles as a function of interior mass are averaged over different angular directions resulting in a fairly constant gas density over the mass zone range  $2.2-5$~\Ms\, at ${\rm t_{ex}=1.29}$\,days, with ${\rm \rho_{gas}(M_r,t_{ex}) \sim 1\times 10^{-7}}$~g/cm$^{-3}$. 

For the sake of simplicity, we assume a constant gas density over the ejecta core and choose the average value derived by \citet{utro17} at position $M_\mathrm{r}=2.1$~\Ms. At $t_\mathrm{ex}$, $\rho_\mathrm{gas}(2.1,t_\mathrm{ex})=6.3\times 10^{-8}$~g/cm$^{-3}$. According to Equation~\ref{den}, the gas density at ${\rm t_0}$ is equal to $1.35\times 10^{-13}$~g/cm$^{-3}$ for the mass zones comprised between 1.78~\Ms\ and 4.14~\Ms. This density value characterises our Standard Case for this study. The zone number densities at $t_0$ calculated from Equation~\ref{den} are consistent with values derived from the analysis of SN light curves, e.g. SN1987A.   

%. ----------------- FIG3 - gas temperature --------
\subsection{Gas temperature}
\label{temp}
%_____________________________________________________________
%                       temperature versus zones
%-------------------------------------------------------------
   \begin{figure}[h!]
   \centering
   \includegraphics[width=\hsize]{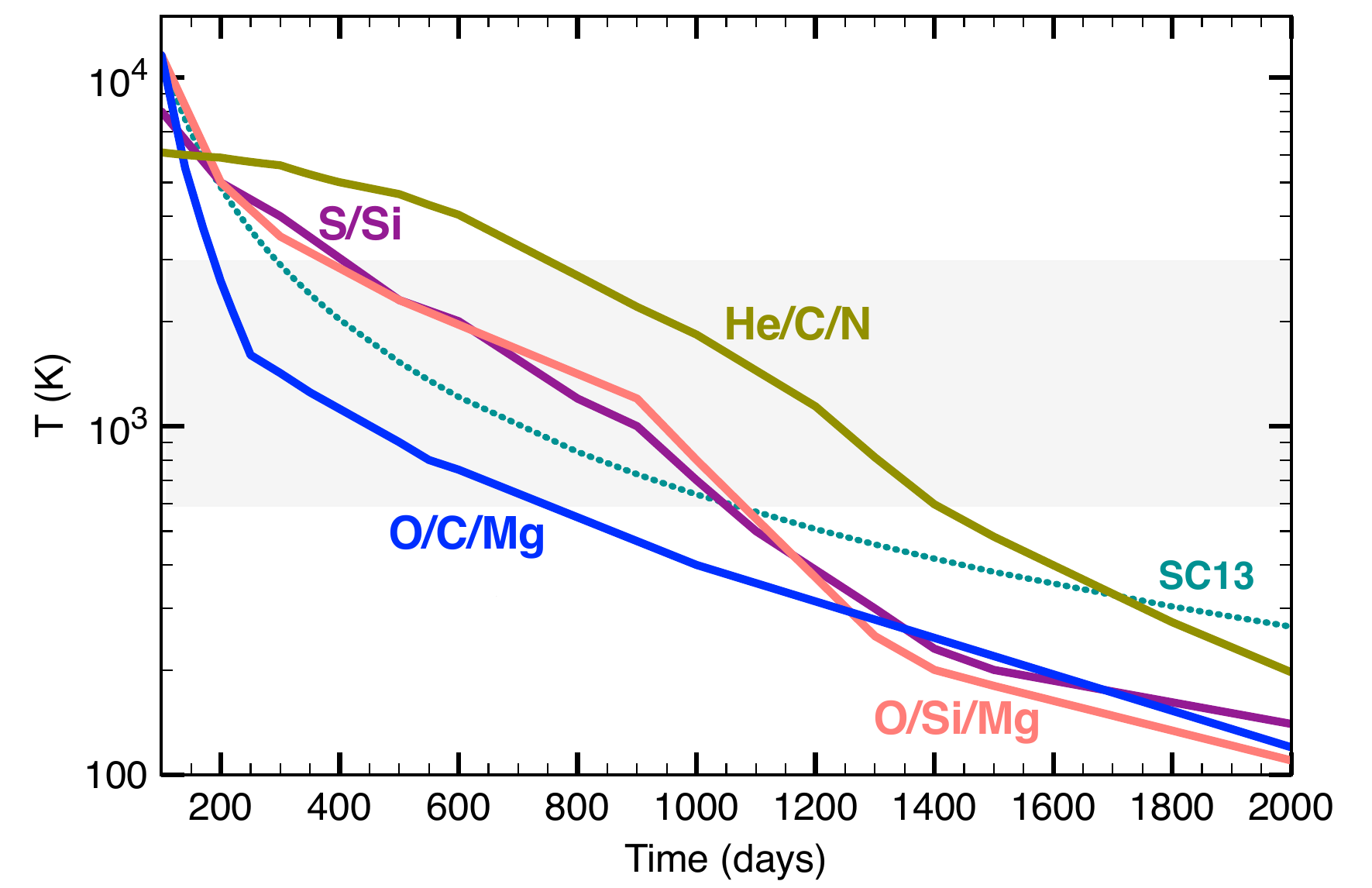}
      \caption{Gas temperature for the ejecta regions defined in Table~\ref{par} as a function of post-explosion time from \citet{koz98} except the O/C/Mg zone where values are from \citet{lil20}. The label SC13 refers to the temperature profile used in \citet{sar13}. The shaded area represents the temperature regime at which dust forms in dust-producing experiments in the laboratory (e.g. high-temperature flame or vapour condensation experiments).}
         \label{fig3}
   \end{figure}
%-----------------------------------------------
%
The temperatures for the various ejecta regions are displayed in Fig.~\ref{fig3} along with the temperature variation that was used in \citet{sar13}. These temperatures were derived for the ejecta of SN1987A by \citet{koz98} who modelled all thermal and non-thermal heating and cooling processes in the ejecta as a function of time. The four regions from Table \ref{par} having different elemental compositions, the temperature profiles drastically vary from one another when going from the inner region outwards. For example, the inner regions are rich in heavy elements compared to the outer carbon-rich region, and thus cool faster through atomic line emissions. By contrast, the He/C/N region will remain hot over the period day~100-1000 as seen in Fig.~\ref{fig3}. 

A special case is the region {\rm O/C/Mg} which produces most of the CO molecules, as shown by \citet{liu92}, \citet{cher09} and \citet{sar13}. The strong cooling resulting from the collision-induced vibrational emission of CO forces the gas temperatures to sharply drop before day~500 so that the region is much cooler than others \citep{liu92,lil20}. We actually choose the temperature profile derived by \citet{lil20} for our model. As pointed out by \citet{liu95}, any of the ejecta regions that forms molecules in relevant quantities experiences strong molecular cooling. This is true for the O/Si/Mg region that forms large amounts of SiO, and for which a gas temperature profile lower than the one shown in Fig. \ref{fig3} is also considered. 

The ejecta temperatures are modelled for SN1987A, which has a progenitor of \env~19~\Ms\ and injected \env~0,075~\Ms\ of \Ni\ in the ejecta. Our SN template has slightly different parameters as seen from Table~\ref{tab1}. Since there exists to date no temperature model for such a progenitor, we adopt the profiles for SN1987A and believe the temperature trends of the ejecta regions hold for other progenitor, specifically the molecular cooling in molecule-forming regions, and the high temperature of the outer ejecta region. The temperature profile of each region as a function of time is fitted by a sum of power laws in order to be implemented within the code. 
% 
% %%%%%%%%%%%%%%%%     CHEMICAL MODEL    %%%%%%%%%%%%%%%%%%%%%%
%
\section{The chemical model}
\label{chem}
Modelling the chemistry active in the ejecta requires to set up a list of relevant species and molecules, including neutrals and ions. Some species have already been observed in SNe while others are closely linked to them chemistry-wise. We also consider molecular species that are observed in other astrophysical environments, e.g. the wind of low- and high mass evolved stars, or planetary atmospheres. We bear in mind that the He core of the SN ejecta being H-free, this condition restricts our choice of species and related chemical processes. As for dust clusters, we consider several groups of solids identified in pre-solar grains of SN origin and their molecular precursors. These solids include silicates, silica, alumina, carbon, silicon carbide and silicon nitride \citep{hop22}. 

\subsection{Method}
\label{met}

We consider formation pathways that are identified in the laboratory or proposed on the basis of established chemical structures and exo/endothermicity of reaction processes. All species we consider are gathered in Table \ref{tab2}. We include 203 chemical species and 1447 reactions in total, which reaction rates are documented or estimated. Most of them are taken from the following chemical databases: National Institute of Standards and Technology Chemical Kinetics Database \citep{man08}, UMIST22 \citep{mil24} and KIDA \citep{wak24}. For non-documented rates, the estimation is based on the calculation of reaction energies that are chosen as activation barriers for endothermic processes. The set of reactions consists of thermal and non-thermal processes, which description and occurrence regime are gathered in Table \ref{tabap1} of the Appendix. 

The temporal variation of the number density $n$ of a molecular species $i$ located at a given mass zone $z$ is described by the following rate equation
%
%----------- ODEs --------------------
\begin{equation}
    \begin{split}
        \frac{dn_i(z,t)}{dt}    &    = P_i - L_i  \\   
                                       &    =  \sum_{j}{k_{j i}(t) n_j(z,t) n_i(z,t)} - \sum_{k}k_{i k}(t) n_i(z,t) n_k(z,t),
    \end{split}
\label{ode}
\end{equation}
%------------------------------
where $P_i$ and $L_i$ are the production ($\equiv$ formation) and loss ($\equiv$ destruction) term, respectively, and $k_{j i}$ is the reaction rate coefficient for a reaction between species $j$ and $i$. The reaction rate is provided in a Arrhenius form as follows
%----------- rate --------------------
\begin{equation}
\label{rate}
   k_{i j}(t) = A_{ij}\ \left[\frac{T(t)}{300}\right]^n \exp(-E_a/T(t)),
\end{equation}
% -----------------------
where $T$ is the gas temperature, $A_{ij}$ the Arrhenius coefficient (in s$^{-1}$ molecule$^{-1}$, cm$^{3}$ s$^{-1}$molecule$^{-1}$ and cm$^{6}$ s$^{-1}$ molecule$^{-1}$ for a unimolecular, bimolecular, and termolecular process, respectively), $n$ reflects the temperature dependence of the reaction, and $E_a$ is the activation energy barrier along the reaction path.

The destruction rate by Compton electrons of species $i$ leading to ionisation or dissociation is defined as \citep{cher09}
%-------- Compton rate ----------
\begin{equation}
 k_{\rm C,i}(t)= \alpha \frac{L_{\rm tot}(t)}{N_{\rm tot} W_i},
 \label{kcom}
 \end{equation}
%-------------------------
 
where  $L_{\rm {tot}}(t)$ i the total deposition rate for \grays\ as given by Equation~\ref{decay}, $N_{\rm tot}$ is the total number of ejecta particles, and $\alpha$ is the fraction of the deposited energy in the ejecta that goes into ionisation and dissociation of atoms and molecules. As in \citet{liu95} and \citet{koz98}, we choose $\alpha \sim 0.35$. Finally, $W_i$ is the mean energy per ion-pair for species i (in eV), and is defined as the ratio of the energy of the incident electron divided by the number of ionisation or dissociation produced by collision with the incident electron until it comes to rest \citep{liu94,dal99}. Values of $W_i$ are taken from \citet{cher09}. 

Modelling the chemistry of the ejecta consists of solving 203 stiff, coupled, ordinary differential equations as described by Equation~\ref{ode} for each mass zone and the gas conditions described in \S \ref{ngas} and \ref{temp}. We first choose a few mass zone positions for each ejecta region and specifically study the chemistry active at these positions to identify the chemical processes at the origin of molecule and dust cluster formation. We thus derive number densities and mass fractions for the 203 species as a function of post-explosion time and identify prominent species formed at specific positions in the ejecta. Secondly, we run the chemistry for each zone across the ejecta regions and sum up the masses of species we obtain to derive the total mass of species produced per ejecta region as a function of time. Finally, we sum up the masses obtained per region for each species to derive the variation of the final mass formed as a function of time across the ejecta. This method provides an exhaustive grasp of the chemistry taking place across the ejecta and the evolution of the mass budget of molecules and dust clusters with post-explosion time. 

One major improvement of this study resides in considering one unique chemical scheme applied to all ejecta regions. Previous studies assume each region is governed by its own chemistry and chemical scheme in line with the initial elemental composition of the region \citep[e.g,][]{cher09,sar13}, and such an assumption is adopted in recent studies \citep{sar22,sar25}. This restriction on the chemistry at play induces a biased outlook on what an ejecta region might produce by limiting the various types of species formed in each region, for example the exclusion of the formation of oxides in carbon-rich regions and C-bearing molecules in O-rich zones. 

We briefly discuss below our choices of species, dust molecular clusters, and the new chemical pathways leading to the formation of silicate, silica, alumina, and carbon dust clusters. 

%
%---------- molecules and dust clusters ----------------------- 
\begin{table*}
\caption{Atoms, molecules, ions, and dust molecular clusters considered in the new chemical model.}              
\label{tab2}      
\centering                         
\begin{tabular}{l l l l l l l l l l}        
\hline\hline  
 \multicolumn{9}{c}{Atoms/Molecules}  \\
 \hline            
O & O{\rm $_2$} & O{\rm $_3$} &OF&MgO&AlO&SiO&PO&SO \\
C & C{\rm $_2$} & C{\rm $_3$} &CN & CO &CO{\rm$_2$} &CF&CP&CS  \\ 
Si & Si{\rm$_2$}& SiC &SiN &SiO{\rm$_2$}& SiO{\rm$_3$} &SiP& SiS &   \\
N & N{\rm $_2$}  & NO & NO$_2$ &NF & NS &  &  &   \\
S& S{\rm$_2$}  & CS{\rm$_2$ } &SO{\rm$_2$ } &OCS &  & & &  \\ 
Mg & Mg$_2$ & MgO$_2$ &MgS& MgS$_2$ &  &  &   &  \\
Al & Al$_2$ &AlO$_2$ & AlO$_3$ &Al$_2$O  &  & & &  \\
Ca& Ca{\rm$_2$}  & CaO& CaO{\rm$_2$ }  &CaS& CaS{\rm$_2$ } & &  &  \\ 
Fe & Fe{\rm$_2$}  &FeO& FeO{\rm$_2$ }  & FeO{\rm$_3$ }  &FeS& &  &  \\
P &P{\rm$_2$}  & PN&  PO{\rm$_2$ }& PS& PS{\rm$_2$ }& & &  \\ 
F & F$_2$ & MgF & AlF& SiF&CaF & & &  \\
He&Ne&Ar &${\rm^{44}}$Ti&${\rm^{56}}$Co&${\rm^{56}}$Ni& & &   \\
\hline
\multicolumn{9}{c}{Atomic and molecular ions} \\
\hline
O{\rm$^+$}&O{\rm $_2^+$} &MgO{\rm $^+$} &AlO{\rm $^+$}& OCS{\rm $^+$} & & & &  \\
C{\rm$^+$}& C{\rm $_2^+$} & C{\rm $_3^+$} &CN{\rm$ ^+$} &CO{\rm $^+$} & CO{\rm $_2^+$}&CS{\rm$ ^+$} & &  \\
Si{\rm$^+$}&SiC{\rm $^+$} &SiN{\rm $^+$} &SiO{\rm $^+$} &SiS{\rm $^+$} & &  & &  \\
N{\rm$^+$}&N{\rm $_2^+$} &NO{\rm $^+$} &NO{\rm $_2^+$} & NS{\rm $^+$} & & & &  \\
S{\rm$^+$}&SO{\rm $^+$} &SO{\rm $_2^+$} & &  & &  & &  \\
Al{\rm$^+$}&ALO{\rm $_2^+$} & &&  & & & &  \\
Ca{\rm$^+$}&CaO{\rm $^+$} & & &  & &  & &  \\
Fe{\rm$^+$}&FeO{\rm $^+$} &FeO{\rm $_2^+$} &FeS{\rm $^+$}  &  & &  & &  \\
P{\rm$^+$}&PO{\rm $^+$} & & &  & &  & & \\
Mg{\rm$^+$}&F{\rm$^+$}&He{\rm$^+$}&Ne{\rm$^+$}&Ar{\rm$^+$} & &  &  &  \\
\hline
\multicolumn{9}{c}{{Dust clusters and ions }} \\
\hline
(Mg)${\rm_{n=3,6}} $& (MgO)${\rm_{n=2,4}}$ & (MgS)${\rm_{n=2,4}}$ &(Fe)${\rm_{n=3,6}} $&(FeO)${\rm_{n=2,4}}$  & (FeS)${\rm_{n=2,4}}$  &(SiO)${\rm_{n=2,20}}$  & (CaO)${\rm_{n=2,4}}$  & (CaS)${\rm_{n=2,4}}$   \\
Al$_2$O$_2$ &Al$_2$O$_3$& Al$_2$O$_4$ &Al$_3$O$_3$ &Al$_4$O$_6$& Si$_2$O$_3$ &Si$_3$O$_4$ &Si$_3$O$_5$&Si$_3$O$_6$  \\
 MgSiO &MgSiO$_2$ &MgSiO$_3$& Mg$_2$SiO$_2$ &Mg$_2$SiO$_3$ &Mg$_2$SiO$_4$&Mg$_2$Si$_2$O$_6$ & Mg$_4$Si$_2$O$_8$ & \\
 (C)${\rm_{n=4,20}} $&(C)${\rm^+_{n=4,10}}$ & & & & & & & \\
\hline  
\end{tabular}
\end{table*}
% ---------------------------------------------

\subsection{Silicates and silica}
\label{sili}

The nucleation of silicate clusters remains an enigma, both experimentally and theoretically. The condensation of Mg-SiO-H$_2$-O$_2$ vapours results in the formation of magnesiosilica smoke where grains are arranged in necklaces and agglomerates. MgO (periclase) and \sili\ (silica) inclusions are also found in these grains of stoichiometric enstatite (\ens) or forsterite (\fors) chemical composition \citep{reit02}. Flash-evaporation experiments exploiting the vapour phase from magnesium and silicon oxide in a mixed atmosphere of Ar and O$_2$ to form silicates, show the grains change from crystalline to amorphous as the Mg-to-Si ratio decreases \citep{kim08}. 

The natural precursor to silicates in circumstellar environments is the molecule SiO, which is abundant in the wind of AGB and supergiants stars, and SN ejecta. Theoretical studies have focussed on identifying chemical structures of relevant clusters and routes linking SiO, SiO dimers and silicate molecular clusters. For example, \citet{esca19} employ global-optimisation methods to find the lowest-energy isomers of (\ens)$_{\rm N}$\ and (\fors)$_{\rm N}$ clusters for ${\rm N=1-10}$. Interestingly, the \ens\ and \fors\ monomers are both kyte-shaped with one external O atom, and are thus reactive molecules. In the context of understanding dust formation in the H-rich wind of AGB stars, \citet{brom12} derive exothermic chemical pathways consisting of the dimerisation of SiO as a bottleneck to the formation of Si$_2$O$_3$, followed by the addition of Mg atoms to form small molecular clusters of the type Mg$_x$Si$_y$O$_z$ with silicate stochiometry. These pathways are adapted by \citet{sar13} for a H-free gas and used to model the formation of silicate clusters in several SN environments \citep{sar15,slu18,sar22,sar222,sar25,puru25}. 

However, a quantum chemical study of SiO clustering by \citet{brom16} shows the dimerisation is too slow a process under the physical conditions (i.e., gas  temperature and pressure) met in circumstellar environments to induce silicate nucleation. Experimental investigations on SiO clustering from vapour, which show very inefficient grain formation, also corroborate the finding that SiO dimerisation is inefficient at triggering the formation of silicate clusters \citep{kim22}. Therefore, both theoretical and experimental studies invalidate the chemical model used so far for the synthesis of silicates in SNe and question the reliability of the results on dust by existing investigations. 

In order to revisit the problem of silicate nucleation, we explore new chemical pathways based on the initial conditions met in a SN ejecta, which we consider H-free, and rich in Mg, O, and Si, like the region O/Si/Mg of Table \ref{par}. The formation of dust is observed in many SN ejecta characterised by various stellar progenitor masses, explosion energies and physical conditions. We thus wish to identify a simple chemical scheme efficient enough to sustain dust formation under different SN environments. Starting with simple di- and triatomic molecules we find are abundant in the ejecta gas (e.g. SiO, MgO, SiO$_2$ and MgO$_2$), we pinpoint all possible exothermic and some slightly endothermic chemical channels that buildup small intermediate clusters Mg${\rm_x}$Si$_{\rm y}$O$_{\rm z}$ with ${\rm x=0-4}$, ${\rm y=0-1}$ and ${\rm z=1-4}$. We characterise these small silicate molecular clusters by deriving stable structures from Density Functional Theory (DFT), and follow their chemical growth to finally form  \ens\ and \fors\ molecules and their dimers. 

The exo- or endothermicity of a reaction is calculated by deriving the reaction energy $\Delta E$ given by 
%--------------- energy reaction --------------
\begin{equation}
\label{ener}
\Delta E=\sum_{\rm products}{E}-\sum_{\rm reactants}{E},
\end{equation}
%---------------------------------------------------
where $E$ is the absolute energy of a species at 0~K and includes zero-point corrections. Energy data on small silica are from \citet{lu03} while we calculate from DFT the energies for silicate intermediates. For a few silicate clusters, we use existing values from \citet{brom12}. All calculations are made at the B3LYP/6-31G(d) level. The relevant new chemical reactions leading to the formation of \ens\ and \fors\ monomers and dimers are listed in Table \ref{siliform} along with their reaction energy, whereas the calculated absolute energies at 0~K of the silicate molecular clusters used in our chemical description are listed in Table \ref{tabap2} of the Appendix.

%\citep[e.g.,][]{cher10,sar13,sar22,sar25} 
Existing chemical models consider that the formation of silica (\sili) occurs as a consequence of the polymerisation of SiO to form (SiO)$_{\rm n}$ clusters. The polymerisation is followed by silicon segregation and formation of \sili\ units in silicon monoxide clusters as the clusters size increases, usually for n > 10 \citep{wang08,reb08}. As we mentioned previously, the polymerisation of SiO is extremely inefficient at forming clusters for circumstellar gas conditions. This implies the synthesis of silica clusters must operate through other channels. Therefore, we investigate the formation of small silica clusters up to (\sili)$_3 \equiv {\rm Si}_3{\rm O}_6$ by considering all exothermic reactions leading to the formation of the trimer and taking into account the structures as provided by \citet{lu03}. These chemical pathways mainly involve the exothermic reaction ${\rm SiO_2 + SiO_2 \longrightarrow Si_2O_3 + Si}$ to form Si$_2$O$_3$ and subsequent exothermic reactions with \sili\ to grow to Si$_3$O$_5$ and \sili\ trimers. 
%
%---------------- silicate reactions -------
\begin{table}[!h]
\caption{Proposed reactions for the nucleation of enstatite and forsterite dimers, and the corresponding reaction energy.}                 % title of Table
\label{siliform}    % is used to refer this table in the text
\centering                        % used for centering table
\resizebox{\columnwidth}{!}{\begin{tabular}{l ll ll ll ll}      % centered columns (4 columns)
\hline\hline               % inserts double horizontal lines
& \multicolumn{3}{c}{Reactants}& &  \multicolumn{3}{c}{Products}&  \multicolumn{1}{c}{$\Delta$E\  \tablefootmark{a}} \\
\hline
R1&MgO&+& MgO&${\longrightarrow}$& MgO$_2$&+&O& - 86.4\\  
R2&MgO&+& SiO&${\longrightarrow}$& SiO$_2$&+&Mg& - 188.3\\
R3&MgO&+& SiO$_2$&${\longrightarrow}$& SiO$_3$&+&Mg& - 16.3\\            
R4&MgO$_2$&+& SiO&${\longrightarrow}$& SiO$_3$&+&Mg& - 98.1\\    
\\
R5&MgO$_2$&+& SiO$_2$&${\longrightarrow}$& MgSiO$_3$&+&O& - 263.4\\    
R6&MgSiO$_3$&+& MgO$_2$&${\longrightarrow}$& Mg$_2$SiO$_4$&+&O& - 269.1\\ 
R7&MgO&+& SiO$_2$&${\longrightarrow}$& MgSiO&+&O$_2$&+54.3\\  
R8&MgO&+& SiO$_3$&${\longrightarrow}$& MgSiO$_3$&+&O& - 342.3\\ 
R9&Mg$_2$O$_2$&+& SiO$_3$&${\longrightarrow}$& Mg$_2$SiO$_4$&+&O& - 355.3\\    
R10&MgO$_2$&+& SiO&${\longrightarrow}$& MgSiO$_2$&+&O& - 377.9\\ 
R11&MgSiO$_2$&+& MgO&${\longrightarrow}$& MgSiO$_3$&+&Mg& - 291.2\\
R12&MgSiO$_2$&+& O$_2$&${\longrightarrow}$& MgSiO$_3$&+&O& +41.8\\     
\\
R13&MgO$_2$&+& SiO&${\longrightarrow}$& MgSiO&+&O$_2$& - 32.1\\ 
R14&MgSiO&+& O$_2$&${\longrightarrow}$& MgSiO$_2$&+&O& - 128.5\\ 
R15&MgSiO$_2$&+& MgO&${\longrightarrow}$& Mg$_2$SiO$_2$&+&O&+73.5\\ 
R16&MgSiO$_2$&+& MgO$_2$&${\longrightarrow}$& Mg$_2$SiO$_2$&+&O$_2$& - 147.5\\ 
R17&Mg$_2$SiO$_2$&+& MgO$_2$&${\longrightarrow}$& Mg$_2$SiO$_4$&+&Mg& - 663.7\\ 
R18&Mg$_2$SiO$_2$&+&O$_2$&${\longrightarrow}$& Mg$_2$SiO$_3$&+&O& - 90.5\\ 
R19&MgSiO$_2$&+& MgO$_2$&${\longrightarrow}$& Mg$_2$SiO$_3$&+&O& - 237.1\\ 
R20&Mg$_2$SiO$_3$&+& MgO$_2$&${\longrightarrow}$& Mg$_2$SiO$_4$&+&Mg& - 323.1\\ 
R21&Mg$_2$SiO$_3$&+&O$_2$&${\longrightarrow}$& Mg$_2$SiO$_4$&+&O& - 2.2\\ 
\\
R22&MgSiO$_3$&+& MgSiO$_3$ &${\longrightarrow}$& Mg$_2$Si$_2$O$_6$&  & &-770.3\\ 
R23&Mg$_2$SiO$_4$&+& Mg$_2$SiO$_4$&${\longrightarrow}$& Mg$_4$Si$_2$O$_8$&  & &-2407.5 \\ 
 \hline
\end{tabular}}
\tablefoot{
\tablefoottext{a} {In kJ mol$^{-1}$.}}
%\tablefoottext{b} {Exothermic reactions according to \citet{brom12}.}
%\tablefoot{Masses are in \Ms.} 
\end{table}
% ------------------------------------------------------------%SiO$_2$nucleation of silicates in condensation experiments has been studied for  and revealed ... Synthesis of Si$_{'rm x}$O$_{\rm y}$ grains in the absence of hydrogen was studied.

%It is usually assumed that most silicate dust has stoichiometric pyroxene or olivine chemical compositions, for which MgSiO3 (enstatite) and Mg2SiO4 (forsterite)  are the corresponding Mg-based examples.
%It is thus likely that the very first species that initiate circumstellar silicate nucleation are, as yet unidentified, extremely stable refractory seeds

\subsection{Alumina}
The structure of \alu\ clusters (\alu)$_{N=1,7}$) were studied by \citet{li12} while structures of aluminum oxide (Al$_{n=2,7}$O$_{m=1,10}$) clusters were extensively investigated by \citet{arm19}. When the \alu\ unit has a kyte-shaped structure, the dimer (\alu)$_2$ is a cage. However, cage structures become instable with increasing cluster sizes and disordered structures are favourable for large clusters. A recent study by \citet{saba21} explores the gas-phase synthesis of alumina from aluminium oxidation.  Based on DFT, they construct a detailed chemical kinetic description of alumina formation in a O$_2$ environment. In the present model, we stop our cluster growth at the formation of alumina dimers ($\equiv$~Al$_4$O$_6$) and base our chemical description on results from \citet{saba21}, \citet{cat00} and \citet{glo16}.  
\subsection{Carbon clusters}
\label{carbon}
The formation of pure ($\equiv$~H-free) carbon clusters was extensively studied 30 years ago after the discovery of the Buckminsterfullerene cage molecule, C$_{60}$ \citep{kro85}. Since then, a consensus was reached in nano-science on the formation of pure carbon cages, which involves the growth of small carbon chains up to C$_9$ into carbon single or multiple rings and open cages, the smallest cage identified in the laboratory being C$_{28}$. The growth of cage lattices occurs through the incorporation of atomic C and C$_2$ \citep{dunk122}. The structure of small carbon chains and rings is then important to better model the formation of carbon grains in circumstellar environments. 

In their study of dust formation in SNe, \citet{sar15} consider the synthesis of carbon clusters up to C$_{28}$. Most of the reaction pathways are derived from \citet{clay99} who consider a very simple carbon chemistry unconnected with other species apart from atomic oxygen. Most processes are temperature independent and their scheme for carbon cluster growth relies on chemical pathways with no energy barriers $E_a$ and with large $A$ factors ($\equiv$ the reactions are fast - see Equation \ref{rate}). Such processes guarantee an "artificial" and efficient formation of carbon clusters in the ejecta over time and the whole temperature span.

In this study, we revisit and improve the formation processes of carbon chains and rings up to C$_{20}$. Following \citet{rem16}, we consider that carbon clusters are chains for C$_{2-5}$ and C$_{7-9}$ and rings for C$_6$ and C$_{10}$. The chemistry of small carbon clusters up to C$_{10}$ is that of \citet{loi14} for reactions between small chains and \citet{clay99} for radiative association processes. For reaction of C$_2$ with small chains ($\equiv$ C$_{\rm n}$ with n$=3-7$), we use DFT to derive energy budgets along the reaction path and assess exo/endothermicity and barriers. We further describe the growth of rings through C incorporation taking into account the ring opening energies derived by \citet{rem16}, while our description of ring growth through C$_2$ addition is based on the study by \citet{schwei95}. The dimerisation of C$_6$ and C$_{10}$ described by \citet{rem16} is also considered. We believe this new description provides more accurate formation pathways for carbon clusters up to C$_{20}$ at the high- and low gas temperatures met in the ejecta.

\subsection{Other clusters}
On top of silicate, silica, alumina, and carbon clusters, we model the formation of silicon diatomic Si$_2$ and sulphur diatomic S$_2$ species as potential first-step products in the formation process of pure Si and S solids. For iron, we model the formation of small ${\rm (Fe)_{n=2,6}}$ clusters based on a study of iron nanoparticle synthesis by condensation of iron vapour at temperatures of $\sim$~1000~K \citep{gie03}. A complementary study by \citet{wen07}, who derive detailed chemical kinetic models for gas-phase synthesis of iron nanoparticles, is also used. Finally, we consider the formation of metal sulphide solids by including the polymerisation of small ${\rm (MgS)_{n=2,4}}$, ${\rm (CaS)_{n=2,4}}$ and ${\rm (FeS)_{n=2,4}}$ clusters, for which we adopt the rates derived for ${\rm (SiO)_{n}}$ polymerisation \citep{brom16}. 

We do not model the formation of small silicon carbide, SiC, and silicon nitride, Si$_3$N$_4$, clusters at this stage but include formation processes for the two parent molecules SiC and SiN to better understand where these molecular precursors form in the ejecta. 

\subsection{Phosphorus and fluorine}
\label{PF}
We see from Fig.~\ref{fig01} that phosphorus, P, is present in the inner O/Si/Mg region, while the mass fraction of fluorine, F, is important in the inner part of the He/C/N region. Phosphorus-bearing molecules were detected in the H-rich circumstellar environments of evolved stars. Phosphorus monoxide PO, was first detected in the wind of the supergiant star VY Cam \citep{ten07} while both PO and PN were observed in the wind of the O-rich AGB star IK Tau \citep{deb13}. Finally, carbon and silicon monophosphide, CP and  SiP respectively, were both identified in the wind of the carbon star IRC+10216 \citep{gue90,koe22}. As for fluorine, AlF was observed again in the wind of IRC10216 at millimetre wavelengths \citep{cer87,ziu94}, but the search for MgF and CaF was unfruitful. 

In order to assess whether SN ejecta produce P- and F-bearing molecules, we include in our new chemical scheme the fluoride and phosphorous species listed in Table \ref{tab2} and the related chemistry.

%%%%%%%%% RESULTS %%%%%%%%%%%%%%%%%%%%%%

\section{Results}
\label{res}
With the purpose of identifying the chemical processes controlling the synthesis of molecules and dust clusters, we study the ejecta by first considering a "Standard Case", defined as follows: 1) the gas number density is derived from the initial density at day~100 given in Table~\ref{tab1}; and 2) the gas temperatures in the various ejecta regions are those illustrated in Fig. \ref{fig3}. We first present the results as a function of ejecta regions in the following sections, discuss the chemistry at play, and provide the final masses of molecules and dust clusters formed over each region. 

We then consider two cases out of our Standard Case to investigate two specific issues: 1)~a high-density case for the He/C/N~region to study the conditions required for the efficient nucleation of carbon dust clusters, and 2)~a low-temperature profile for the O/Si/Mg region to assess the impact of lower temperatures on the synthesis of silicates, silica and alumina. Finally, all molecular and cluster mass values for the entire ejecta and the Standard Case are gathered in Table~\ref{finmas}. 
\subsection{The Si/S/Ca region}
\label{SiSreg}
%_____________________________________________________________
%                              FIG 4 - Mol in Si/S/Ca zone 1.85 Msun
%-------------------------------------------------------------
  \begin{figure}
%   \begin{subfigure}{.5\textwidth}
  %   \centering
\resizebox{\hsize}{!}{\includegraphics{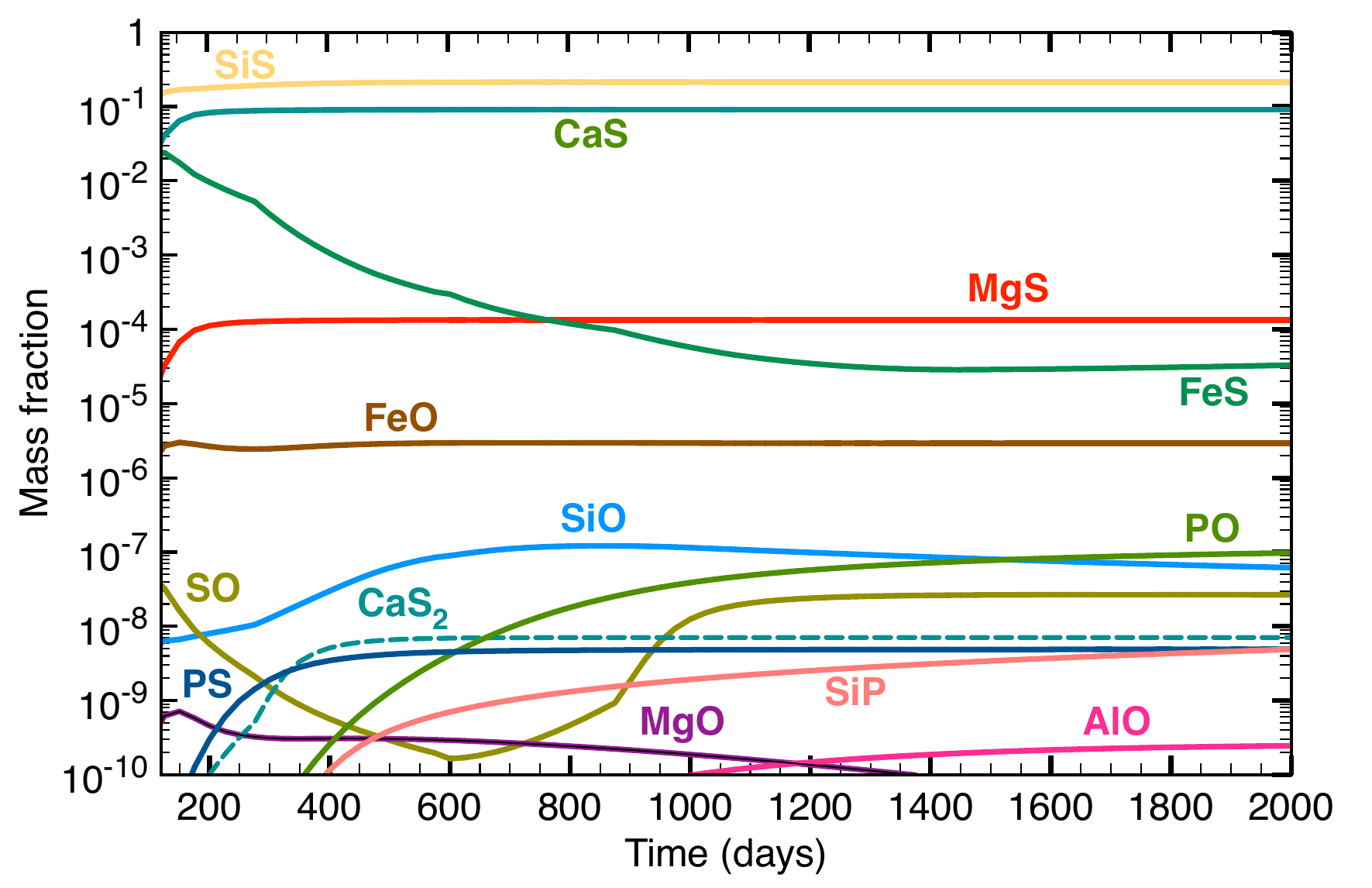}}
\caption{Mass fractions of molecules produced in the Si/S/Ca ejecta region at z=1.85~\Ms\ as a function of post-explosion time for the Standard Case. The zone mass is $4.64 \times 10^{-3}$ \Ms. }
 \label{fig4}
\end{figure}
%------------

In this region, we choose the zone position $z=1.85$~\Ms\ to investigate the chemistry since the mass fraction of oxygen is still small. As seen from Fig.~\ref{fig4}, the prominent molecules are SiS, CaS, FeS and MgS and they readily form. The dominant formation pathways for these molecules are through reaction with S$_2$ and radiative association as follows
%------
\begin{equation}
\label{SiS}
Si + S_{2} \rightarrow SiS + S,
\end{equation}
\begin{equation}
\label{CaS}
Ca + S \rightarrow CaS + h\nu,
\end{equation}
\begin{equation}
\label{FeS}
Fe + S_{2} \rightarrow FeS + S,
\end{equation}
\begin{equation}
\label{MgS}
Mg + S \rightarrow MgS + h\nu.
\end{equation}
The rate for Reaction \ref{SiS} has not yet been studied while reaction of C with S$_2$ to form CS is well documented \citep{mit84,smith04}. As in \citet{cher09}, we use the isovalence of Si and C to attribute a rate to Reaction \ref{SiS}. A similar approach is used to estimate the rate for Reaction \ref{FeS} by assuming isovalence of O and S and the rate of FeO formation \citep{smir12}. Likewise, we use the isovalence of O and S, and Mg and Ca, for determining the rates of Reactions \ref{CaS} and \ref{MgS}, since the radiative association reaction between Mg and O is studied by \citet{bai21}.

Because the reaction rates for the dominant formation processes of these four molecules are estimated, we want to test whether their large mass fractions are sensitive to our choice of rates. We decrease by a factor 100 the rates and obtain unchanged results for the SiS, CaS, FeS and MgS mass fractions. The robustness of these results is due to the combination of the limited chemistry controlling the Si/S/Ca~region compared to other ejecta regions, and the initial elemental composition that lacks oxygen. 

\begin{figure}
\resizebox{\hsize}{!}{\includegraphics{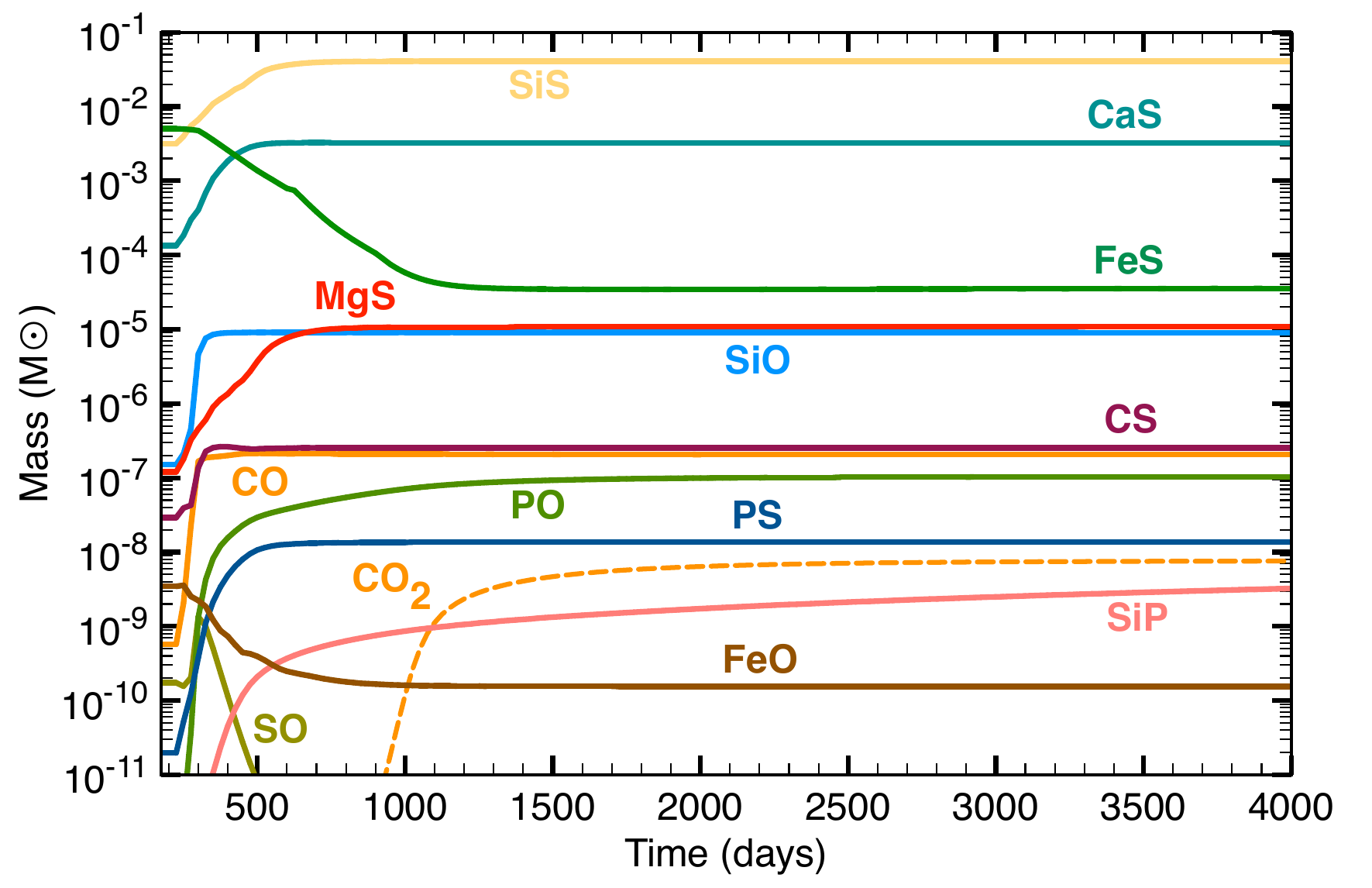}}
%\resizebox{\hsize}{!}{\includegraphics{fig43}}
%    \includegraphics{width=8cm}{fig41}
%     \includegraphics{width=8cm}{fig42}
%    \includegraphics[width=8cm]{fig41}
%  \end{subfigure}
 \caption{Total masses of molecules produced over the Si/S/Ca~region as a function of post-explosion time for the Standard Case. The region mass is 0.111 \Ms\ (see Table \ref{par}).}
 \label{fig5}
    \end{figure}

The prominent molecules formed over the entire Si/S/Ca~region are SiS and CaS, and to a lesser extent, FeS and MgS, as seen from Fig.~\ref{fig5} where the total molecular masses summed over the Si/S/Ca~region are presented as a function of post-explosion time. Towards the outer part of the region where the oxygen mass fraction starts rising (see Fig. \ref{fig01}), some O-bearing species form in small quantities (e.g. SiO and PO). While the molecular budget represents $\sim 39.7$~\%\ of the region's mass, none of the dust clusters present in our model are synthesised in relevant quantities in this region.

%_____________________________________________________________
%                              FIG 5 - SI/S/Ca mol mass
%-------------------------------------------------------------
 %----------------------------

\subsection{The O/Si/Mg region (Inner oxygen region)}
\label{inO}
%_____________________________________________________________
%                              FIG 6 - Mol in O/Si/Mg zone 2.15 Msun
%------------------------------------------------------------- 
\begin{figure*}
\begin{subfigure}{.5\textwidth}
  \centering
  \includegraphics[width=\linewidth]{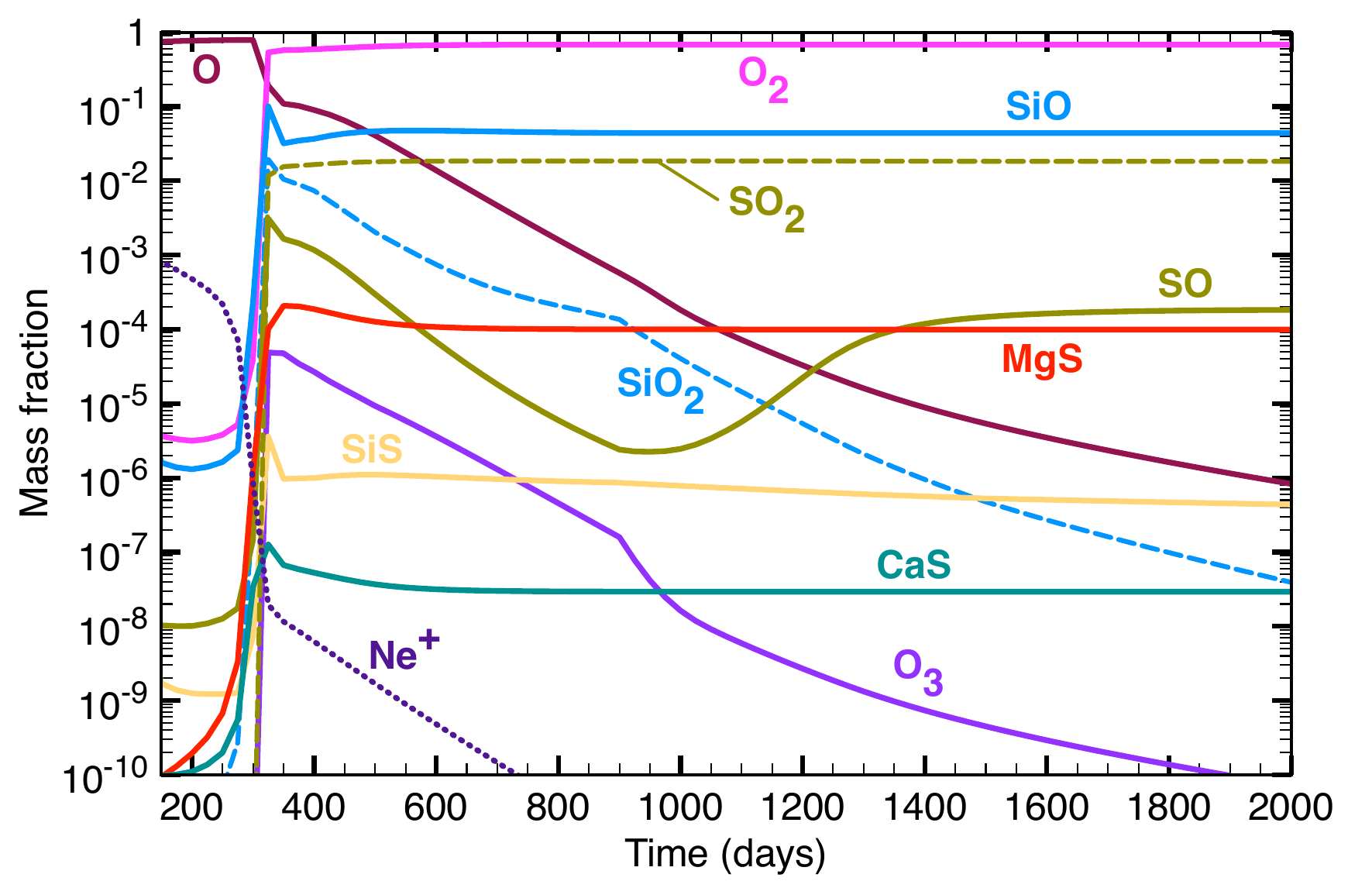}
%  \caption{Main molecules}
\end{subfigure}%
\begin{subfigure}{.5\textwidth}
  \centering
  \includegraphics[width=\linewidth]{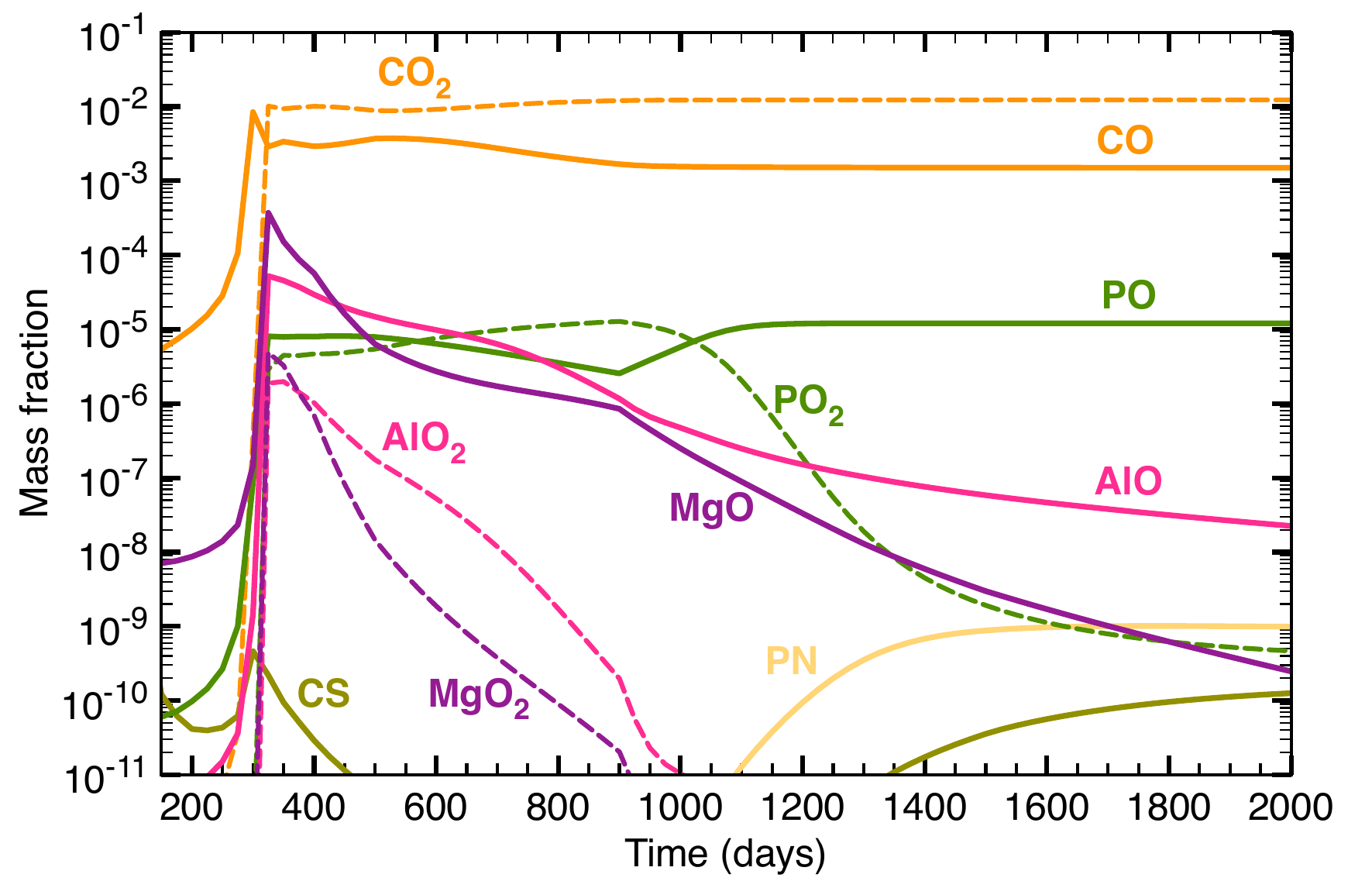}
%  \caption{Main molecules}
  \end{subfigure}
 \caption{Mass fractions of main molecules produced in the O/Si/Mg region at $z=2.15$ \Ms\ for the Standard Case. The zone mass is $5.42 \times 10^{-3}$ \Ms.}
  \label{fig6}
  \end{figure*}
%--------------------------------------------------
The O/Si/Mg region contains mostly atomic oxygen and large mass fractions of atomic silicon, magnesium, and sulphur (see Fig.~\ref{fig01}). Therefore, this region forms most of the silicate, silica, and alumina clusters in the ejecta. As  mentioned in \S~\ref{PF}, phosphorous is also present, although in much lesser quantities. A significant mass fraction of Ne characterises this region but it remains always smaller than that of atomic oxygen. Compton electrons ionise Ne to create a population of \nep\ ions that dissociates molecules and return to the neutral state. However, it is prominently molecules like O$_2$ that drive the chemistry of the region. The situation is drastically different in the He/C/N region where \hep\ controls the time of molecular formation. 

\subsubsection{Results for $z=2.15$ \Ms.}
%
%_____________________________________________________________
%                              FIG 7 - DUST in O/Si/Mg zone 2.15 Msun
%------------------------------------------------------------- 
\begin{figure} 
  \centering
  \includegraphics[width=\linewidth]{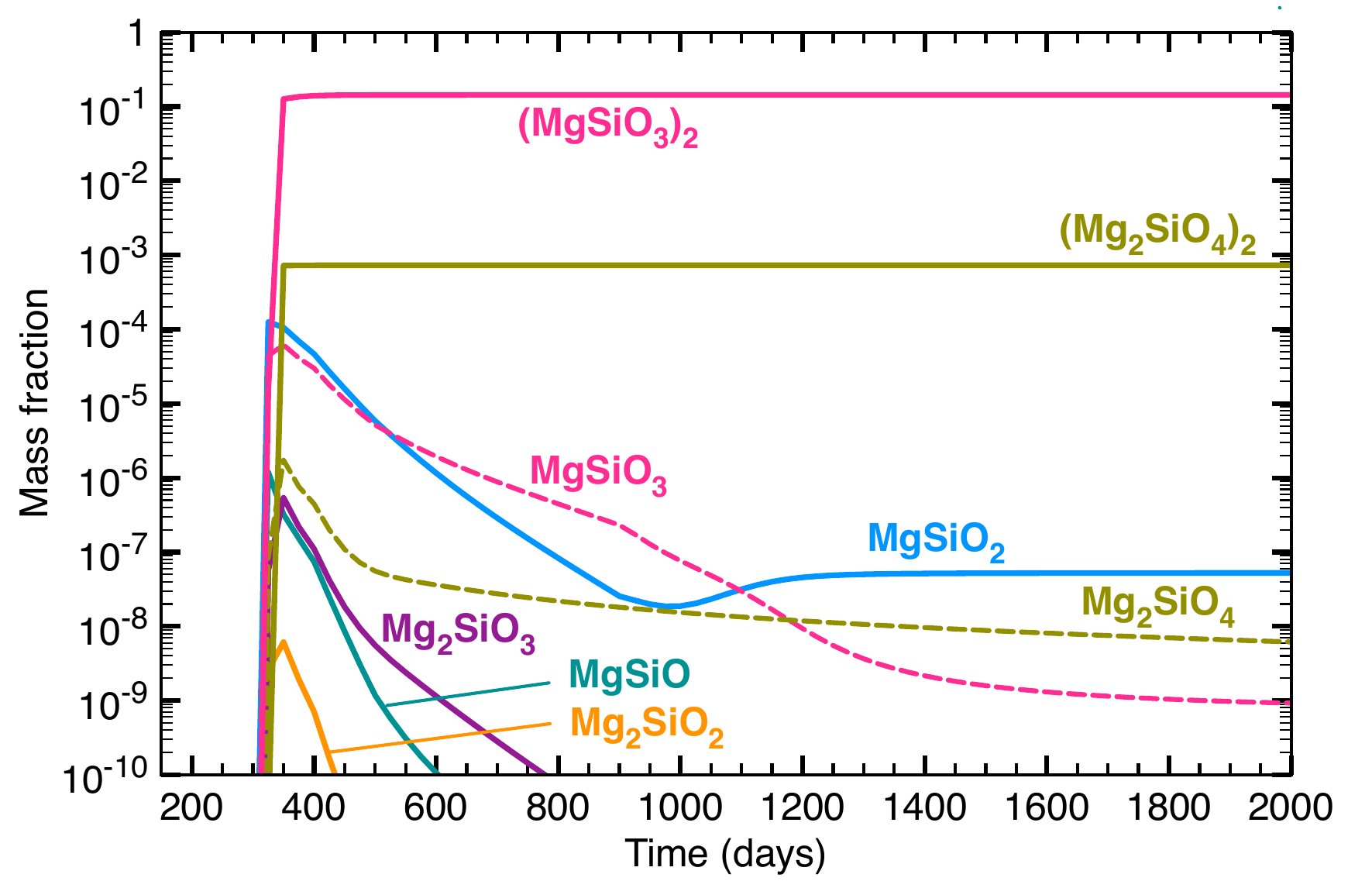}
  \includegraphics[width=\linewidth]{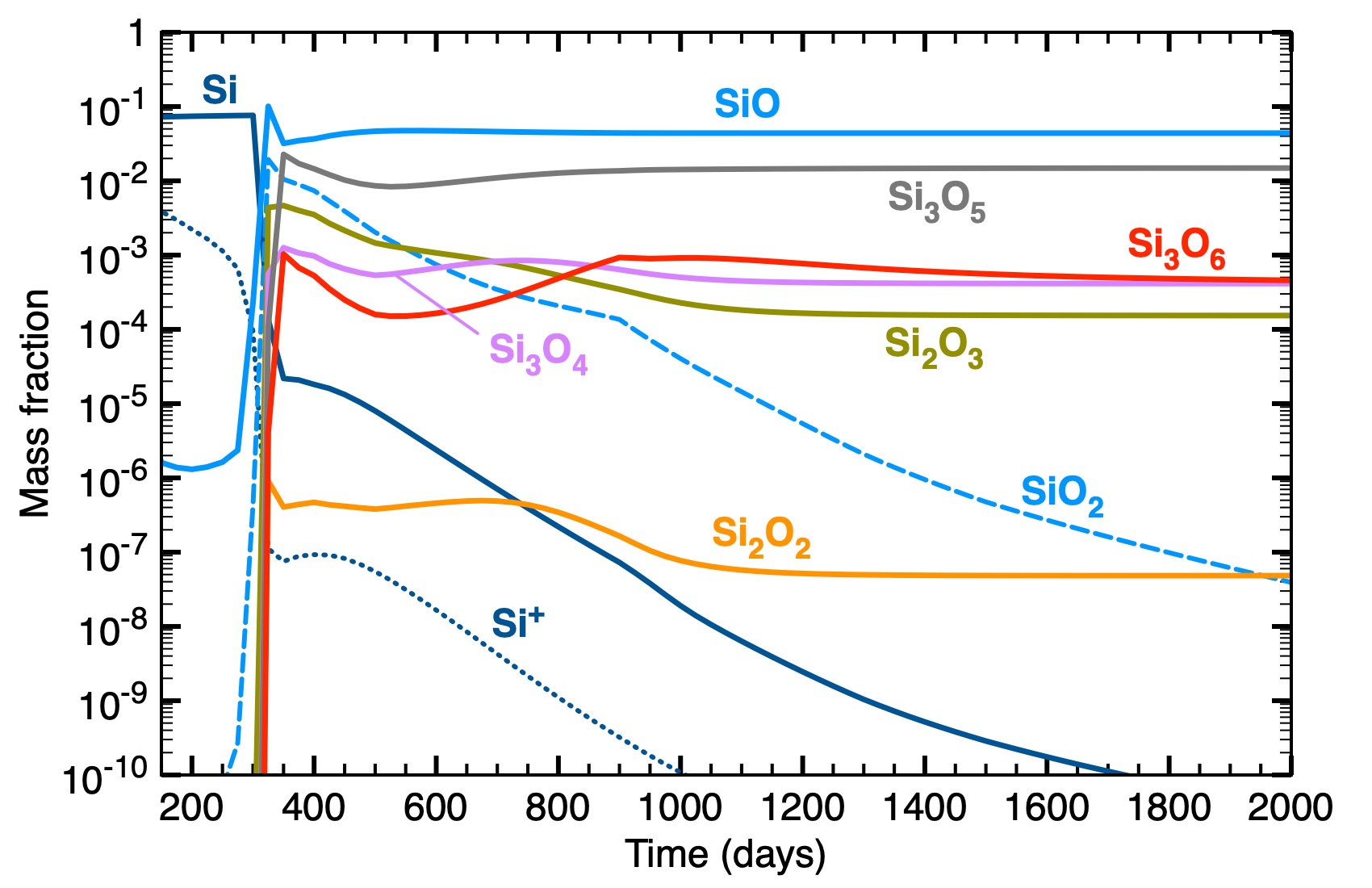}
  \includegraphics[width=\linewidth]{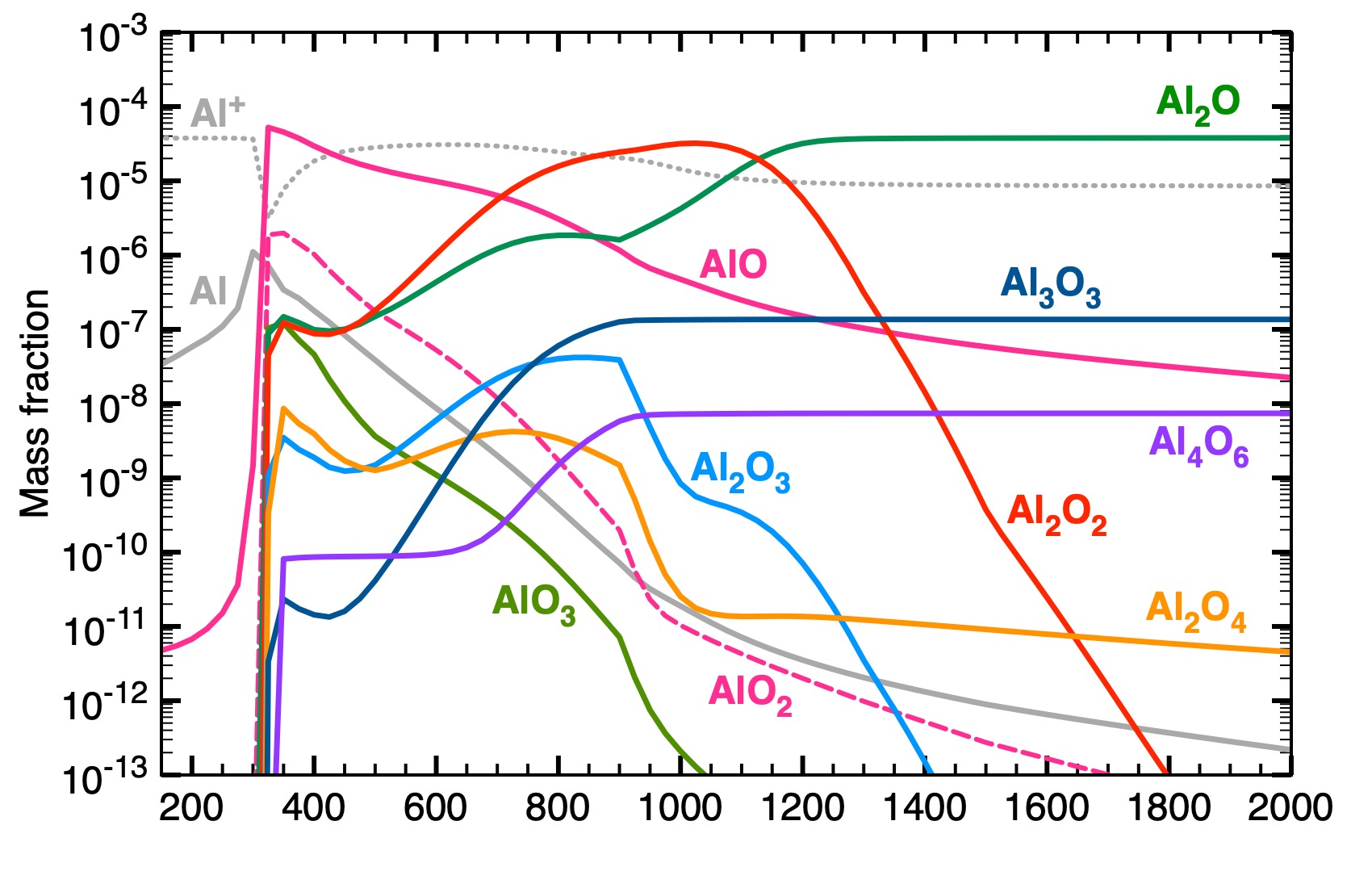}
\caption{Mass fractions of dust clusters formed in the O/Si/Mg region at $z=2.15$~\Ms\ versus post-explosion time for the Standard Case (zone mass = $5.42 \times 10^{-3}$ \Ms). Top: silicates; Middle: silica (quartz); Bottom: alumina.}
\label{fig7}
\end{figure}
% ---------------------------------------------------------

We present results for the zone position $z=2.15$~\Ms, which shows high O, Si, and Mg initial mass fractions, as well as a P maximal initial mass fraction. The position has thus optimal conditions to conduce to silicate formation and results are shown for molecules in Fig.~\ref{fig6} and dust clusters in Fig.~\ref{fig7}. We see in Fig.~\ref{fig6} that O is over-abundant at early times so that O$_2$ forms according the radiative association 
\begin{equation}
\label{O21}
O + O \rightarrow O_{2} + h\nu.
\end{equation}
The formation of O$_2$ continues mainly through the reaction
\begin{equation}
\label{O22}
MgO + O \rightarrow O_{2} + Mg,
\end{equation}
and to a lesser extent, 
\begin{equation}
\label{O23}
SiO + O \rightarrow O_{2} + Si,
\end{equation}
while the reverse processes provide the prominent destruction routes. When the gas temperatures drops below $\sim {\rm 3500 K}$, molecular formation is boosted over a 40 day time-span at $\sim$~day~300, with O$_2$ capturing most of atomic O and controlling the molecular phase. Indeed, molecules like SiO, SO$_2$, CO$_2$, and CO form in large quantities from oxidation reactions of atomic species like Si or C with O$_2$. For example, SO forms out of 
\begin{equation}
\label{SO}
S + O_2 \rightarrow SO + O,
\end{equation}
but is efficiently destroyed by its conversion into SO$_2$ through 
\begin{equation}
\label{SO2}
SO + O_2 \rightarrow SO_2 + O. 
\end{equation}

Interestingly, PO and PO$_2$ are the only abundant P-bearing species before day 1000 experiencing similar formation processes 
\begin{equation}
\label{PO}
P + O_2 \rightarrow PO + O,
\end{equation}
and
\begin{equation}
\label{PO2}
PO + O_2 \rightarrow PO_2 + O. 
\end{equation}
While Reaction \ref{PO} is slightly exothermic, Reaction~\ref{PO2} is slightly endothermic and thus proceeds at high gas temperatures. The reaction stops at around day~1000 when the gas temperature drops below $\sim 800$~K, resulting in a decrease in PO$_2$ mass fraction. 

Finally, some MgS, SiS and CaS form at this position, albeit in smaller quantities.  While MgS and CaS form out of radiative association over the entire time-span, SiS formation relates to SiO and SO$_2$  through
\begin{equation}
\label{SiS1}
S + SiO  \rightarrow SiS + O,
\end{equation}
and
\begin{equation}
\label{SiS2}
Si + SO_2 \rightarrow SiS + O_2. 
\end{equation}

As for dust clusters, inspection of Fig.~\ref{fig7} reveals the efficient synthesis of silicates through various intermediates as soon as the ejecta gas reaches the molecular regime with O$_2$ formation. The enstatite monomer \ens\ mainly forms from the reaction sequence

%---------------------------------- SIL SEQUENCES ---------------------------
%
\begin{table}[!h]
\label{ens1}
\resizebox{\columnwidth}{!}{\begin{tabular}{ccc ccc ccc}
MgO &+& SiO$_2$ & $\xlongrightarrow{\text{R13}}$ & & & & & \\
   & & & &MgSiO  &$\xrightarrow{\text{R14}}$ &MgSiO$_2$&$ \xrightarrow{\text{R12}}$ &MgSiO$_3$, \\
MgO$_2$ &+&SiO &$\xlongrightarrow{\text{R7}}$ & & & & & \\
\end{tabular}}
\end{table}
%\end{equation}
%

while for the \fors\ monomer, the reaction sequence is as follows 

%
%\begin{equation}
\begin{table}[!h]
\label{fors1}
\resizebox{\columnwidth}{!}{\begin{tabular}{ccc ccc ccc}
%\resizebox{\linewidth}{!}{
%\begin{array}{ccc ccc ccc}
MgSiO$_2$ &+ & MgO$_2$ & $\xlongrightarrow[]{\text{R16}}$&  &&  & & \\
   & & & & Mg$_2$SiO$_2$&  $\xrightarrow{\text{R18}}$& Mg$_2$SiO$_3$ &$\xrightarrow{\text{R21}}$& Mg$_2$SiO$_4$, \\
 MgSiO$_2$ &+& MgO &$\xlongrightarrow[]{\text{R15}}$ & & & & & \\
%\end{array}}
%\end{equation}
\end{tabular}}
\end{table}
%-----------------------------------------------------------------------------
where the reaction labels correspond to processes listed in Table~\ref{siliform}. At the high gas temperatures around day~300, the moderate endothermicities of reactions R7, R12, and R16 are easily overcome. Finally, the dimers of enstatite and forsterite are formed from reactions R22 and R23 at collisional rate. The \ens\ dimers are by far the most abundant in the zone by at least two orders of magnitude compared to \fors\ dimers. These two sequences involve simple and abundant molecules that induce the efficient formation of MgSiO and growth of larger intermediate clusters on short time scales of a few dozen days. 

We see from Fig.~\ref{fig7} that next in mass fraction are clusters of silica, with large mass fraction of Si$_3$O$_5$, followed by Si$_3$O$_6$ (the trimer of the SiO$_2$ unit), and Si$_3$O$_4$. For these three species, the energetically favorable structures are rhombus chains, with adjacent rhombuses perpendicular to each other \citep{chu01,lu03}. While Si$_3$O$_5$ has one peripheral oxygen atom attached to one rhombus, Si$_3$O$_6$ has one peripheral oxygen on each rhombus, implying these two molecules are reactive species that foster chain growth. As mentioned in \S~\ref{sili}, the sequence for quartz cluster formation starts with the reaction of two SiO$_2$ molecules to give Si$_2$O$_3$ clusters. Subsequent additions of SiO$_2$, starting with Si$_2$O$_3$, lead to Si$_3$O$_5$ and Si$_3$O$_6$, while reaction of the latter with atomic O leads back to Si$_3$O$_5$. All these processes are exothermic and proceed quickly ensuring these rhombic chains can easily grow by SiO and SiO$_2$ addition. 
%the double oxygen bridged chains with adjacent rhombuses oriented perpendicular to one another 

Finally, the least abundant dust clusters are of alumina, as seen from Fig.~\ref{fig7}. The most abundant species is the linear Al$_2$O molecule followed by the kyte-shaped cluster Al$_3$O$_3$, and the \alu\ dimer. The formation sequence starts with two AlO molecules reacting together to give Al$_2$O, which reacts with O$_2$ to form Al$_2$O$_2$, of rhombus structure \citep{arm19}. The latter further reacts with O$_2$ to give the kyte-shaped \alu\ cluster. Once formed, \alu\ reacts with O$_3$ to give Al$_2$O$_4$, which also forms from Al$_2$O$_2$ reacting with AlO$_3$. Finally, the \alu\ dimer mainly forms out of \alu\ reacting with Al$_2$O$_4$. The most abundant final products Al$_2$O, Al$_3$O$_3$ and Al$_4$O$_6$ have fragmentation energies above $\sim$ 5 eV and are therefore stable, the most stable being Al$_4$O$_6$ with its cage structure \citep{arm19}. However, it appears that for our Standard Case, the mass fractions of small alumina clusters remain modest compared to other dust clusters. 

\subsubsection{Total molecule and cluster masses}
\label{innertot}
%_____________________________________________________________
%                              FIG 8 - O/Si/Mg MOL+DUST mass
%-------------------------------------------------------------
\begin{figure*}
   \centering
\begin{subfigure}{0.48\textwidth}
        \includegraphics[width=1.\textwidth]{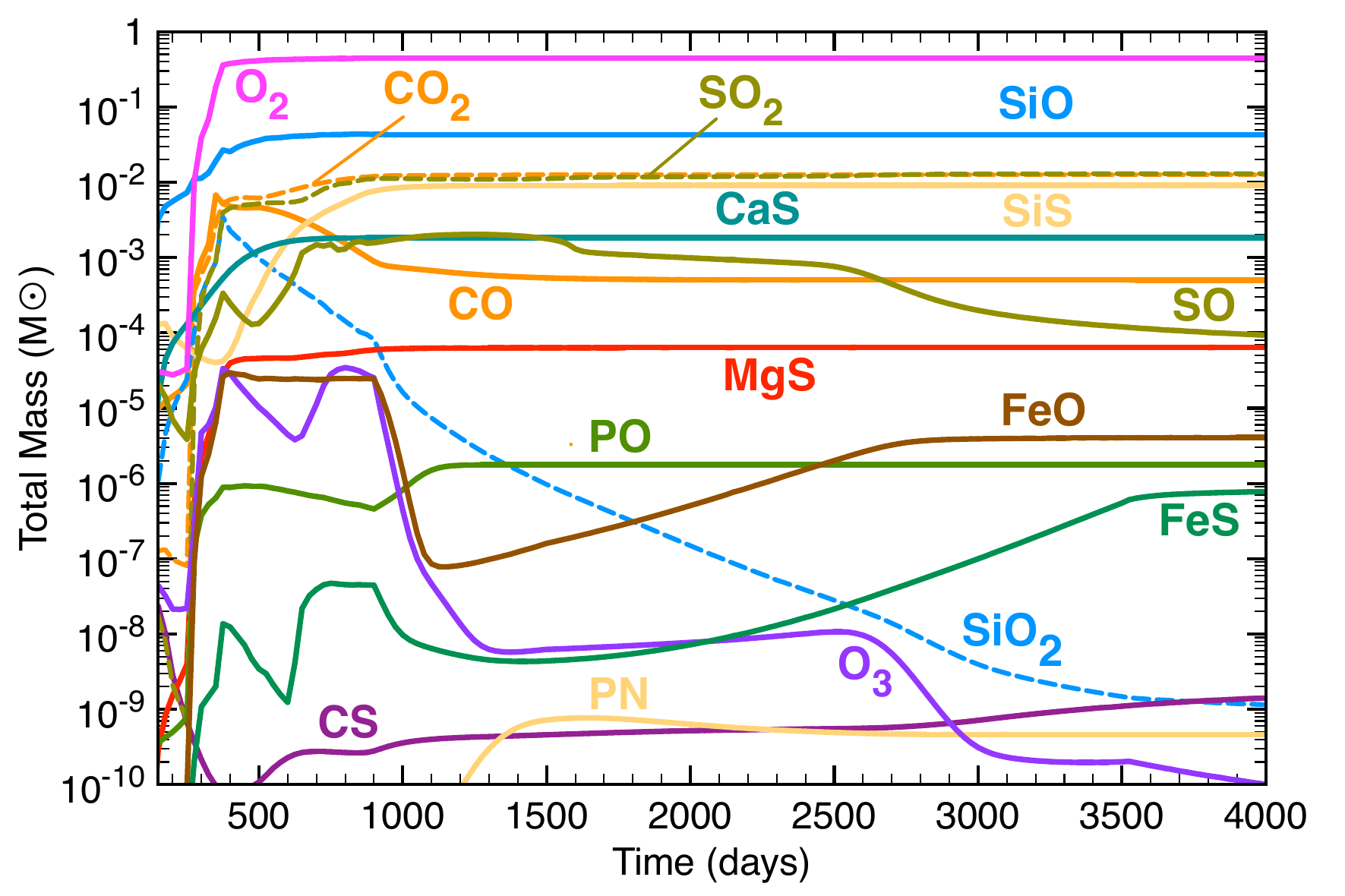}
%        \caption{ }
        \end{subfigure}
\hfill
\begin{subfigure}{0.48\textwidth}
        \includegraphics[width=1.\textwidth]{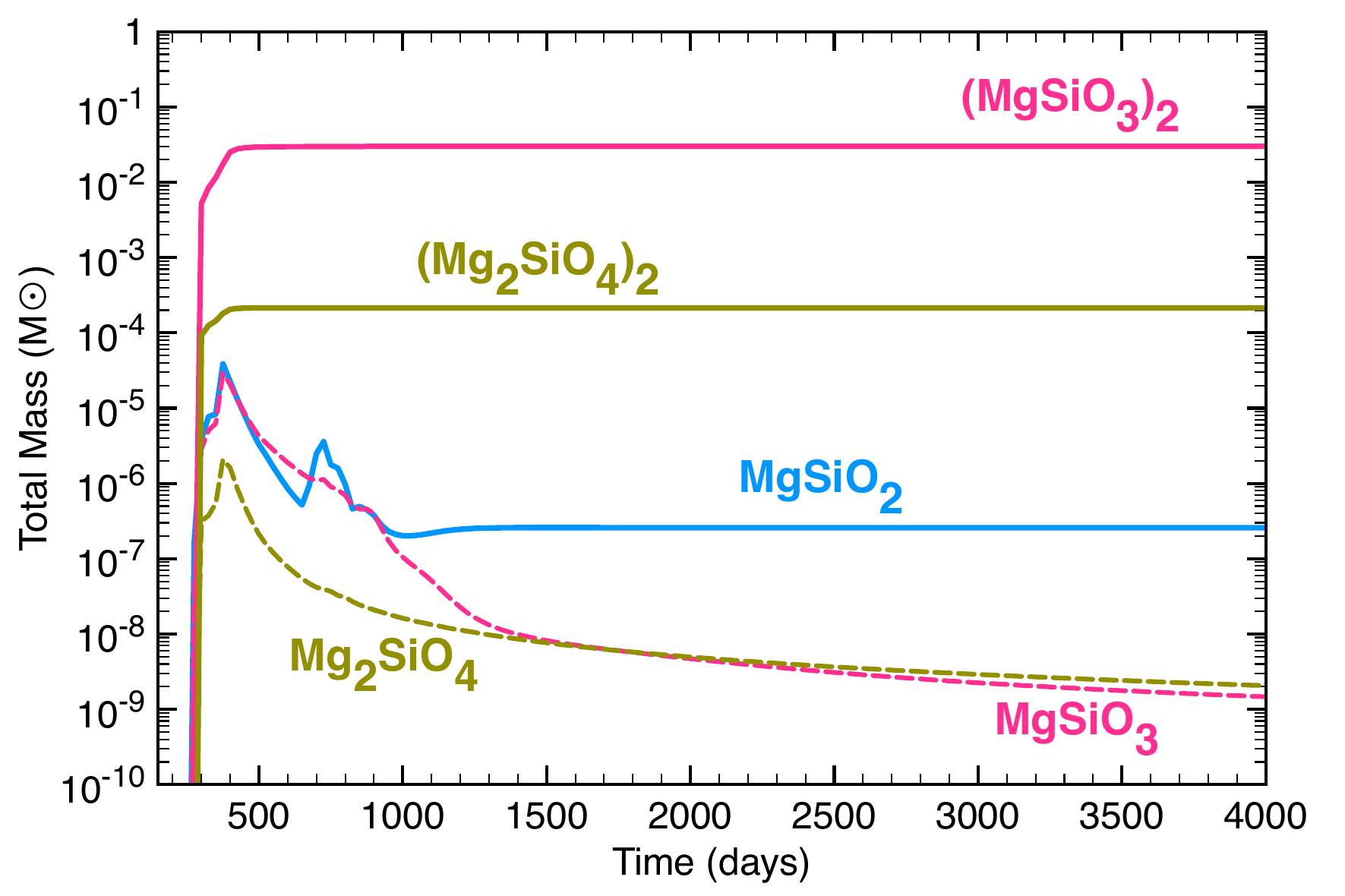}
%         \caption{ }
        \end{subfigure}
\hfill
\begin{subfigure}{0.48\textwidth}
        \includegraphics[width=1.\textwidth]{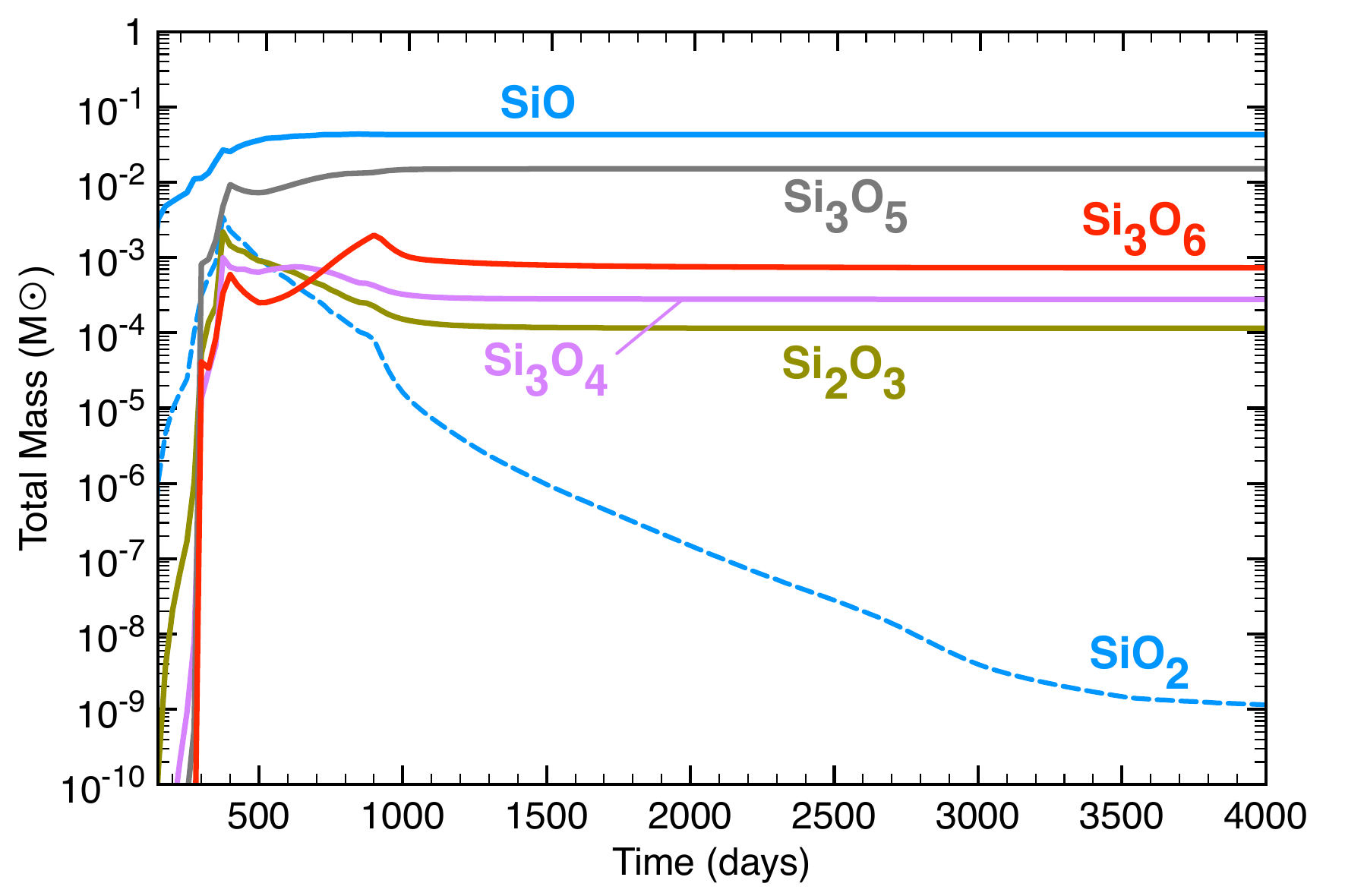}
%         \caption{ }
        \end{subfigure}
\hfill
\begin{subfigure}{0.48\textwidth}
        \includegraphics[width=1.\textwidth]{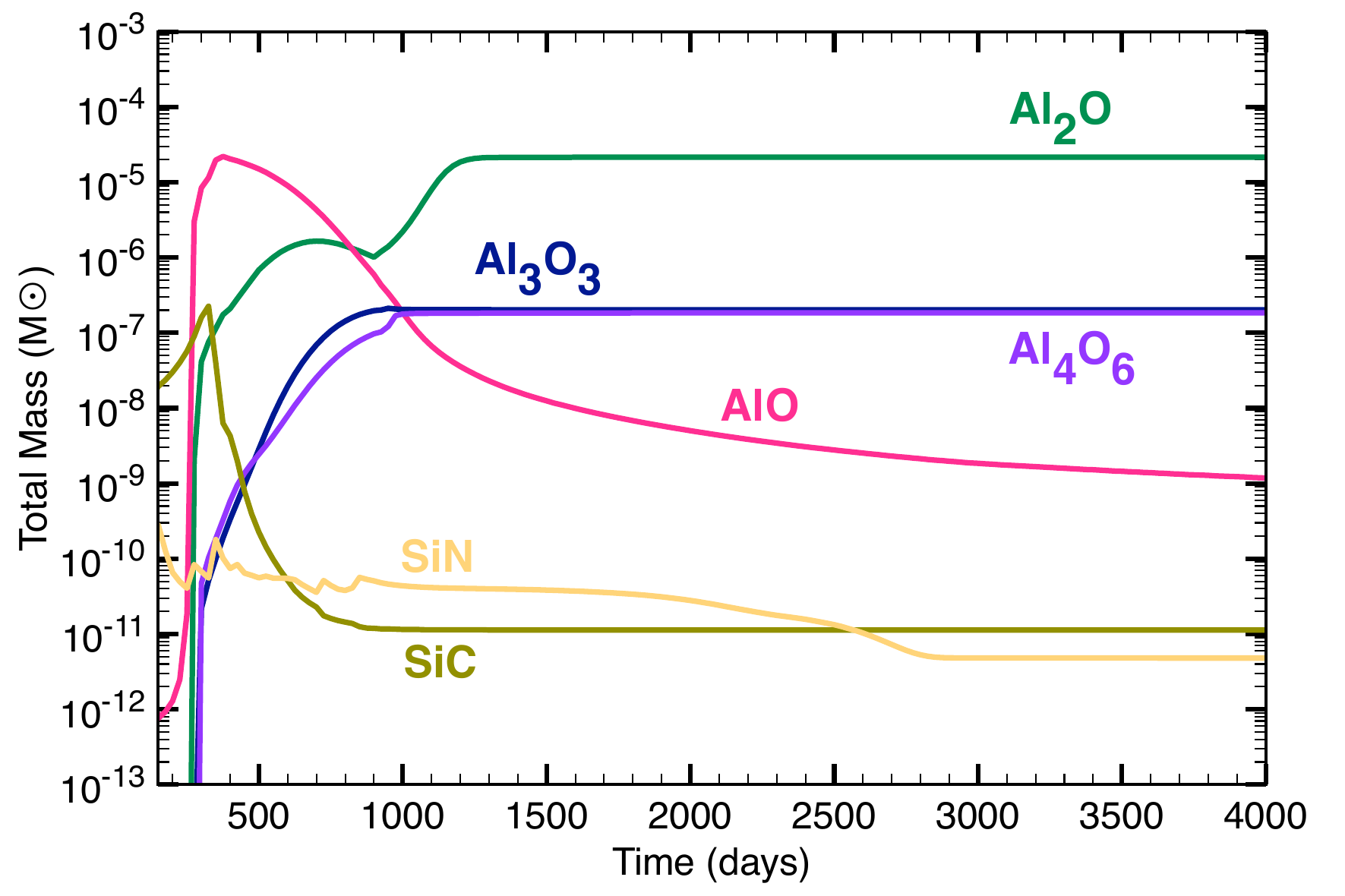}
%         \caption{ }
        \end{subfigure}
 \caption{Total mass of molecules and dust clusters produced in the O/Si/Mg region as a function of post-explosion time for the Standard Case. Top left: molecules; Top right: silicates; Bottom Left: silica; Bottom right: alumina. The region mass is 0.729~\Ms.}
       \label{fig8}
  \end{figure*}
% -----------------------------------------------------
%_____________________________________________________________
%                              FIG 9 - O/Si/Mg mol + silicate MASS DISTRIB in region 
%-------------------------------------------------------------
\begin{figure} 
  \centering
  \includegraphics[width=1.0\linewidth]{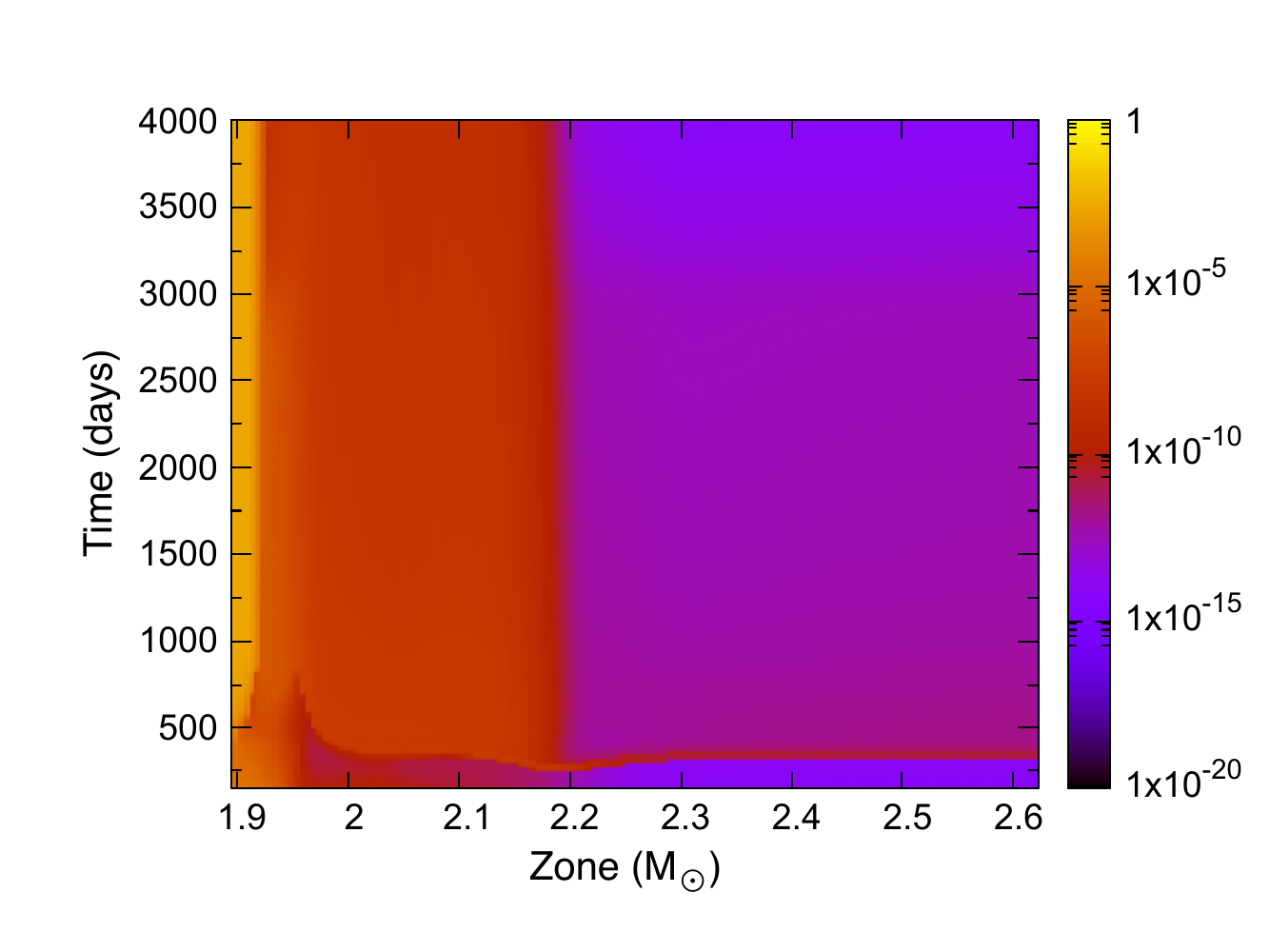}
  \includegraphics[width=1.0\linewidth]{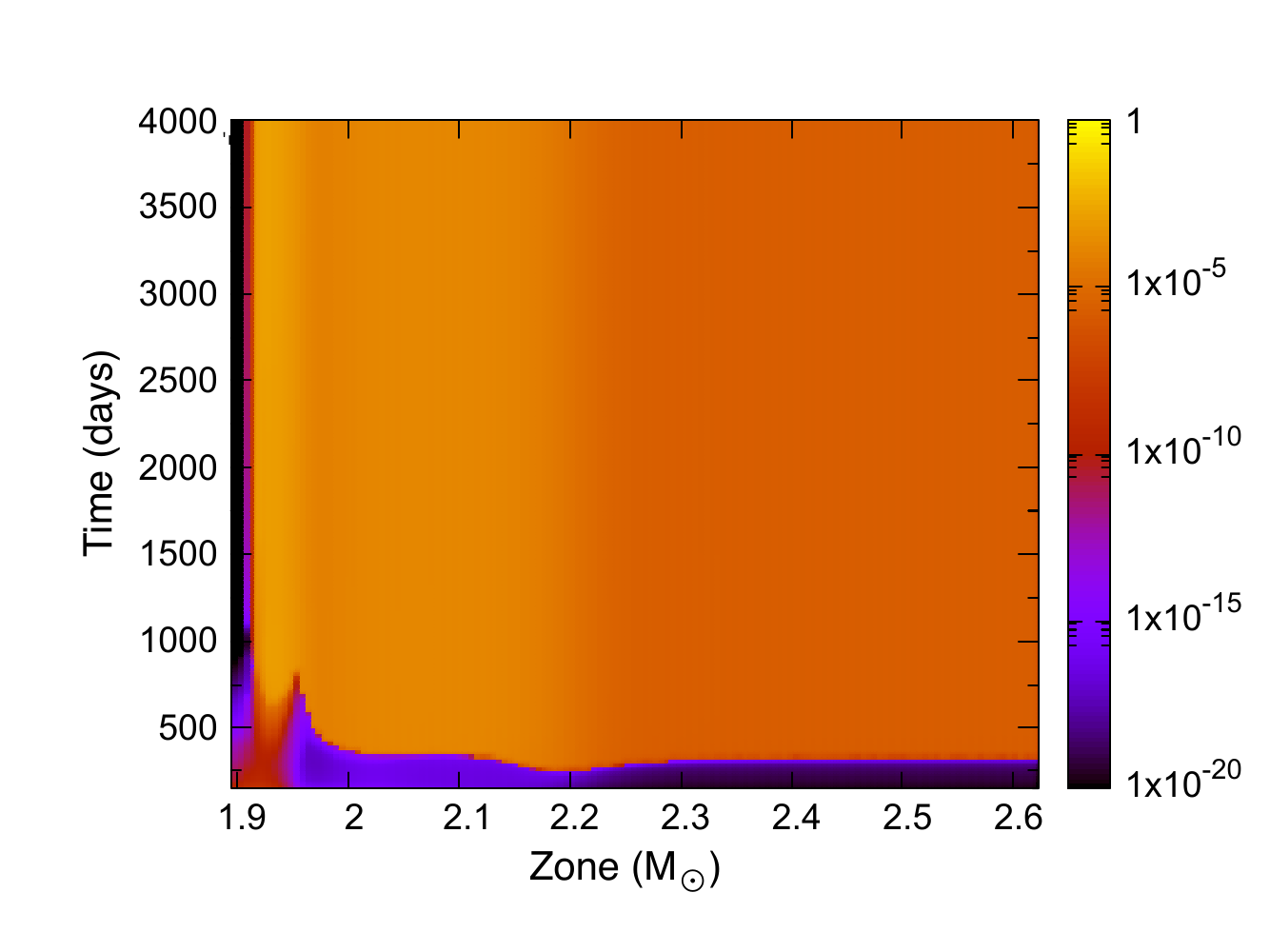}
  \includegraphics[width=1.0\linewidth]{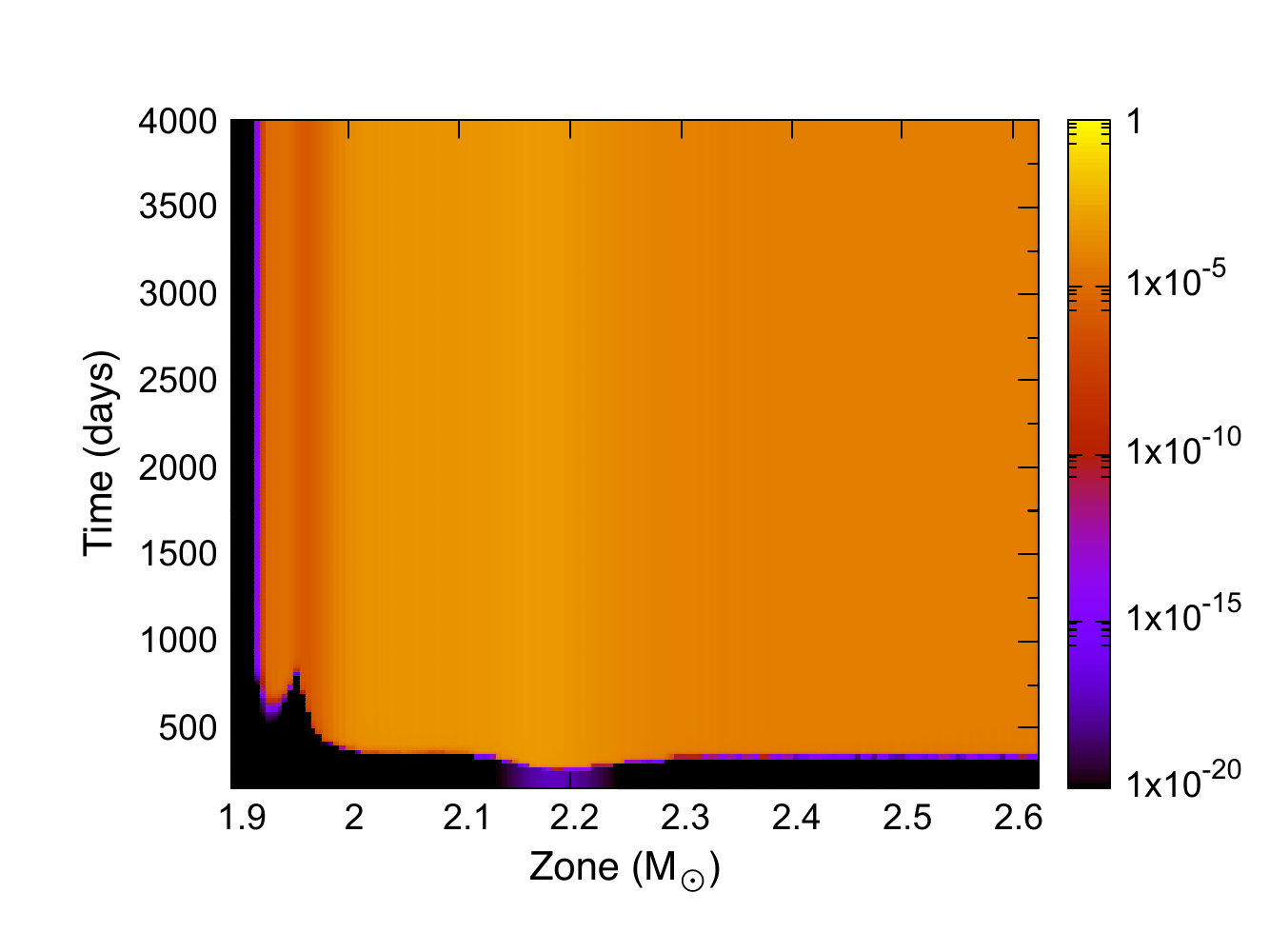}
\caption{Mass maps of molecules and enstatite dimers produced in the O/Si/Mg region as a function of time and position $z$ in the region: From Top to Bottom:  SiS, SO$_2$ and (\ens)$_2$. }
\label{fig9}
\end{figure}
% -----------------------------------------------------------------

The total masses of molecules and dust clusters formed over the 145 zones of the O/Si/Mg region are presented in Fig.~\ref{fig8} and values at day 4000 are gathered in Table~\ref{finmas}. The formation processes for the various molecules and dust clusters at the zone position $z=2.15$~\Ms\ also operate at other positions in the O/Si/Mg region, albeit with varying efficiencies owing to the different gas conditions and initial elemental compositions. 

It emerges that molecules and dust cluster formation is not homogenous in the region as seen from Fig.~\ref{fig9}, where maps of the masses for SiS, SO$_2$, and enstatite dimers are shown as a function of  time and zone position in the region. According to Fig.~\ref{fig8}, SiS does form with quite large masses in the region. However, we see from Fig.~\ref{fig9} that its formation locus is restricted to zones spanning from the region inner boundary to $\sim 1.92$~\Ms, where the mass fractions of atomic Si and S are still high. The formation of SiS there is residual of the large production of SiS in the Si/S/Ca region, and the molecule cannot be considered as typical of the O/Si/Mg~region. By contrast, the formation of SO$_2$ and (\ens)$_2$ extends quite homogeneously over the entire region although favoured zones exist for optimal synthesis. For SO$_2$, the maximum formation occurs between 1.92~\Ms\ and 1.96~\Ms\ where atomic S and O are abundant before the gradual decrease of atomic sulphur. The formation efficiency gradually drops over the entire region for $z> 1,96$~\Ms. For enstatite dimers, formation is precluded in the very inner zones where SiS is at maximum, while it is favoured from $z=1.96$~ \Ms\ and  $z=2.2$~ \Ms. The maximum formation efficiency lies between 2.1~\Ms\ and 2.2~\Ms, where both atomic O, Si and Mg are simultaneously abundant, and drops after  $z=2.2$~ \Ms\ to stay constant for the rest of the region. 
%A similar case applies to CaS although the zone of maximum formation extends until $\sim 1.95$~\Ms, because of the combined presence of atomic Ca and S, while the mass fraction of atomic oxygen starts increasing significantly. 

From Table~\ref{finmas}, we see the O/Si/Mg region is extremely efficient at forming molecules and dust clusters. The prominent molecules are O$_2$, SiO, SO$_2$, and CO$_2$, and the total molecular mass is $0,525$~\Ms, which represents $\sim 72$~\% of the region's mass. As for dust clusters, the region forms essentially silicate dimers and silica trimers for a total mass of $0.046$~\Ms, equivalent to $\sim 6.3$~\% of the region's mass. 
%
% --------------- O/C/Mg outer O zone ---------------------
%
\subsection{The O/C/Mg region (Outer Oxygen region)}
\label{exO}
%_____________________________________________________________
%                              FIG 10 - MOL+DUST in O/C/Mg zone 2.7 Msun
%------------------------------------------------------------- 
\begin{figure} 
  \centering
  \includegraphics[width=\linewidth]{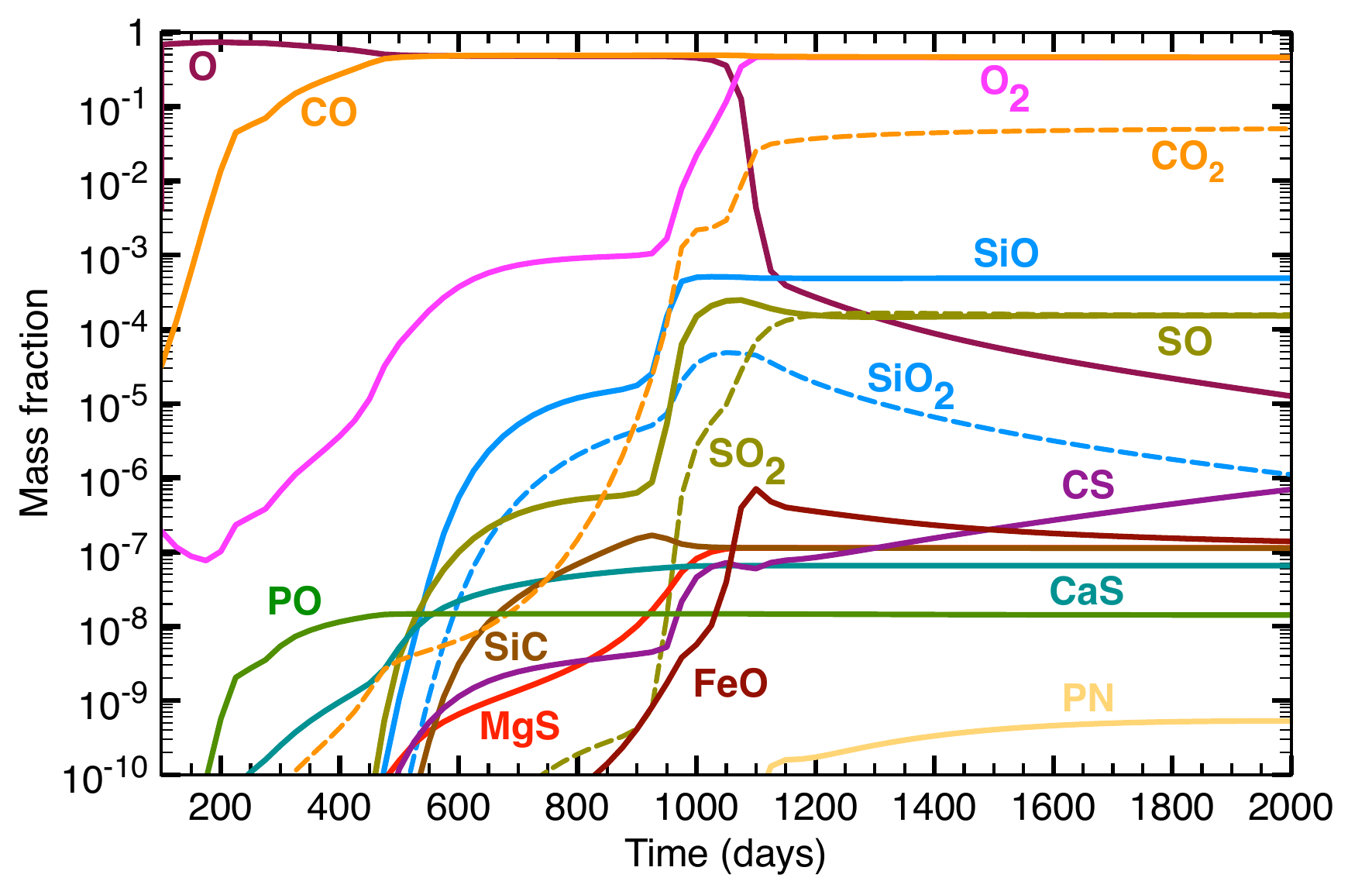}
  \includegraphics[width=\linewidth]{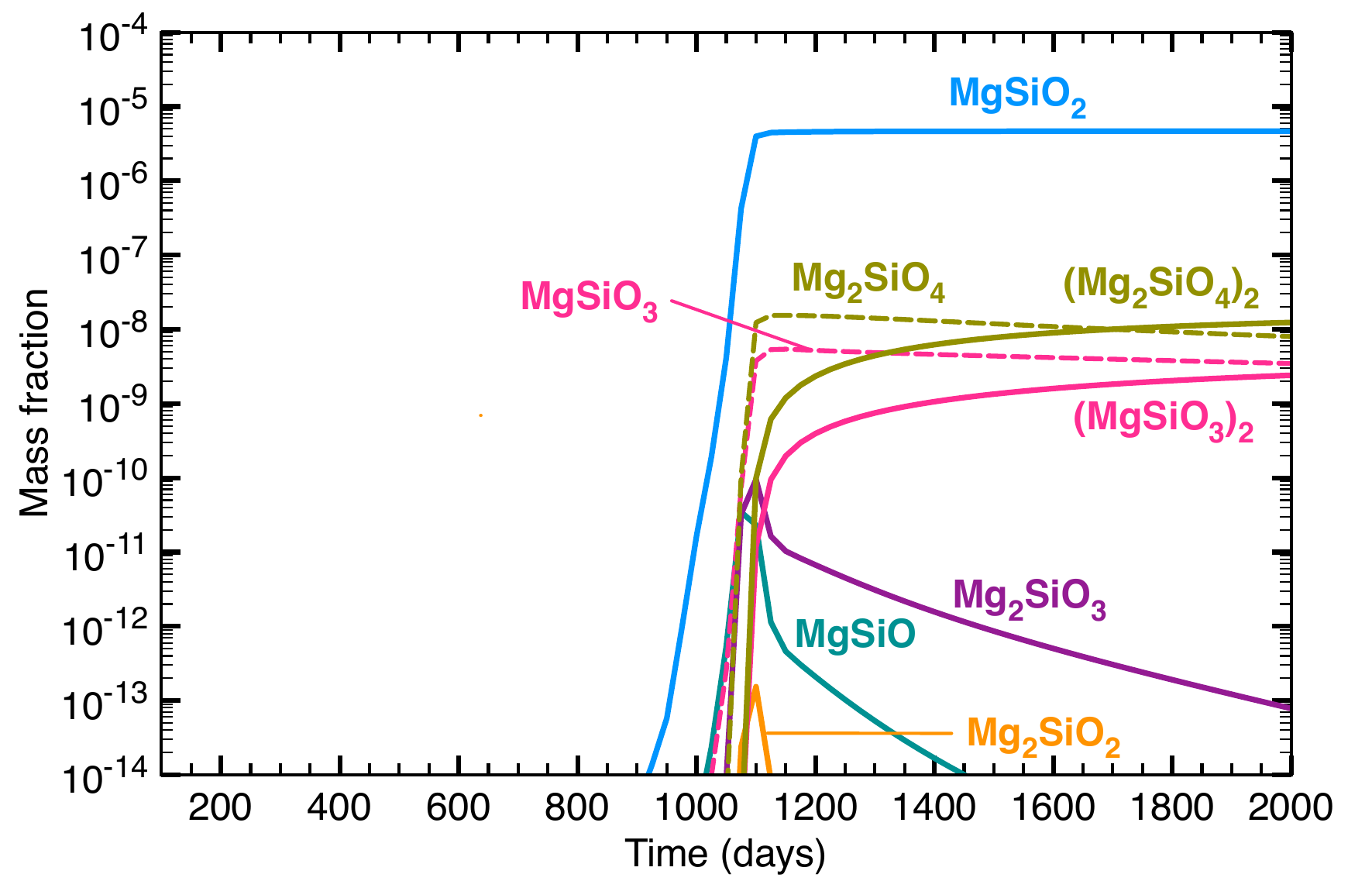}
\caption{Mass fractions of molecules and dust clusters formed in the O/C/Mg region at $z=2.7$~\Ms\ versus post-explosion time for the Standard Case (zone mass = $5.42 \times 10^{-3}$ \Ms). Top: molecules; Bottom: silicates.}
\label{fig10}
\end{figure}
% ---------------------------------------------------------
The region is characterised by high O and C initial mass fractions while Mg is still present and the quantity of Si drastically decreases compared to the Inner Oxygen region. We study the chemistry in the zone located at $z=2.7$~\Ms\ for which the mass fractions of molecules and silicate clusters are presented in~Fig.~\ref{fig10}. Because of the large initial content of atomic C, the formation of CO controls the overall chemistry and freezes other molecular synthesis until day 900. Indeed, until day 300, CO is formed out of the fast, temperature-independent charge-exchange reaction 
\begin{equation}
\label{F1}
CO^+ + O\rightarrow CO + O^+, 
\end{equation}
and destroyed directly by Compton electrons mainly through the process
\begin{equation}
\label{D1}
CO + e_C \rightarrow CO^+ + e^- + e_C. 
\end{equation}
The destruction of CO by atomic O,
\begin{equation}
\label{D2}
CO + O\rightarrow O_2 + C,
\end{equation}
gradually grows in strength with time to insure the efficient synthesis of O$_2$ at the expense of CO. At day 700, Reaction~\ref{D2} is the dominant formation process for O$_2$ while it is destroyed by the formation of SiO following
\begin{equation}
\label{D3}
O_2 + Si \rightarrow SiO + O. 
\end{equation}
After day 900, the gas then switches to a molecular formation regime controlled by reactions with O$_2$. At that time, dust clusters start forming through the processes discussed in \S~\ref{inO}. However, we see from Fig.~\ref{fig10} that the mass fraction of silicates is quite low ($\sim 10^{-8}$) at day 4000. Similarly, the mass fraction of silica is $\sim 3\times 10^{-5}$ when that of alumina clusters is $\sim 1.3\times 10^{-7}$.

%_____________________________________________________________
%                              FIG 11 - MOL+DUST in O/C/Mg region
%------------------------------------------------------------- 
\begin{figure} 
  \centering
  \includegraphics[width=1.0\linewidth]{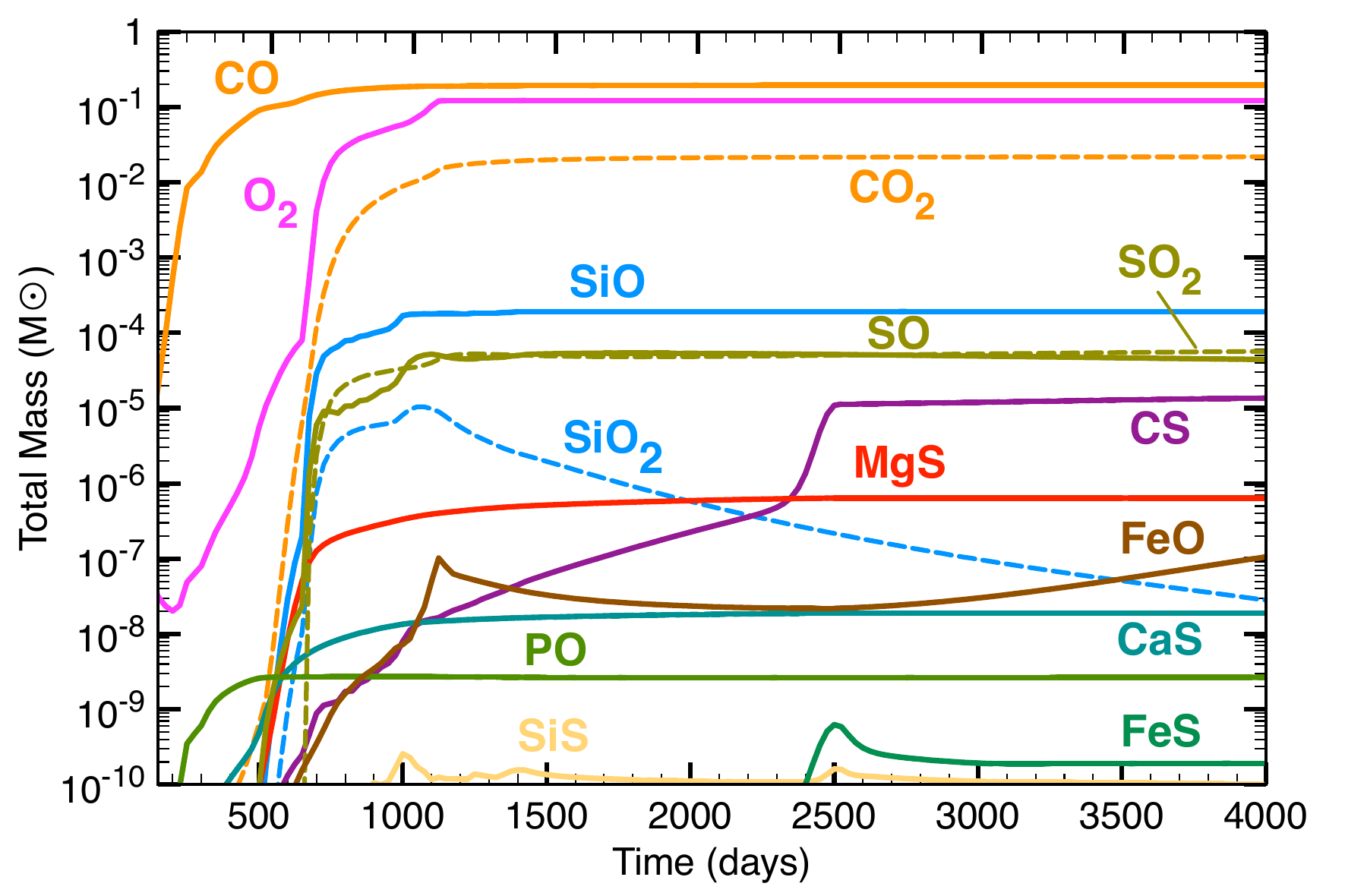}
\caption{Total mass of molecules produced in the O/C/Mg region as a function of post-explosion time for the Standard Case. The region mass is 0.39~\Ms.}
\label{fig11}
\end{figure}
%------------------------------------------------------------- 

The masses of molecules formed over the entire region are shown in Fig.~\ref{fig11}. The prominent species produced in the region are CO, O$_2$ and CO$_2$. This region synthesises the largest masses of CO and CO$_2$ in the ejecta as already found in previous studies \citep[e.g.][]{cher09}, with CO mass growing from day 150 to 400 from $\sim 10^{-4}$~\Ms\ to over $10^{-1}$~\Ms\ in good agreement with existing ALMA and recent JWST observations \citep{kam13,med25}. Interestingly, the region also produces some SiO, SO, and SO$_2$ with masses at day 4000 of $\sim 2\times 10^{-4}$~\Ms, $4\times 10^{-5}$~\Ms\ and $5\times 10^{-5}$~\Ms, respectively. 

However, the region contributes modest amounts of dust clusters, the most abundant being silica, followed by alumina and silicates. As in the O/Si/Mg region, no carbon dust clusters form. We find that molecules represent 87 \% of the region's mass, with a total mass of 0,339 \Ms, while the dust clusters have a modest mass of $1.46\times 10^{-5}$~\Ms\ $\equiv$ to $3,74 \times 10^{-3}$~\% of the region's mass. Although the region has the greatest efficiency at synthesising molecules, it is second to the O/Si/Mg region in terms of molecule masses because the region is of smaller size.  

%Along with the efficient synthesis of CO and {\rm CO$_2$} in this zone, the ion {\rm CO$^+$} forms at early times between 100 and 300 days post-explosion to drastically decrease afterwards. This result is consistent with observations of SN1987A where the first overtone emission of {\rm CO$^+$} at 2.26 \mic~was possibly detected between day 192 and 255 \citep{spyro88,meik89}, along with emission of the first overtone of CO. They derive a [CO + {\rm CO$^+$}] mass of $~ 4\times10^{-5}$ \Ms, in agreement with our results (see Fig. 1000), despite our {\rm CO$^+$}/CO ratio is smaller than that derived from observations by a factor $10^3$. 

\subsection{The He/C/N region (Carbon zone)}

The He/C/N region expands from $3.013$~\Ms\ to $4.141$~\Ms\ and is unique for four reasons. Firstly, it has a C/O ratio greater than $10$ until $z\sim 3.8$~\Ms. Secondly, the most abundant species is atomic He over the entire region. Thirdly, the region is characterised by a high nitrogen initial mass fraction, N being even more abundant than O and C from $z\sim3.8$~\Ms\ outwards. Finally, fluorine, F, is present at position $z\sim 3.05-3.6$~\Ms. The only possibility for a SN ejecta to form carbon dust resides in this region, since no carbon dust clusters form in the Si/S/Ca, the O/Mg/Si and the O/C/Mg regions, as we saw previously. 

We investigate the chemistry at position $z=3.3$~\Ms\ where ${\rm C/O\sim16}$ to highlight the processes conducive to carbon cluster formation. We then study the inner region depleted in N at $z=3.03$. Finally, we derive the masses of molecules and dust clusters over the entire region.  

%Carbon clusters and CN and CS are tracers of over-densities in the ejecta. In the Standard Case, most of the carbon rings C$_{20}$ form in the zones between 3.013 and 3.05 \Ms. 
\subsubsection{Molecules and dust clusters at $z=3.3$~\Ms.}
\label{3.3}
%_____________________________________________________________
%                              FIG 12 - MOL+DUST in He/C/N zone 3.3 Msun
%------------------------------------------------------------- 
\begin{figure} 
  \centering
  \includegraphics[width=\linewidth]{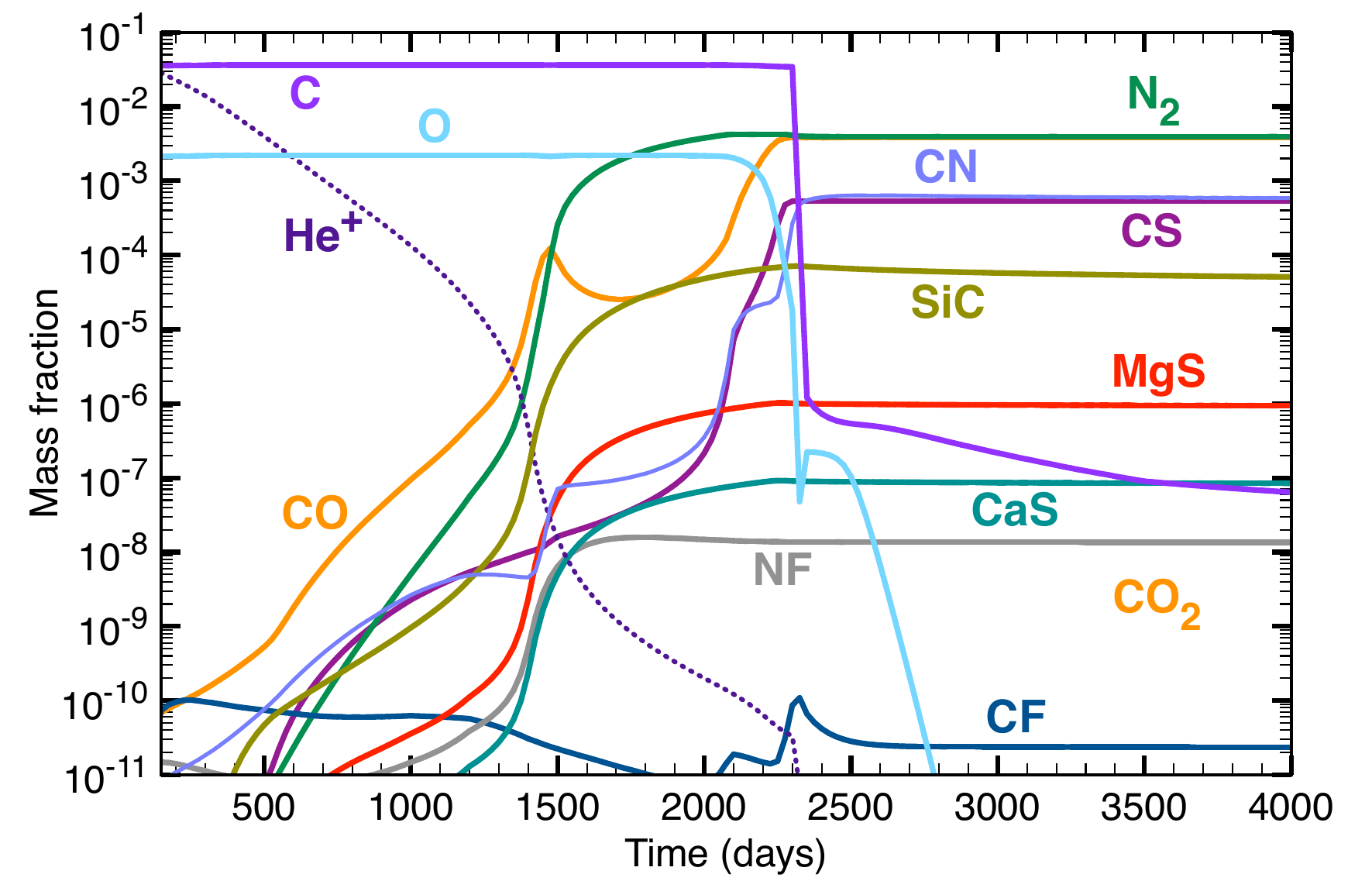}
  \includegraphics[width=\linewidth]{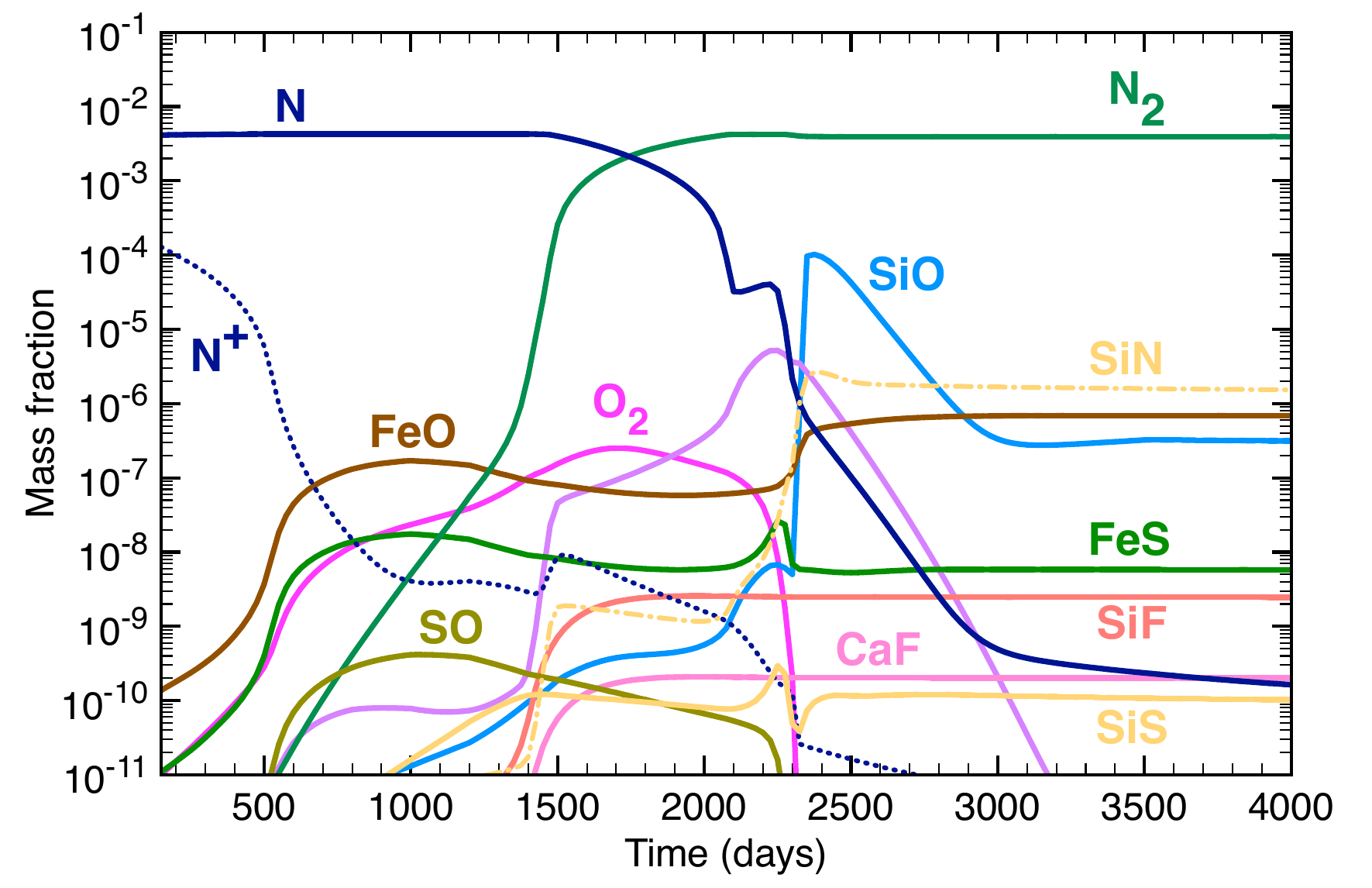}
   \includegraphics[width=\linewidth]{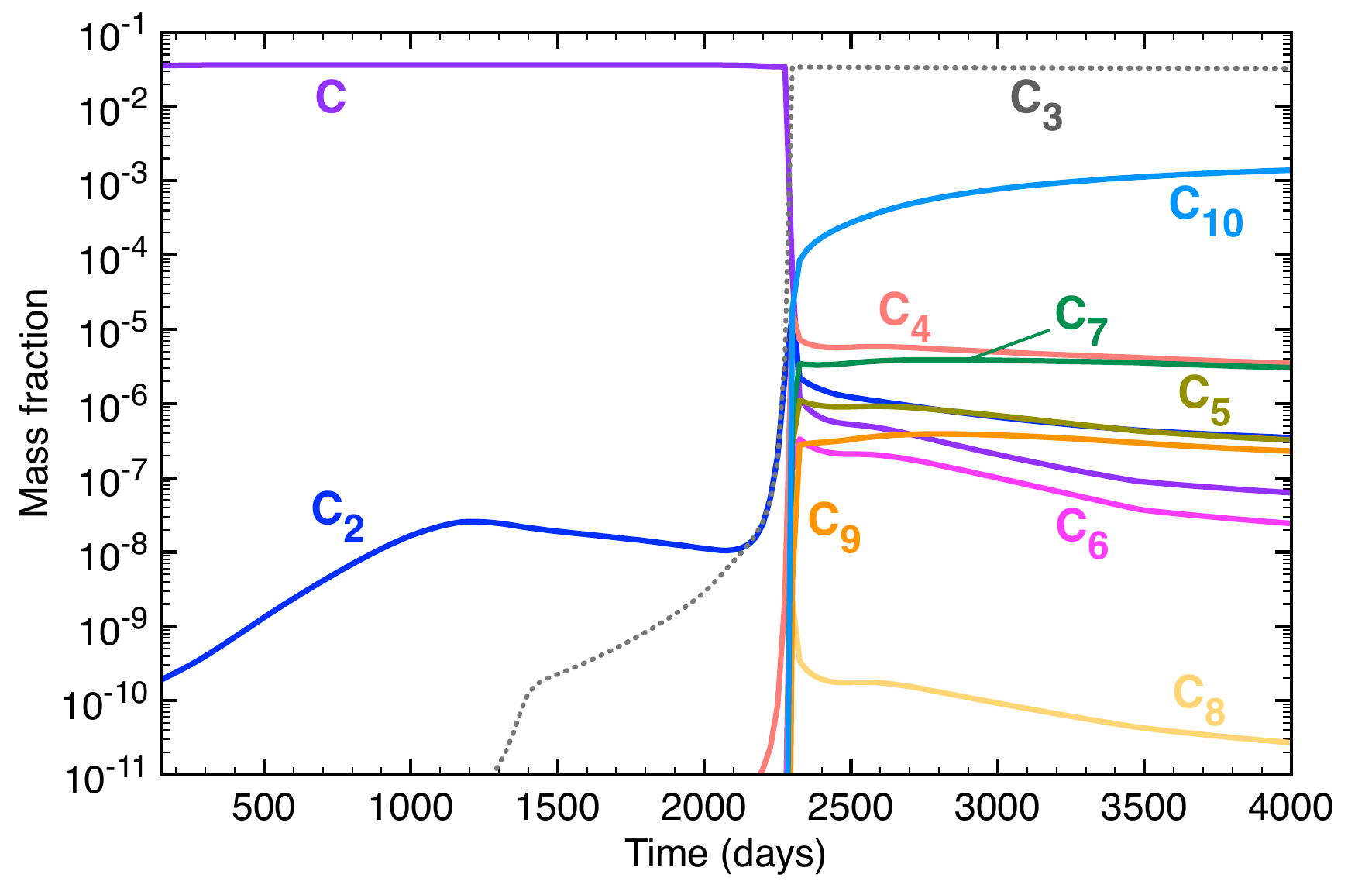}
\caption{Mass fractions of molecules and dust clusters formed in the He/C/N region at $z=3.3$~\Ms\ versus post-explosion time for the Standard Case (zone mass = $1.35 \times 10^{-2}$ \Ms). Top and Middle: molecules; Bottom: small carbon clusters.}
\label{fig12}
\end{figure}
% ---------------------------------------------------------
Results for molecules and carbon dust clusters are shown in Fig.~\ref{fig12} and reflect the non-straightforward chemistry of the zone. Inspection of Fig.~\ref{fig12} reveals three chemical regimes: the He$^+$-controlled regime until day 1400, the C/N/O chemistry regime from day 1400 to 2300, and finally, the regime dominated by carbon chain formation once CO has fully formed. 

Until day 1400, the formation of molecules (e.g. CO, N$_2$, C$_2$, and others) is hampered because of the strong destruction by \hep. The formation of the radicals CN and NO is then responsible for the destruction of CO and N$_2$ through the reaction
\begin{equation}
\label{CON2}
CO + N_2\rightarrow CN + NO,
\end{equation}
while reformation of CO occurs via the following reactions
\begin{equation}
\label{FCO}
C + NO\rightarrow CO + N,
\end{equation}
along with reaction of atomic C with O$_2$. The prominent formation process for N$_2$ is
\begin{equation}
\label{FN2}
N + CN \rightarrow C + N_2. 
\end{equation}
When CO synthesis is completed around day 2300, the excess in atomic carbon contributes the rapid formation of C$_2$ through the radiative association of two atomic C, while C$_3$ forms through reaction
\begin{equation}
\label{FC3}
C_2 + C_2 \rightarrow C_3 + C. 
\end{equation}
The carbon chain cycle is then activated whereby small chains grow through radiative association reactions and processes similar to Reaction~\ref{FC3}, as described by \citet{loi14}. 

Results for small carbon chains and rings up to C$_{10}$ are shown in Fig.~\ref{fig12}. The synthesis and growth of larger rings from C$_{11}$ to C$_{20}$ is hindered by the low gas temperatures at day 2300 (${\rm T_{gas} \sim 125 K}$) and no carbon rings larger than C$_{10}$ form in significant amount. As explained in \S~\ref{carbon}, the growth of rings larger than C$_{10}$ involves the insertion of C and C$_2$ units, implying ring opening. These processes have high energy barriers and only proceed at high enough gas temperatures. Furthermore, the ejecta gas number density after day 2300 is low ($n_{\rm {gas}} \le 1.5\times 10^6$~\cmc, which is at least three orders of magnitude less than ${\rm n_{gas}}$ at day~100). A low temperature combined to a low number density renders the growth of large carbon clusters unlikely, and a fortiori, the synthesis of carbon dust for the Standard Case of this study. Most of atomic carbon recycles into C$_3$ molecules as an end-product, which has a high mass fraction at day 4000. A high number density case for the carbon zone is studied and discussed in \S~\ref{HD}. 

\subsubsection{Molecules and dust clusters at $z=3.03$~\Ms.}
\label{303}
%_____________________________________________________________
%                              FIG 13 - MOL+DUST in He/C/N zone 3.03 Msun
%------------------------------------------------------------- 
\begin{figure} 
  \centering
  \includegraphics[width=\linewidth]{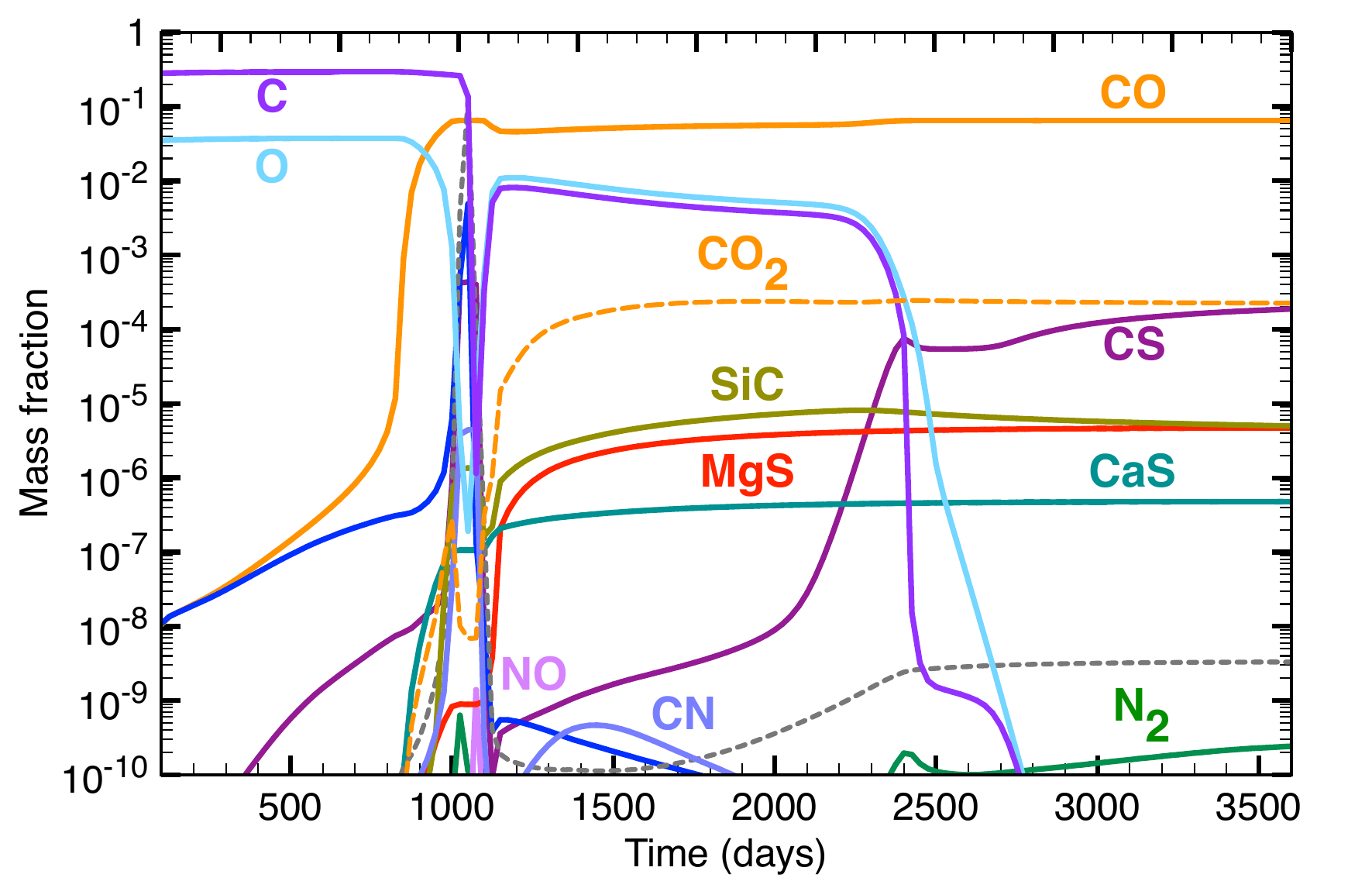}
  \includegraphics[width=\linewidth]{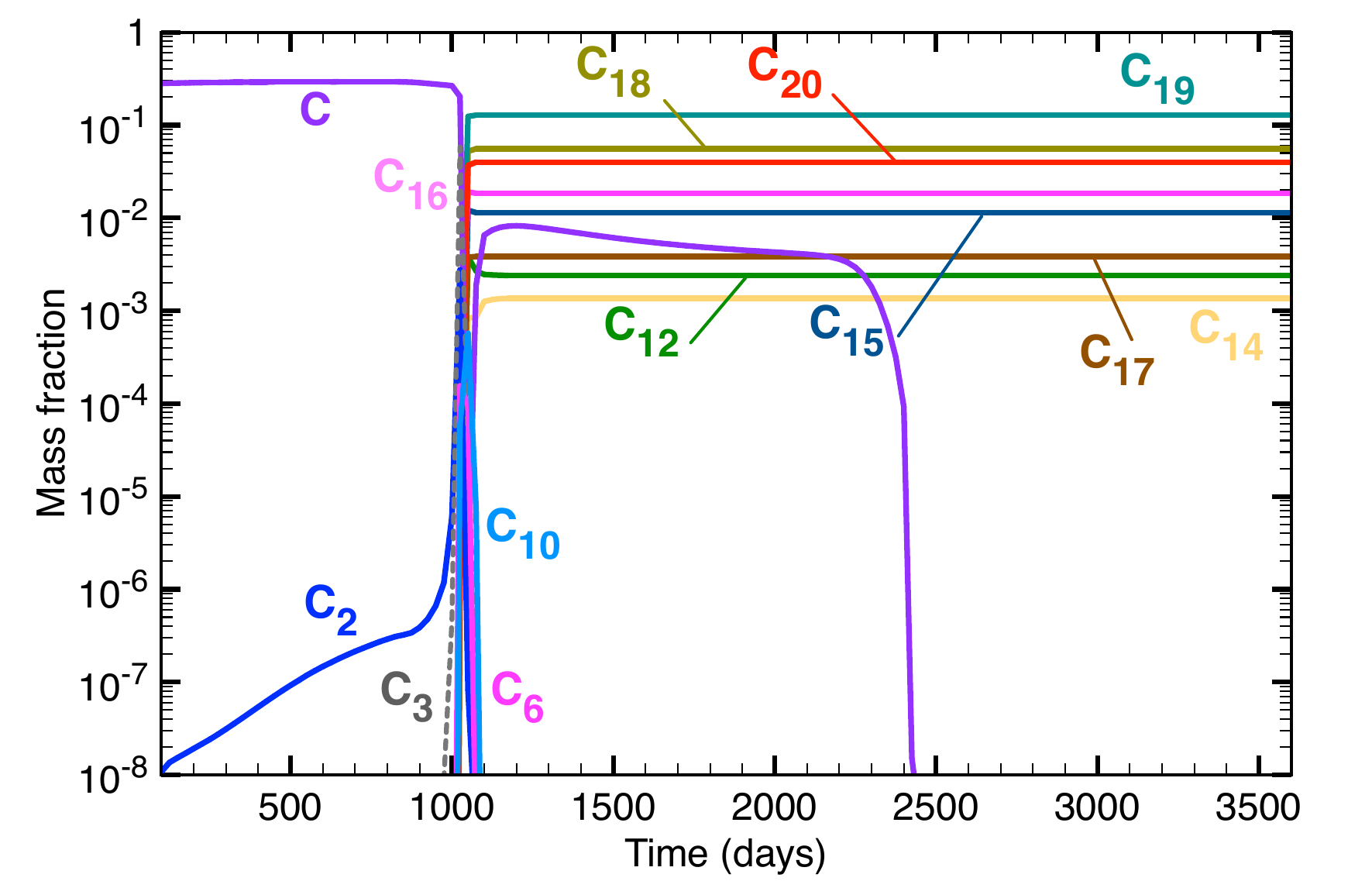}
\caption{Mass fractions of molecules and dust clusters formed in the He/C/N region at $z=3.03$~\Ms\ versus post-explosion time for the Standard Case (zone mass = $1.35 \times 10^{-2}$ \Ms). Top: molecules; Bottom: carbon clusters.}
\label{fig13}
\end{figure}
% --------------------------------------------------------- 

We see from Fig.~\ref{fig01} that there exists a zone range $z=3.013 -3.04$~\Ms, for which atomic C and O have large mass fractions while atomic He mass fraction has not yet reached full value and atomic N mass fraction is low. This small $z$~range corresponds to three zones in total and experiences a different chemistry owing to the peculiar chemical composition. 

We show in Fig.~\ref{fig13} the mass fractions for the main molecules and carbon clusters. Results are drastically different from those at $z=3.3$~\Ms\ in terms of timing for molecule and dust cluster production. The combination of a reduced amount of \hep\ and a small N initial content, which shuts down the C/N/O chemistry regime discussed in \S~\ref{3.3}, leads to molecule formation around day~1050, where $T_{\rm {gas}} \sim 1600$~K and $n_{\rm {gas}} \sim 1.5\times 10^7$~\cmc. In contrast with position $z=3.3$~\Ms, the high gas temperatures allow carbon rings larger than C$_{10}$ to open and grow, resulting in the formation of C$_{20}$ with a mass fraction of $\sim 4.0\times 10^{-2}$. For our Standard Case, the formation of carbon rings over the He/C/N region is reduced to these small inner zones, as we see below.  

\subsubsection{Total mass of molecules and dust for the He/C/N region.}
%_____________________________________________________________
%                              FIG 14 - MOL+DUST in O/C/Mg region
%------------------------------------------------------------- 
\begin{figure} 
  \centering
  \includegraphics[width=1.0\linewidth]{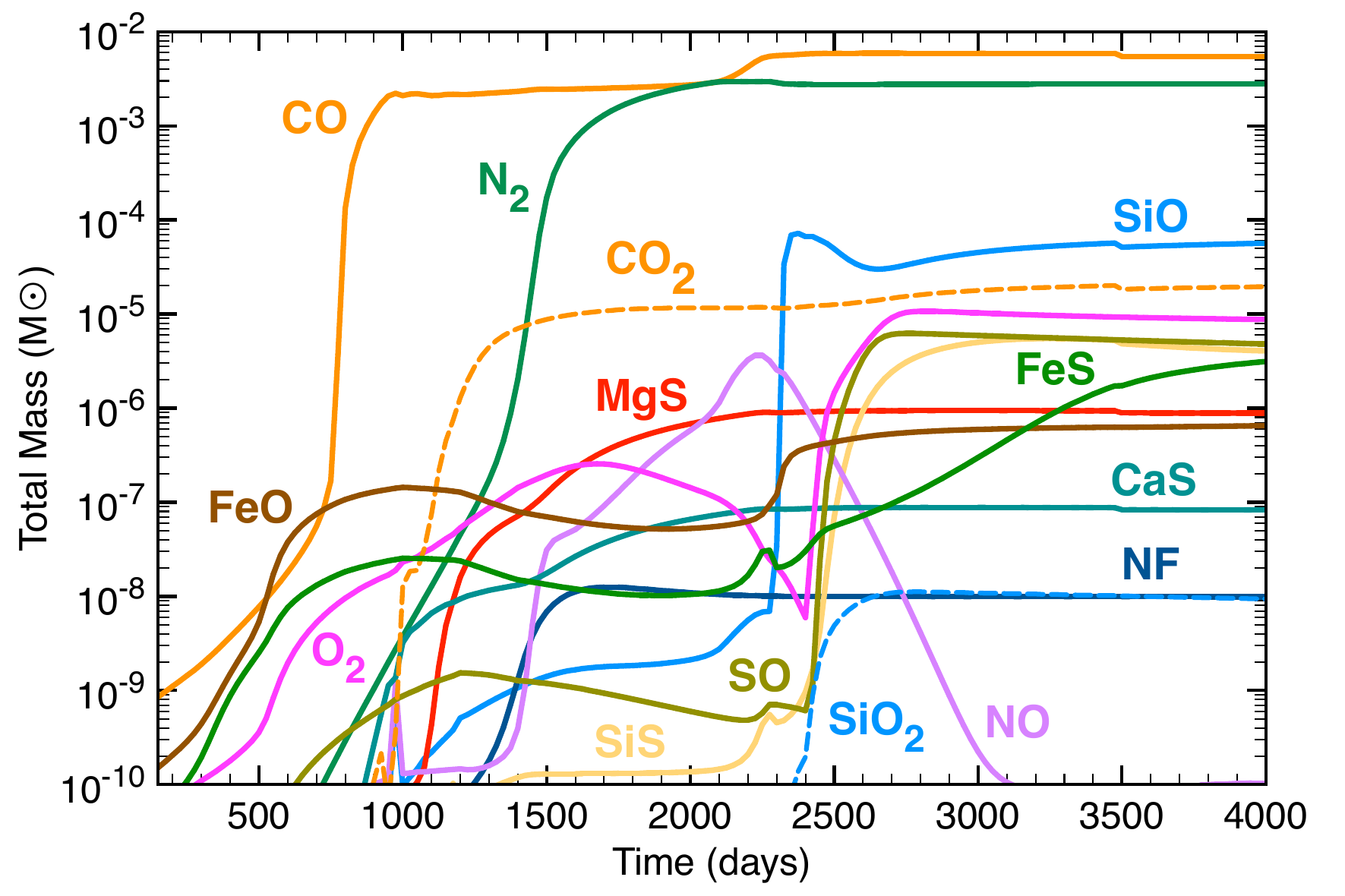}
    \includegraphics[width=1.0\linewidth]{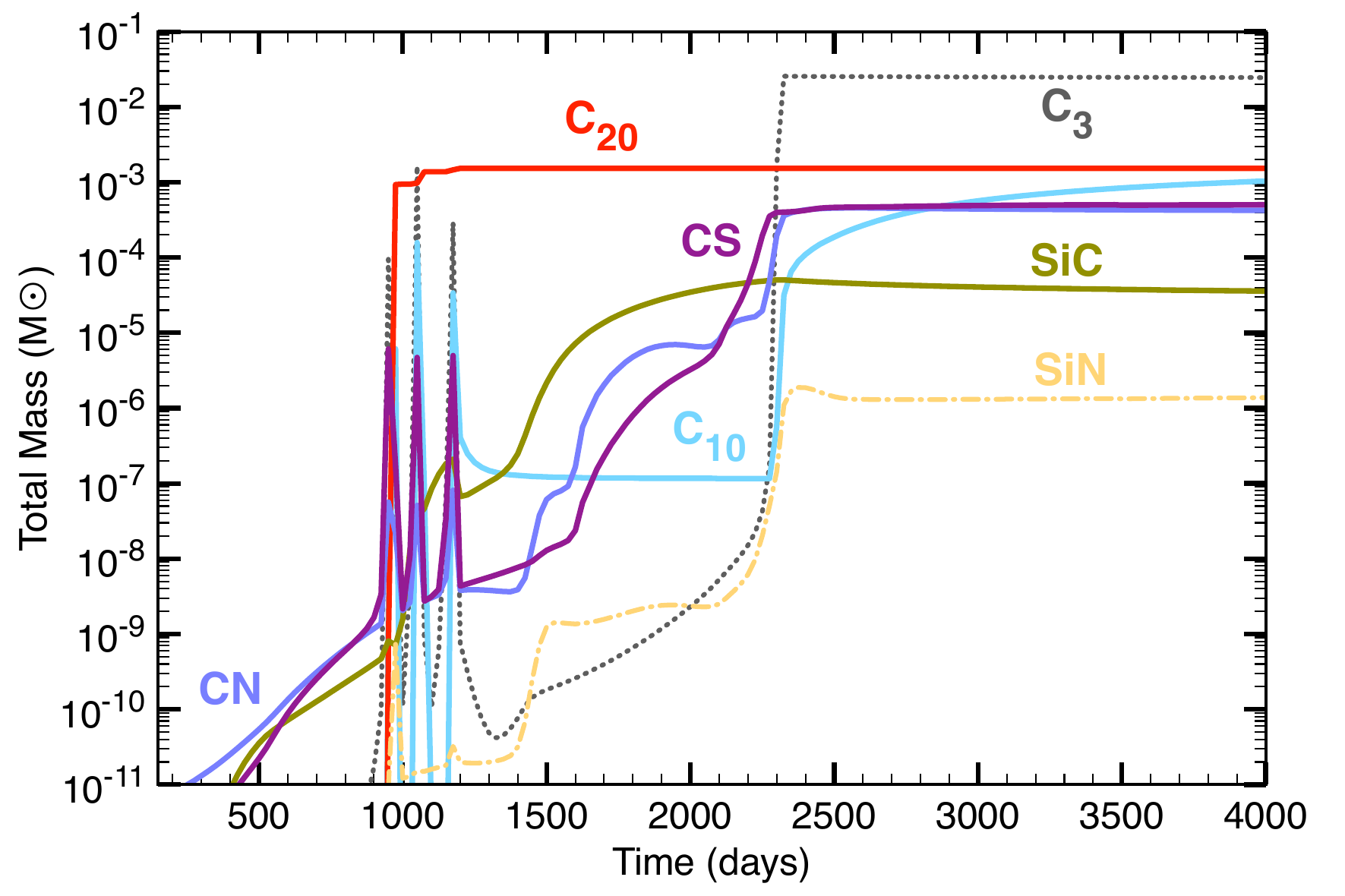}
\caption{Total mass of molecules produced in the He/C/N region versus post-explosion time for the Standard Case Top: molecules; Bottom: carbon clusters, C-bearing radicals, SiC and SiN. The region mass is 1.128~\Ms.}
\label{fig14}
\end{figure}
%------------------------------------------------------------- 
%_____________________________________________________________
%                              FIG 15 - He/C/N region maps
%-------------------------------------------------------------
\begin{figure} 
  \centering
  \includegraphics[width=1.0\linewidth]{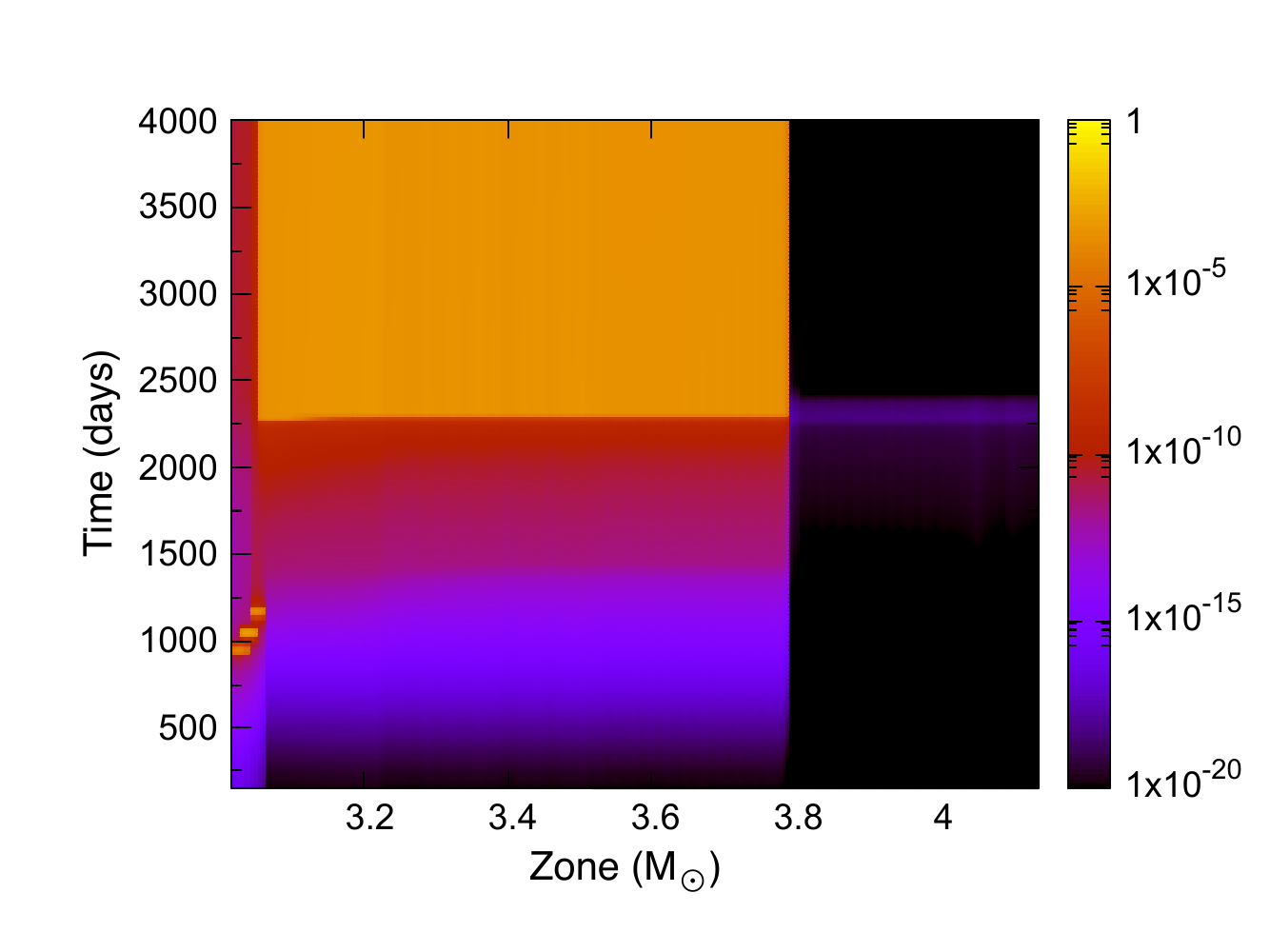}
  \includegraphics[width=1.0\linewidth]{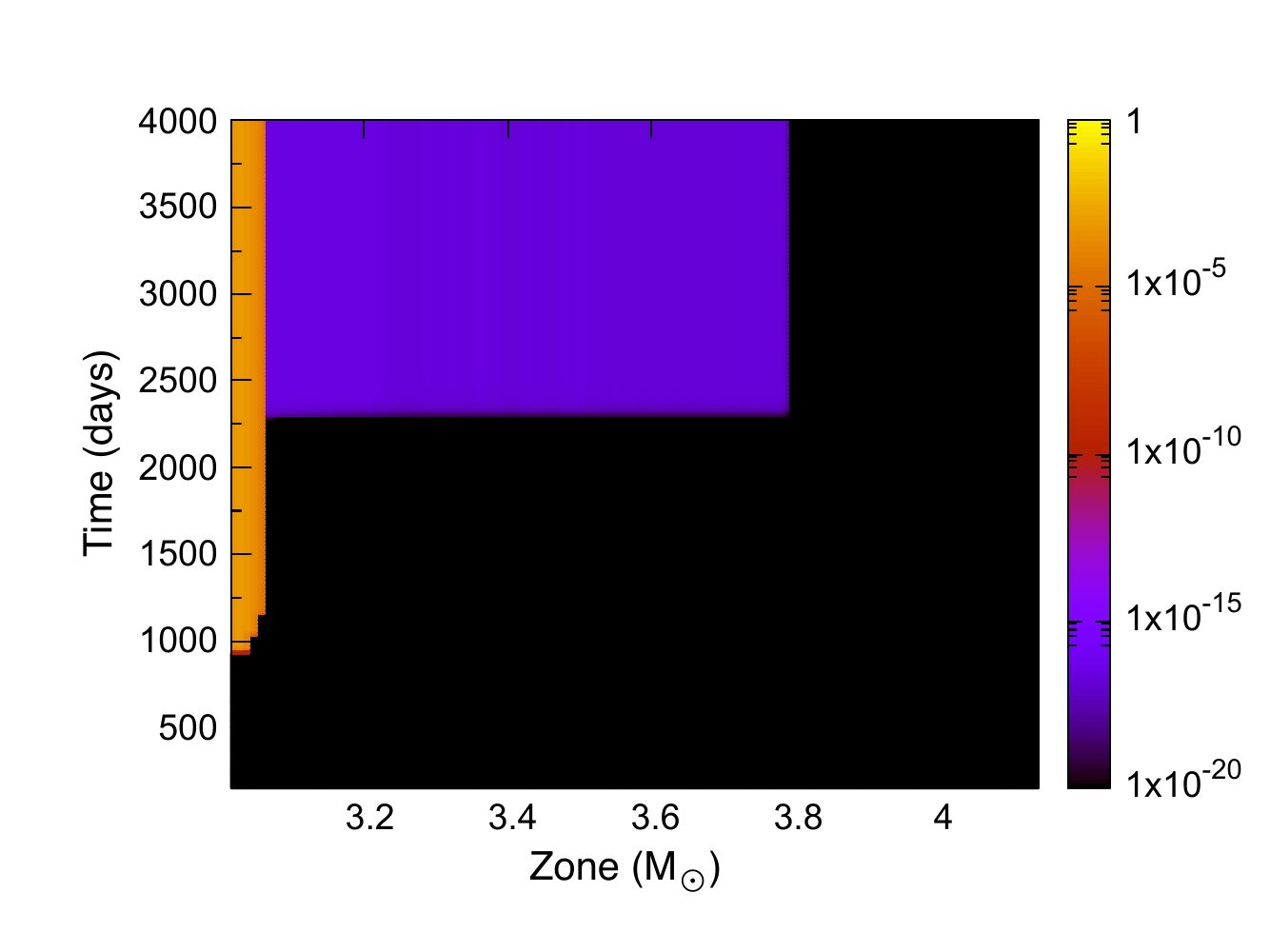}
\caption{Mass maps of small carbon clusters produced in the He/C/N region as a function of time and position $z$ in the region for the Standard Case. Top: C$_3$; Bottom:  C$_{20}$. }
\label{fig15}
\end{figure}
% -----------------------------------------------------------------
The total masses of molecules formed in the He/C/N~region as a function of post-explosion time are presented in Fig.~\ref{fig14}. The prominent species are CO, N$_2$, and SiO and molecular formation does not proceed before day $\sim 800$. Contribution from the three inner zones around $z=3.03$~\Ms\ are clear for the CO mass, which starts rising at day 800. A new increase appears in the CO mass around day 2300 due to contribution to the other zones (e.g. at $z=3.3$~\Ms\ and Fig.~\ref{fig12}). N$_2$ forms in large amount but not before day 1500, while SiO synthesis is delayed to day 2300 in the low-temperature ejecta. We notice the only relevant fluorine-bearing species in the region is NF, which forms along with N$_2$  and reaches a total mass of $1\times 10^{-8}$~\Ms\ at day 4000. 

Results for small carbon clusters also reflect the contribution from the three inner zones. The mass of produced C$_{20}$ originates solely from these zones whereas the C$_3$ mass is primarily produced by the rest of the region until $z\sim 3.8$~\Ms\ after day 2300, as seen from Fig.~\ref{fig15}, where mass maps for C$_3$ and C$_{20}$ are displayed as a function of time and position $z$ in the region. 

Despite the fact the final C$_{20}$ mass reaches the reasonable value of $\sim 1.5\times 10^{-3}$~\Ms\ at day 4000 for our Standard Case, the formation of the largest carbon clusters is inefficient and restricted to the three inner zones present in the 15~\Ms\ model. However, we notice similar inner zones do exist in the He/C/N~region of the 19~\Ms\ progenitor \citep{rau02}. To circumvent this peculiarity and identify the conditions conducive to efficient carbon dust production, we present below results for a case with high gas density in the He/C/N~region. 

%-----------------------------  TABLE: Final masses -------------------------------

\begin{table*}
\caption{Final masses of important molecules and dust clusters (in \Ms) at day 4000 for the full ejecta (Standard Case). The equivalent fraction of the ejecta mass in \% is indicated in the last column.}                 % title of Table
\label{finmas}    % is used to refer this table in the text
\centering                        % used for centering table
\begin{tabular}{c c c c c c l}      % centered columns (4 columns)
\hline\hline               % inserts double horizontal lines
Species/Region& Si/S/Ca & O/Si/Mg & O/C/Mg & He/C/N & Total  ejecta& \% Ejecta mass \\         % table heading
\hline                      % inserts single horizontal line
 \multicolumn{7}{c}{Detected molecules \tablefootmark{a}}\\
\hline 
  {\rm CO} & {\rm $2.05\times 10^{-7}$} & {\rm $5.03\times 10^{-4}$} & {\rm $1.95\times 10^{-1}$}& {\rm $5.86\times 10^{-3}$}& {\rm $2.01\times 10^{-1}$} & 8.52\\ 
   {\rm SiS} & {\rm $4.08\times 10^{-2}$} & {\rm $9.10\times 10^{-3}$} &{\rm $1.00\times 10^{-10}$}& {\rm $4.13\times 10^{-6}$}& {\rm $4.98\times 10^{-2}$}  &2.11  \\   
  {\rm SiO} & {\rm $9.04\times 10^{-6}$} & {\rm $4.30\times 10^{-2}$} & {\rm $1.92\times 10^{-4}$}& {\rm $6.04\times 10^{-5}$}& {\rm $4.32\times 10^{-2}$} & 1.83   \\ 
  {\rm SO$_2$} & $ - $& {\rm $1.30\times 10^{-2}$} & {\rm $5.67\times 10^{-5}$}&$ - $& {\rm $1.31\times 10^{-2}$}  &0.56 \\ 
  {\rm CS} & {\rm $2.54\times 10^{-7}$} & {\rm $1.41\times 10^{-9}$} & {\rm $1.36\times 10^{-5}$}& {\rm $5.02\times 10^{-4}$}& {\rm $5.15\times 10^{-4}$}  &0.02  \\ 
   {\rm SO} & $ - $ & {\rm $9.31\times 10^{-5}$} & {\rm $4.43\times 10^{-5}$}& {\rm $4.80\times 10^{-6}$}& {\rm $1.42\times 10^{-4}$}  &$6.02 \times 10^{-3}$  \\     
\hline                                  %inserts single line
 \multicolumn{7}{c}{Potentially detectable molecules \tablefootmark{a}}\\
 \hline
  {\rm O$_2$} & $ - $ &{\rm $4.45\times 10^{-1}$} & {\rm $1.22\times 10^{-1}$}& {\rm $8.74\times 10^{-6}$}& {\rm $5.67\times 10^{-1}$} & 24.05\\ 
  {\rm CO$_2$} & {\rm $7.57\times 10^{-9}$} &{\rm $1.26\times 10^{-2}$} & {\rm $2.18\times 10^{-2}$}& {\rm $2.13\times 10^{-5}$}& {\rm $3.44\times 10^{-2}$} & 1.46\\ 
  {\rm C$_{3}$} & $-$ &$ - $& $ - $ & {\rm $2.45\times 10^{-2}$} & {\rm $2.45\times 10^{-2}$} &  1.04 \\  
  {\rm CaS} & {\rm $3.26\times 10^{-3}$} & {\rm $1.83\times 10^{-3}$} & {\rm $1.89\times 10^{-8}$}& {\rm $8.65\times 10^{-8}$}& {\rm $5.09\times 10^{-3}$}  &{\rm 0.22}  \\ 
   {\rm N$_2$} & $ - $ &$ - $  & $ - $ & {\rm $2.77\times 10^{-3}$}& {\rm $2.77\times 10^{-3}$} & 0.12 \% \\   
 {\rm MgS} & {\rm $1.08\times 10^{-5}$} & {\rm $6.40\times 10^{-5}$} & {\rm $6.37\times 10^{-7}$}& {\rm $9.25\times 10^{-7}$}& {\rm $7.73\times 10^{-5}$}  &{\rm $3.28\times 10^{-3}$}  \\  
   {\rm CN} & $ - $ &$ - $  & $ - $ & {\rm $4.23\times 10^{-4}$}& {\rm $4.23\times 10^{-4}$} & {\rm $1.79\times 10^{-2}$} \\   
   {\rm Al$_2$O} &$-$& {\rm $2.15\times 10^{-5}$}& {\rm $1.10\times 10^{-6}$}&$-$ & {\rm $2.26\times 10^{-5}$}& {\rm $9.58\times 10^{-4}$}  \\
 {\rm PO} & {\rm $1.03\times 10^{-7}$} & {\rm $1.77\times 10^{-6}$} & {\rm $2.67\times 10^{-9}$}& $ - $ & {\rm $1.88\times 10^{-6}$}  & $7.97 \times 10^{-5}$  \\ 
  {\rm FeS} & {\rm $3.56\times 10^{-5}$} & {\rm $7.79\times 10^{-7}$} & {\rm $1.92\times 10^{-10}$}&  {\rm $3.11\times 10^{-6}$} & {\rm $3.95\times 10^{-5}$}  & $1,68 \times 10^{-3}$  \\ 
  {\rm FeO} & {\rm $1.55\times 10^{-10}$} & {\rm $4.08\times 10^{-6}$} & {\rm $1.06\times 10^{-7}$}&  {\rm $6.48\times 10^{-7}$} & {\rm $4.84\times 10^{-6}$}  & $2.06 \times 10^{-4}$  \\ 
   {\rm SiO$_2$} & $-$ & {\rm $1.15\times 10^{-9}$}& {\rm $2.80\times 10^{-8}$}  &{\rm $9.41\times 10^{-9}$}&{\rm $3.86\times 10^{-8}$}   &{\rm $1.64\times 10^{-6}$}   \\ 
     {\rm NF} & $ - $ &$ - $  & $ - $ & {\rm $9.92\times 10^{-9}$}& {\rm $9.92\times 10^{-9}$} & {\rm $4.21\times 10^{-7}$}\\  
      {\rm NO} & $ - $ &$ - $  & $ - $ & {\rm $1.05\times 10^{-10}$}& {\rm $1.05\times 10^{-10}$} & {\rm $4.45\times 10^{-9}$} \\  
\hline
Total mass & {\rm $4.41\times 10^{-2}$} & {\rm $5.25\times 10^{-1}$} &  {\rm $3.39\times 10^{-1}$}&  {\rm $3.42\times 10^{-2}$}& {\rm $9.43\times 10^{-1}$}& 39.95 \\
 \hline
  \multicolumn{7}{c}{Dust clusters \tablefootmark{a}}\\
 \hline
   {\rm Mg$_2$Si$_2$O$_6$} - enstatite& $-$ & {\rm $2.99\times 10^{-2}$} & {\rm $9.07\times 10^{-10}$} & $-$& {\rm $2.99\times 10^{-2}$} & 1.27 \\  
   {\rm Mg$_4$Si$_2$O$_8$} - forsterite & $-$ & {\rm $2.01\times 10^{-4}$} & {\rm $3.00\times 10^{-9}$} &$-$& {\rm $2.10\times 10^{-4}$} & {\rm $8.91\times 10^{-3}$}  \\  
   {\rm Si$_3$O$_5$} - quartz & $-$ & {\rm $1.51\times 10^{-2}$} & {\rm $1.02\times 10^{-5}$} &$-$& {\rm $1.51\times 10^{-2}$} & 0.64 \\ 
    {\rm Si$_3$O$_6$} - quartz & $-$ & {\rm $7.31\times 10^{-4}$} & {\rm $8.29\times 10^{-7}$} &$-$ &{\rm $7.32\times 10^{-4}$} & {\rm $3.10\times 10^{-2}$}  \\  
    {\rm Al$_4$O$_6$} - alumina & $-$ & {\rm $1.84\times 10^{-7}$} & {\rm $1.90\times 10^{-10}$} & $-$ & {\rm $1.84\times 10^{-7}$}&$7.81\times 10^{-6}$ \\  
    {\rm C$_{20}$} - carbon & $-$ &$ - $& $ - $ & {\rm $1.53\times 10^{-3}$} & {\rm $1.53\times 10^{-3}$} & {\rm $6.49\times 10^{-2}$}  \\  
     {\rm SiC} - silicon carbide& $-$& $-$& {\rm $3.54\times 10^{-6}$} & {\rm $3.56\times 10^{-5}$} & {\rm $3.91\times 10^{-5}$} & {\rm $1.66\times 10^{-3}$} \\ 
     {\rm SiN} - silicon nitride& $-$&$-$ & $-$ & {\rm $1.38\times 10^{-6}$} & {\rm $1.38\times 10^{-6}$} & {\rm $5.85\times 10^{-5}$} \\ 
  \hline
  Total mass & $ - $&{\rm $4.59\times 10^{-2}$} &{\rm $1.46\times 10^{-5}$} &{\rm $1.57\times 10^{-3}$} &$4.75\times 10^{-2}$  & $2.02$   \\  
  \hline
\end{tabular}
\tablefoot{
\tablefoottext{a}{Masses less than $10^{-10}$ \Ms\ are labelled as $ - $. } }
\end{table*}
% ------------------------------------------------------------
% --------- HD CASE -------------------------
\subsubsection{High-density case in the He/C/N region}
\label{HD}
%_____________________________________________________________
%                              FIG 16 - MOL+DUST in He/C/N zone HD case  3.03 Msun
%------------------------------------------------------------- 
\begin{figure} 
  \centering
  \includegraphics[width=\linewidth]{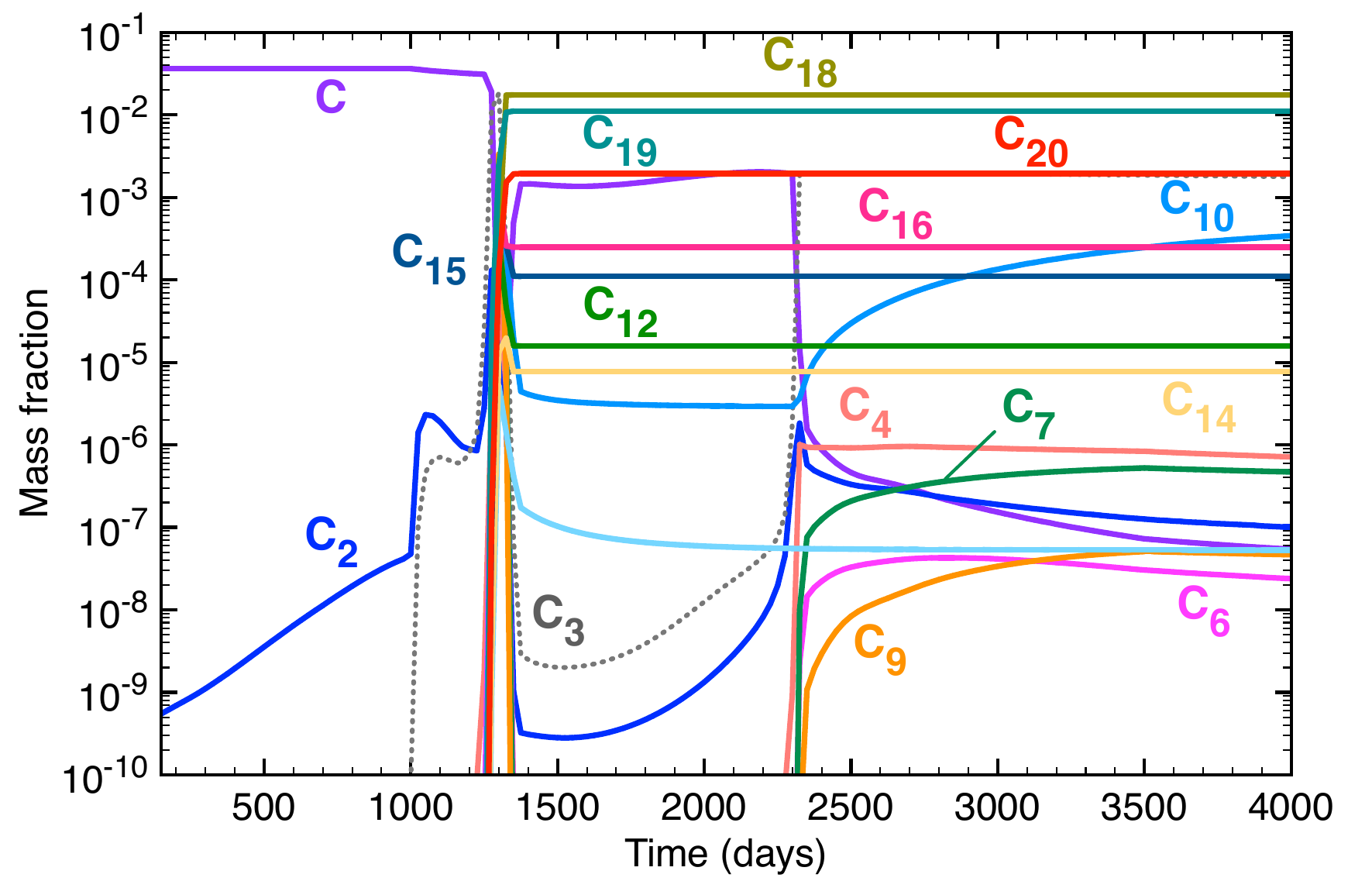}
  \includegraphics[width=\linewidth]{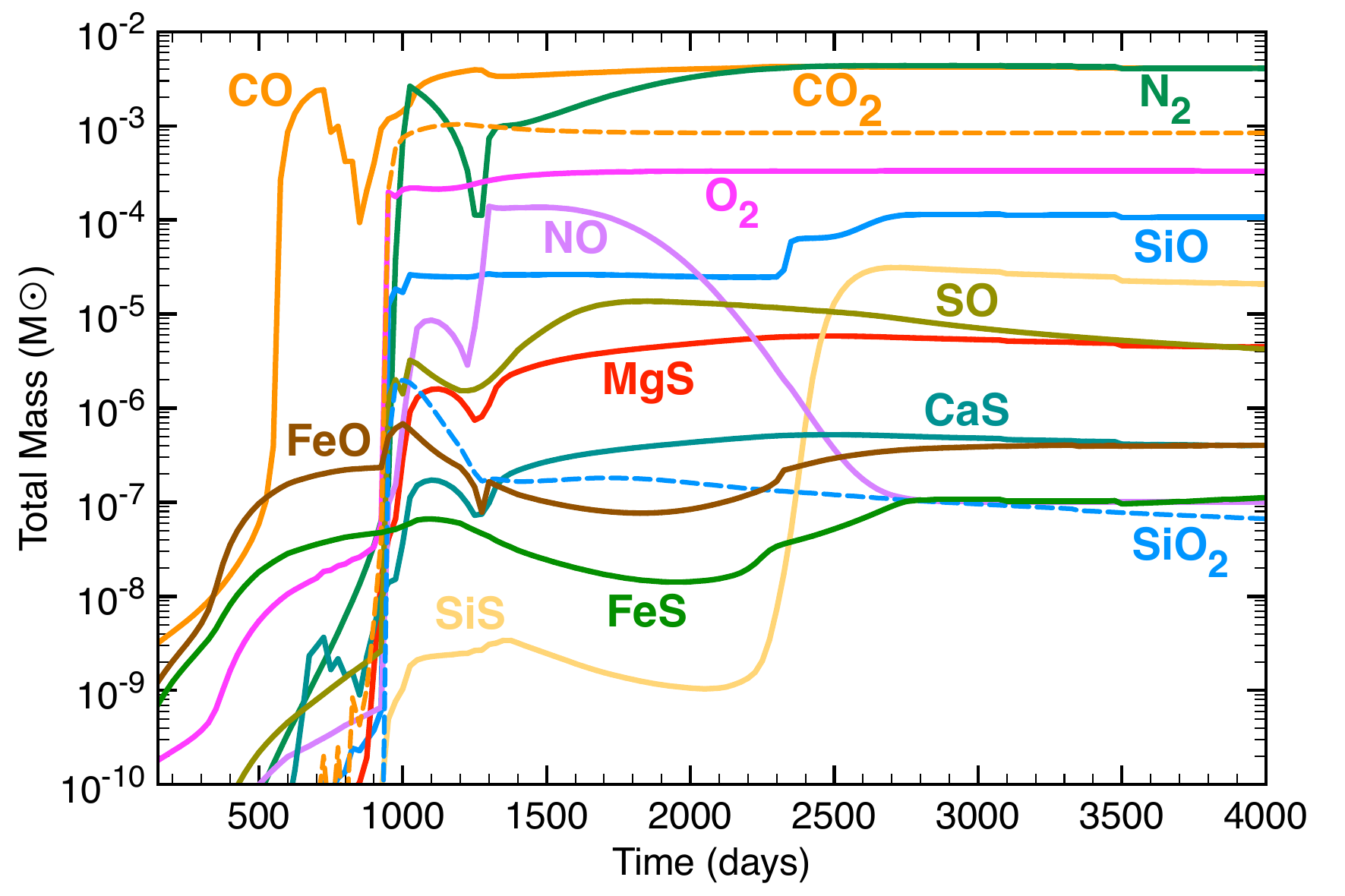}
   \includegraphics[width=\linewidth]{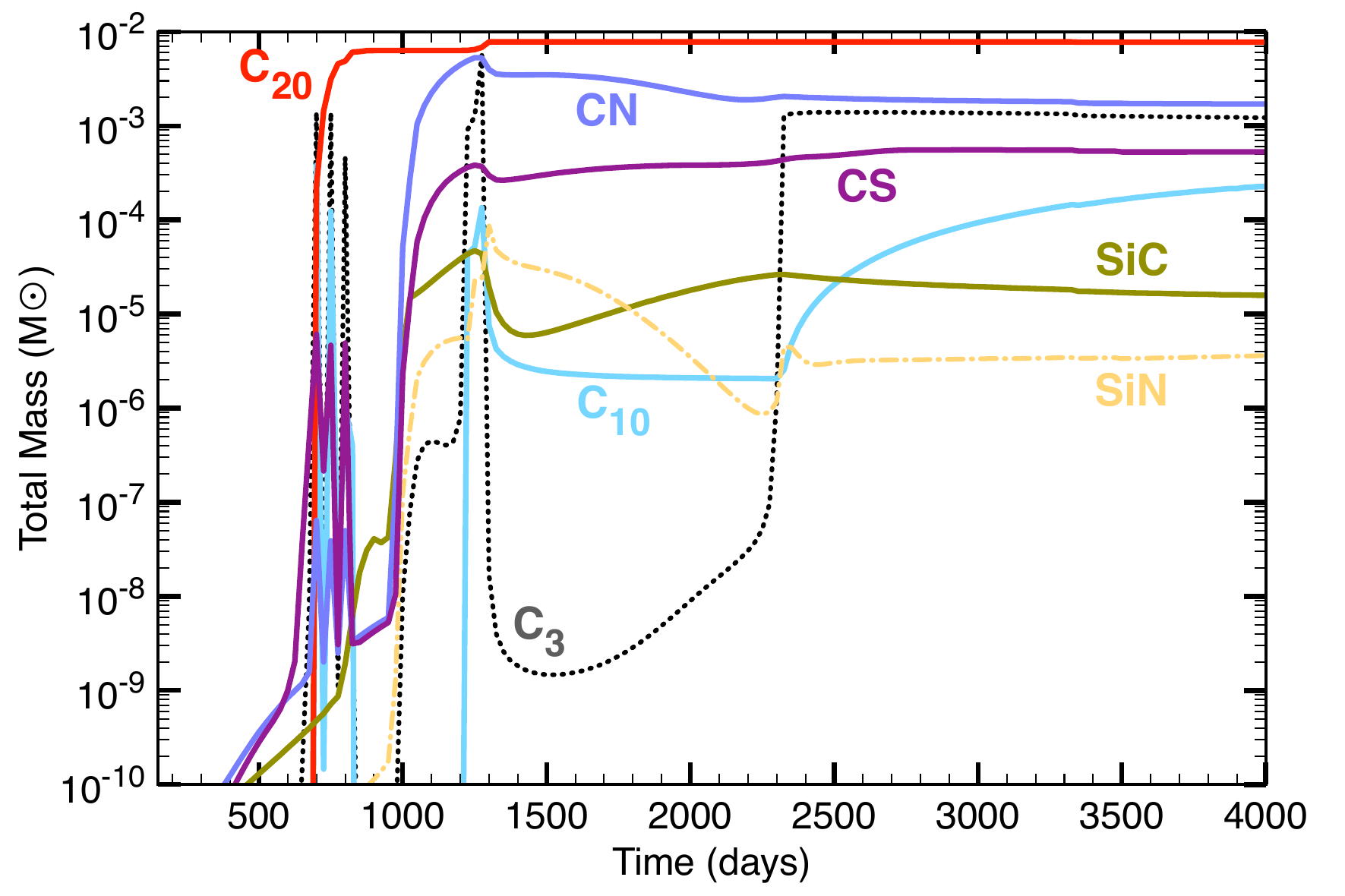}
\caption{Top: Mass fractions of carbon clusters formed in the He/C/N region at $z=3.3$~\Ms\ versus post-explosion time for the high-density case (zone mass = $1.35 \times 10^{-2}$ \Ms). Middle: total masses of molecules formed in the He/C/N region as a function of post-explosion time for the high-density case ; Bottom: total masses of carbon clusters formed in the He/C/N region as a function of post-explosion time for the high-density case}
\label{fig16}
\end{figure}
%--------------------------------------------
We investigate a High Density (HD) case where the He/C/N~region has an initial gas density at day 100 ten times that of the Standard Case, i.e., $\rho_{HD}(100)=10\times \rho_{SC}(100)= 1.35132\times 10^{-12}$ g\ \cmc. A factor 10 is a reasonable choice of over-density factor assuming clumping in the ejecta with a volume filling factor $f_v = 0.1$. We recall the density in the clump $\rho_g$ verifies the equation 
\begin{equation}
\label{fil}
\rho_g \times f_v = \rho_c,
\end{equation}
where $\rho_c$ is the core gas density given by Equation~\ref{rhoc} \citep{tru99}. The HD case corresponds to a number density at day 100 of $n_{\rm {gas}}(100) \sim 2\times 10^{11}$~\cmc\ for the ejecta gas in the He/C/N~region. Such a value is comparable to values derived at day 100 for the clumpy ejecta of SN1987A and SN2005af \citep{sar22,sar25}. However, the present n$_{\rm gas}$(100) value is about two orders of magnitude smaller than the number density for the clumpy case of \citet{sar15}. 

Results for carbon clusters at $z=3.3$~\Ms\ are presented in Fig.~\ref{fig16} (Top) and have to be compared with results for the Standard Case of Fig.\ref{fig12}. The higher gas densities allow a faster recombination of \hep\ so that molecules start forming at day 1000, and squeeze the C/N/O chemistry regime discussed in \S~\ref{3.3} to the time range day~1000-1250. After day 1250, characterised by $T{\rm _g\sim 1000}$~K and $n{\rm _g \sim 1\times 10^9}$~\cmc, carbon clusters form quickly and grow efficiently to carbon rings, including C$_{20}$, because ring opening and growth through C and C$_2$ inclusion is made possible at these gas temperatures. 

\begin{figure} 
  \centering
   \includegraphics[width=1.0\linewidth]{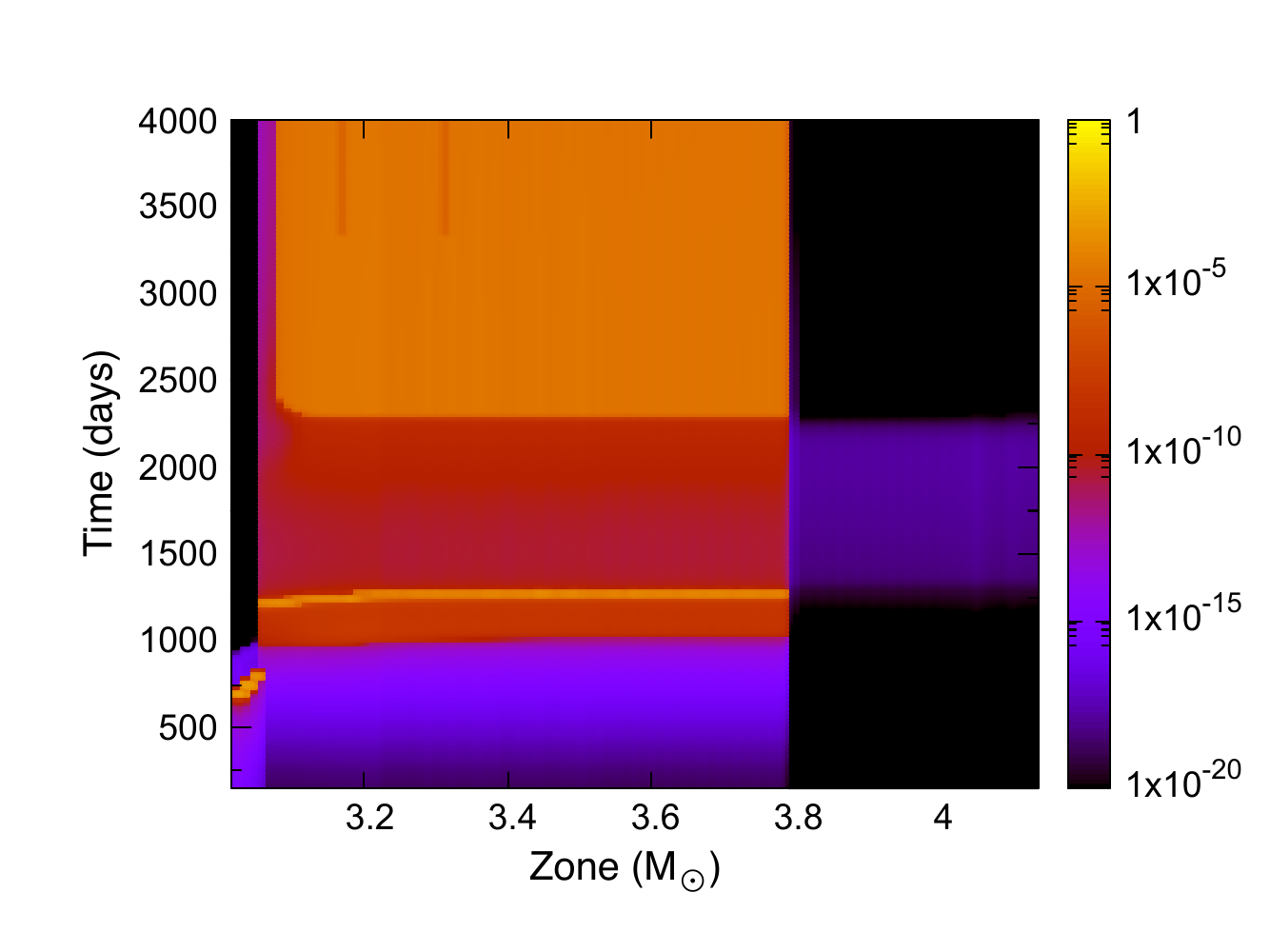}
  \includegraphics[width=1.0\linewidth]{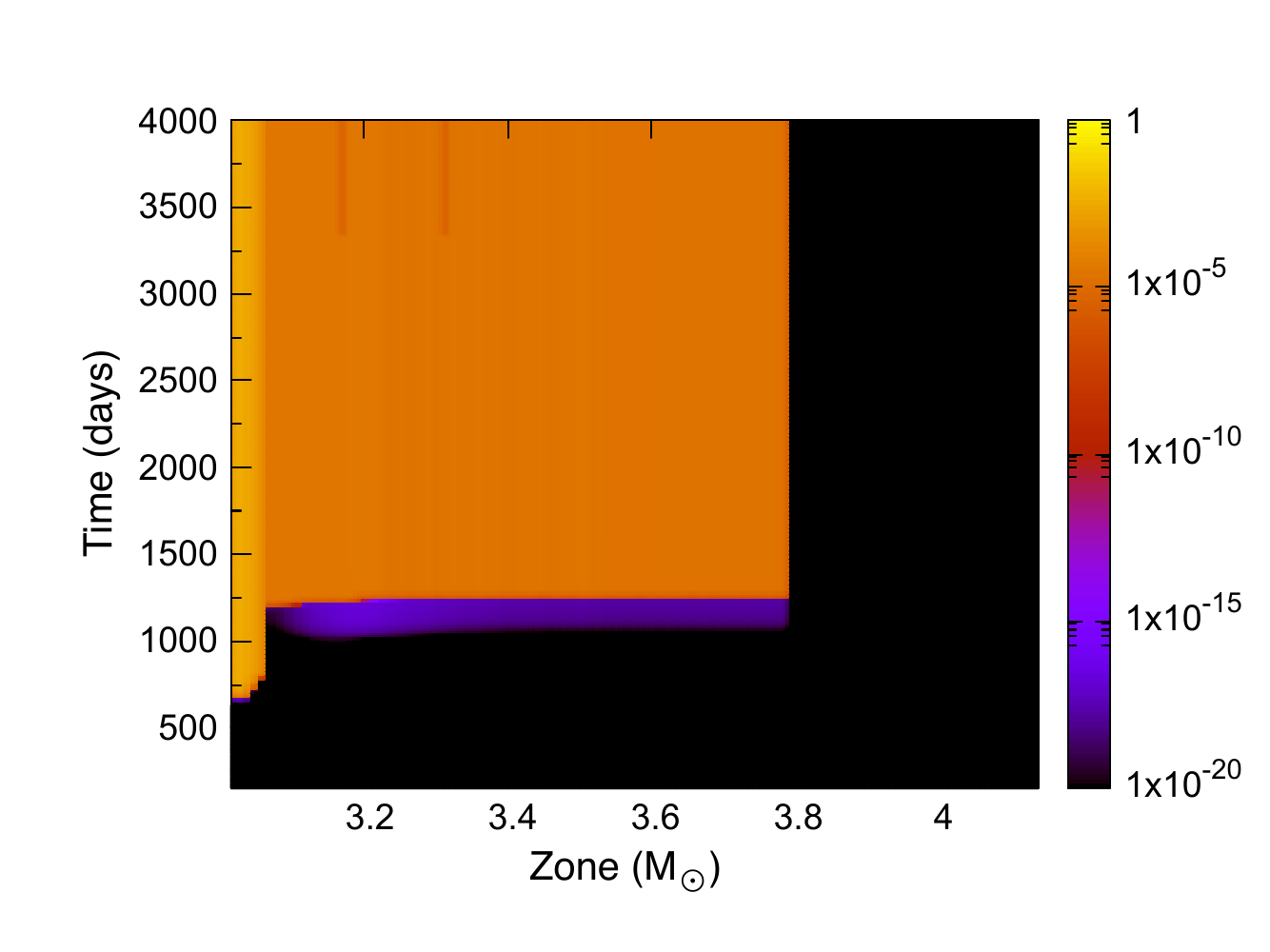}
\caption{Mass maps of small carbon clusters produced in the He/C/N region as a function of time and position $z$ in the region for the high-density case. Top: C$_3$; Bottom:  C$_{20}$. }
\label{fig17}
\end{figure}

The total masses of molecules and carbon clusters are also shown in Fig.~\ref{fig16} (Middle and Bottom, respectively), while we gather all molecule and cluster masses for the HD~case as a function of region and at day 4000 in Table~\ref{tabap3} of the Appendix. The masses of most molecules have increased because the formation of species now occurs earlier and at higher densities. The increase is particularly strong for N$_2$ and CN, which mass at day 4000 is $\sim$~ten~times larger than for the Standard Case. As for carbon clusters and large rings in particular, we now see two contributions, one by the three inner zones at day~$700-800$, and a strong contribution at day~$1200-1300$ that double the mass of C$_{20}$. This double contribution is well evidenced in Fig.~\ref{fig17} where mass maps of both C$_3$ and C$_{20}$ for the high-density case are shown. The mass of C$_{20}$ is now distributed from the inner edge of the region to $z=3.8$~\Ms, the starting point of the N-rich sub-region. In the end, the density enhancement has increased the total mass of C$_{20}$ at day 400 by a factor five. Interestingly, small carbon chains and rings do form after day~2300, but do not grow larger than C$_{10}$, as a result of the low-temperature chemistry characterising the formation of small carbon chains \citep{loi14}.  
%_____________________________________________________________
%                              FIG 17 - He/C/N region maps  carbons - HD ase
%-------------------------------------------------------------
% -----------------------------------------------------------------

We conclude the synthesis of carbon dust is greatly favoured in dense clumps as opposed to other types of dust like silicates or silica, which form efficiently for the gas conditions of the Standard Case. The existence of the narrow carbon-cluster-forming region at the inner edge of the He/C/N~region is present in both explosion models for 15 and 19~\Ms\ progenitors by \citet{rau02}. Therefore, we expect a similar early carbon cluster synthesis from a few inner zones in the ejecta of a 19~\Ms\ progenitor. However, this specific region may not be present in the initial elemental composition of SN explosion models with less or more massive progenitors. This situation does not invalidate our finding that carbon clusters require dense gas to form in significant amount in the ejecta. Consequently, carbon dust synthesis cannot occur at very large explosion times characterised by low gas temperatures, as proposed by certain studies, e.g. \citet{sar22}. Furthermore, we expect the small chain C$_3$ to always be present in the gas at day 4000. Indeed, despite the fact the molecule participates to the formation of carbon clusters at earlier times, there is a residual C$_3$ population at day 4000 as a result of the low-temperature chemistry responsible for small carbon chain production. The lesser the gas density, the larger the C$_3$ population so that the molecule can be seen as a tracer of diffuse carbon-rich ejecta gas. 

\subsubsection{Low-temperature case in the O/Si/Mg region}
\label{LT}

In their study of the temperature of the oxygen core in SN~1987A, \citet{liu95} were the first to assess cooling by vibrational emission lines of CO in the O/C/Mg~region. This cooling is again investigated in SN~1987A through an exhaustive radiative transfer study by \citet{lil20}, who show the gas temperature drops by at least ${\rm 1000~K}$ compared to cases without molecular cooling, as for example in \citet{koz98}. We recall we use in this study the temperature profile derived by \citet{lil20} for the O/C/Mg region of the ejecta. Cooling might occur in the O/Si/Mg zone as well, specifically through vibrational emission of SiO, which is very abundant in this region, as first mentioned by \citet{liu95} who expected the impact on the gas temperature to be close to that of CO in the O/C/Mg region. 

We therefore consider a case of low gas temperature (LT~case) in the silicate-forming region by applying the temperature profile of the O/C/Mg region as derived by \citet{lil20} to the O/Si/Mg region. In the Appendix, the results are presented in Fig.~\ref{figALT} and molecule and cluster masses at day~4000 are listed in Table~\ref{tabap3}. When molecular masses are slightly enhanced, we see the most important changes involve dust clusters: the mass of produced silicates is smaller than that for the Standard Case (see Fig.~\ref{fig8}) by a factor $\sim 300$, while the mass of silica clusters slightly increases by $\sim 6$~\%, and that of alumina drastically drops by $\sim$ five~orders of magnitude. The total cluster mass budget is $\sim$ three~times lower than for the Standard Case. The main reason to the strong decrease in silicate production is due to Reaction~R12 of our proposed new scheme for silicate formation (see  Table~\ref{siliform}). R12 is endothermic with a moderate barrier of $\sim 5000$~K and the main channel to the formation of \ens\ monomers. The lower gas temperatures between day 200 and 500 thus reduce the efficiency of the process. Since \fors\ monomers form out of \ens\ units, masses of both silicate dimers are drastically reduced. We deduce that reasonably high gas temperatures are required to boost silicate formation according to our new proposed scheme. 

From our two tests on gas density and temperature, we infer high gas densities and temperatures are a prerequisite to efficient dust synthesis in the gas and imply dust must form at early times in the evolution of the ejecta, as already shown by previous models \citep[e.g.][]{cher08,cher10,sar15}. Our findings are in conflict with a late dust formation scenario for the remnant of SN1987A as suggested by \citet{wes15} to fit multi-wavelength astronomical data. They also question the late carbon dust formation scenario at $t > 2400$ days for SN1987A as proposed by \citet{sar22} to fit the same near- and mid-infrared data. Our results show no carbon clusters in the form of large rings that act as coagulation seeds, can form at these late times and low temperatures. On the other hand, our model allows for the formation of carbon clusters with large masses at day 500 and day 1200 for gas densities and temperatures similar to those used by \citet{sar22}.
%
%They find carbon dust form at \env~2400 days for the ejecta spherical model, where the gas densities are $\sim 1\times 10^{6}$ \cmc\ and the gas temperature is $\sim 100$K.   
% --------------------- OBSERVATIONS -------------------------
%
\section{Comparison with observations}
\label{obs}
\subsection{CO}
\label{CO}
%_____________________________________________________________
%                              FIG 18 - CO maps SC
%-------------------------------------------------------------
\begin{figure*}
   \centering
\begin{subfigure}{0.495\textwidth}
        \includegraphics[width=1.\textwidth]{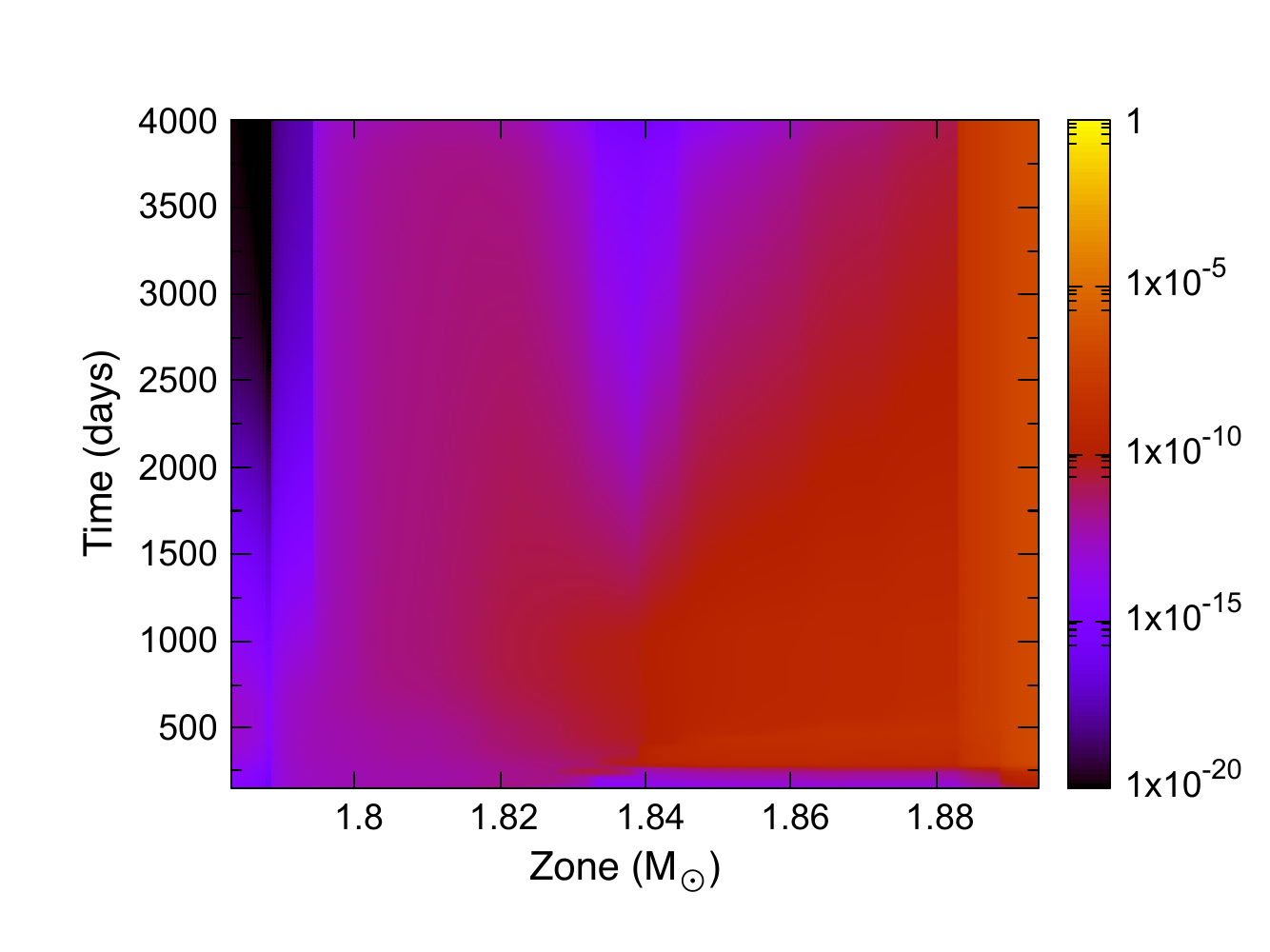}
%        \caption{ }
        \end{subfigure}
\hfill
\begin{subfigure}{0.495\textwidth}
        \includegraphics[width=1.\textwidth]{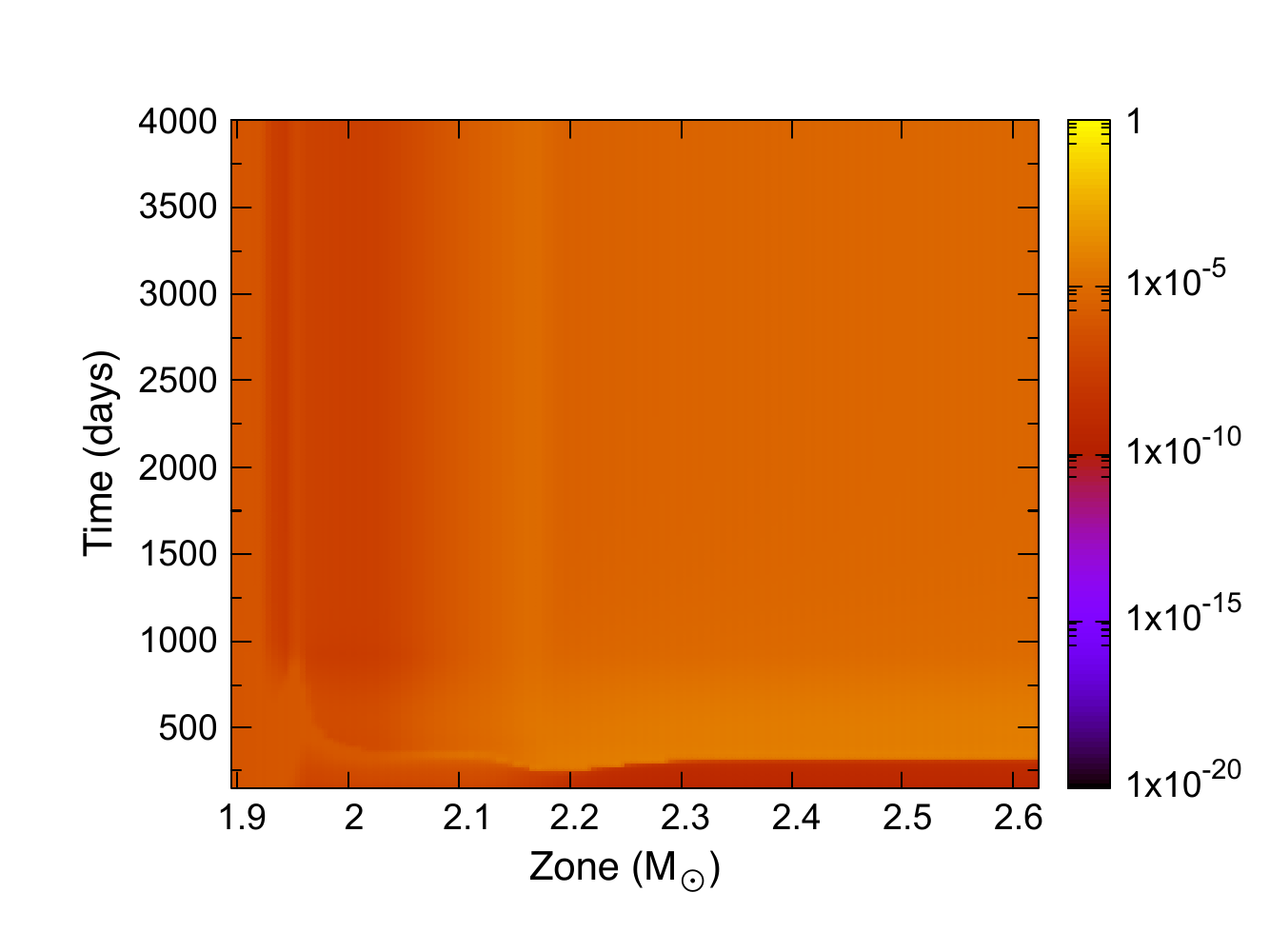}
%         \caption{ }
        \end{subfigure}
\hfill
\begin{subfigure}{0.495\textwidth}
        \includegraphics[width=1.\textwidth]{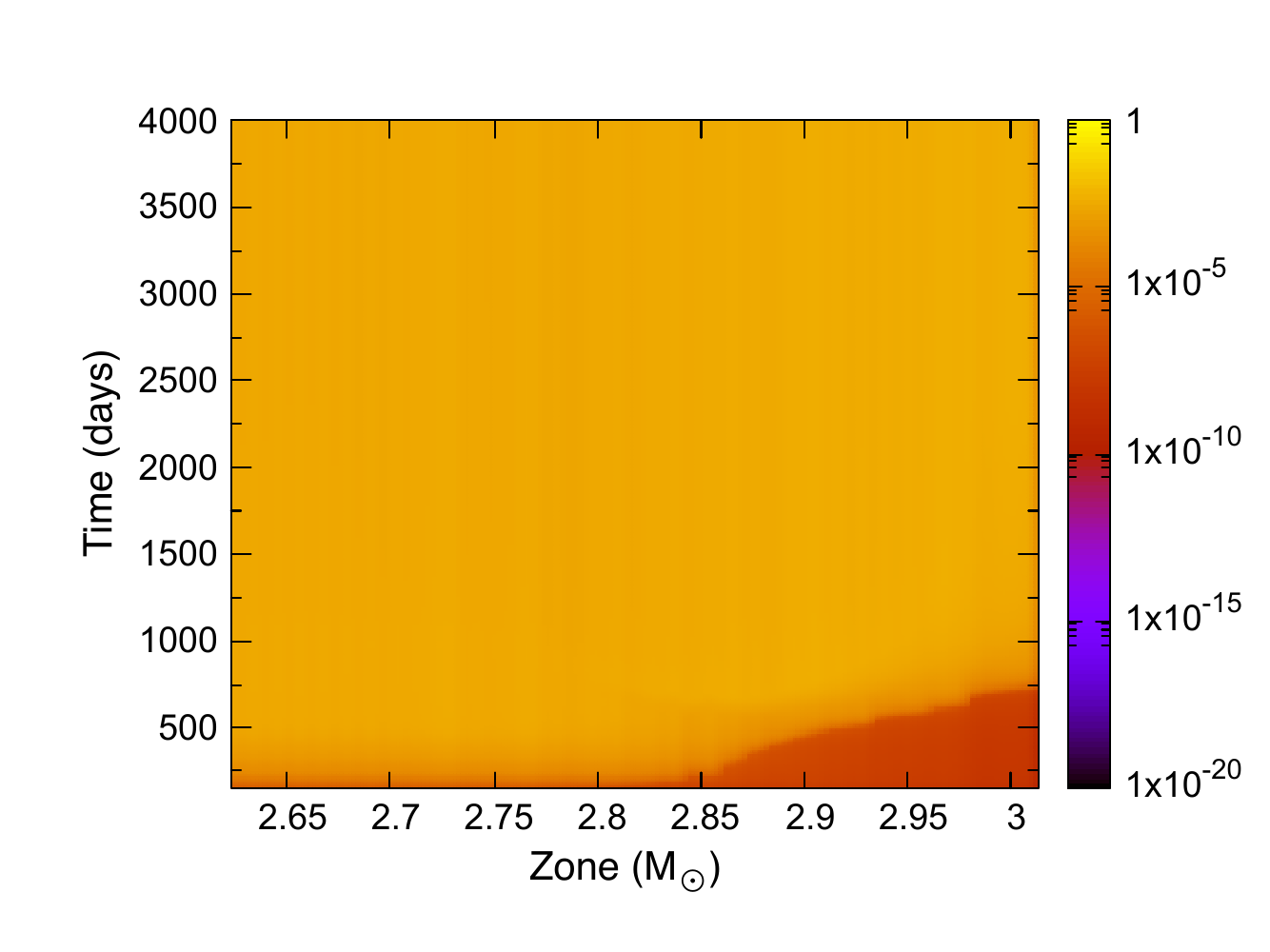}
%         \caption{ }
        \end{subfigure}
\hfill
\begin{subfigure}{0.495\textwidth}
        \includegraphics[width=1.\textwidth]{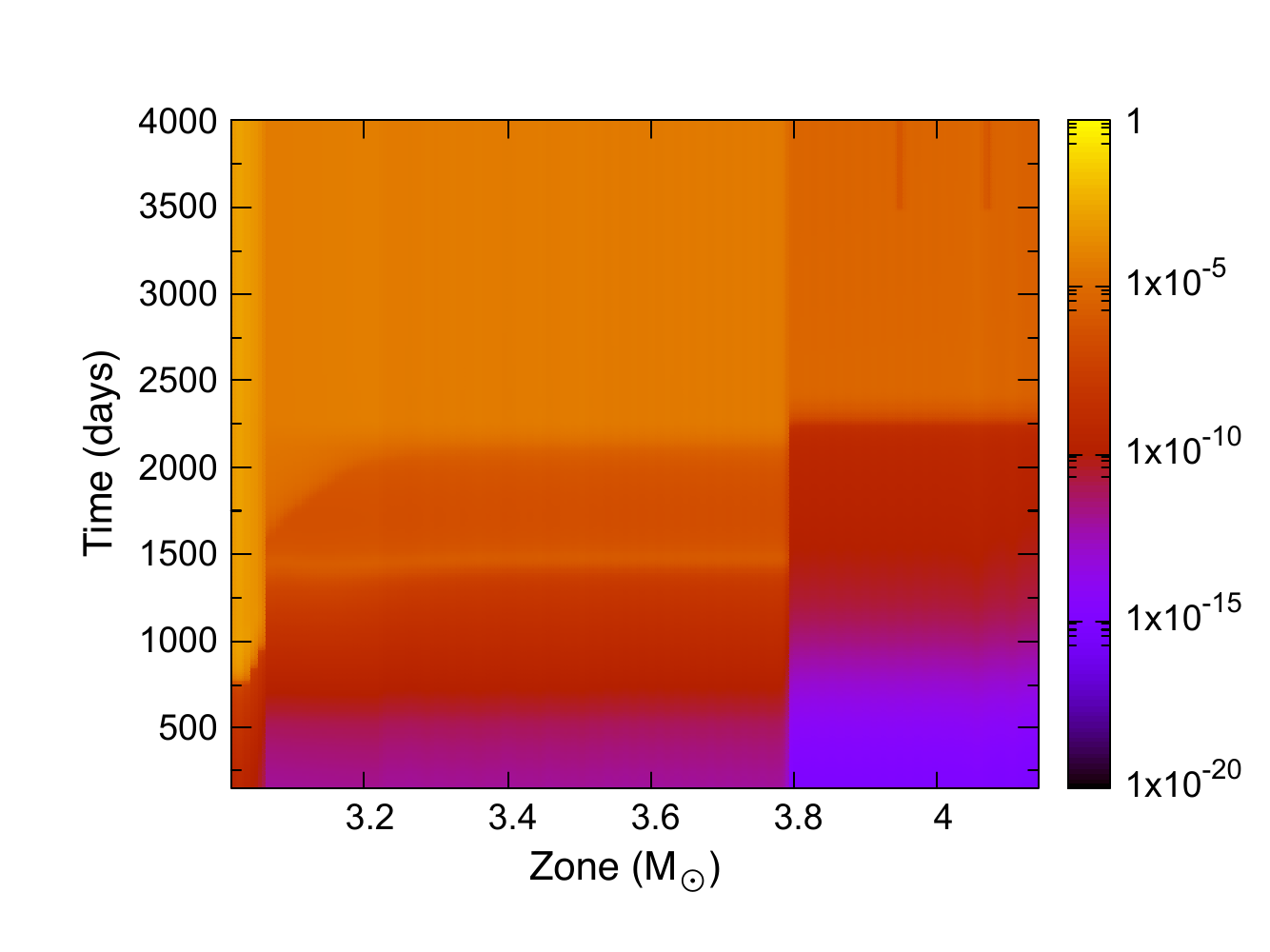}
%         \caption{ }
        \end{subfigure}
 \caption{CO mass maps for the Standard Case. Top left: SiS/Ca region Top right: O/Si/Mg region; Bottom Left: O/C/Mg region; Bottom right: He/C/N region. }
       \label{fig18}
  \end{figure*}
% -----------------------------------------------------
%_____________________________________________________________
%                              FIG 19 - SiO maps SC
%-------------------------------------------------------------
\begin{figure*}
   \centering
\begin{subfigure}{0.495\textwidth}
        \includegraphics[width=1.\textwidth]{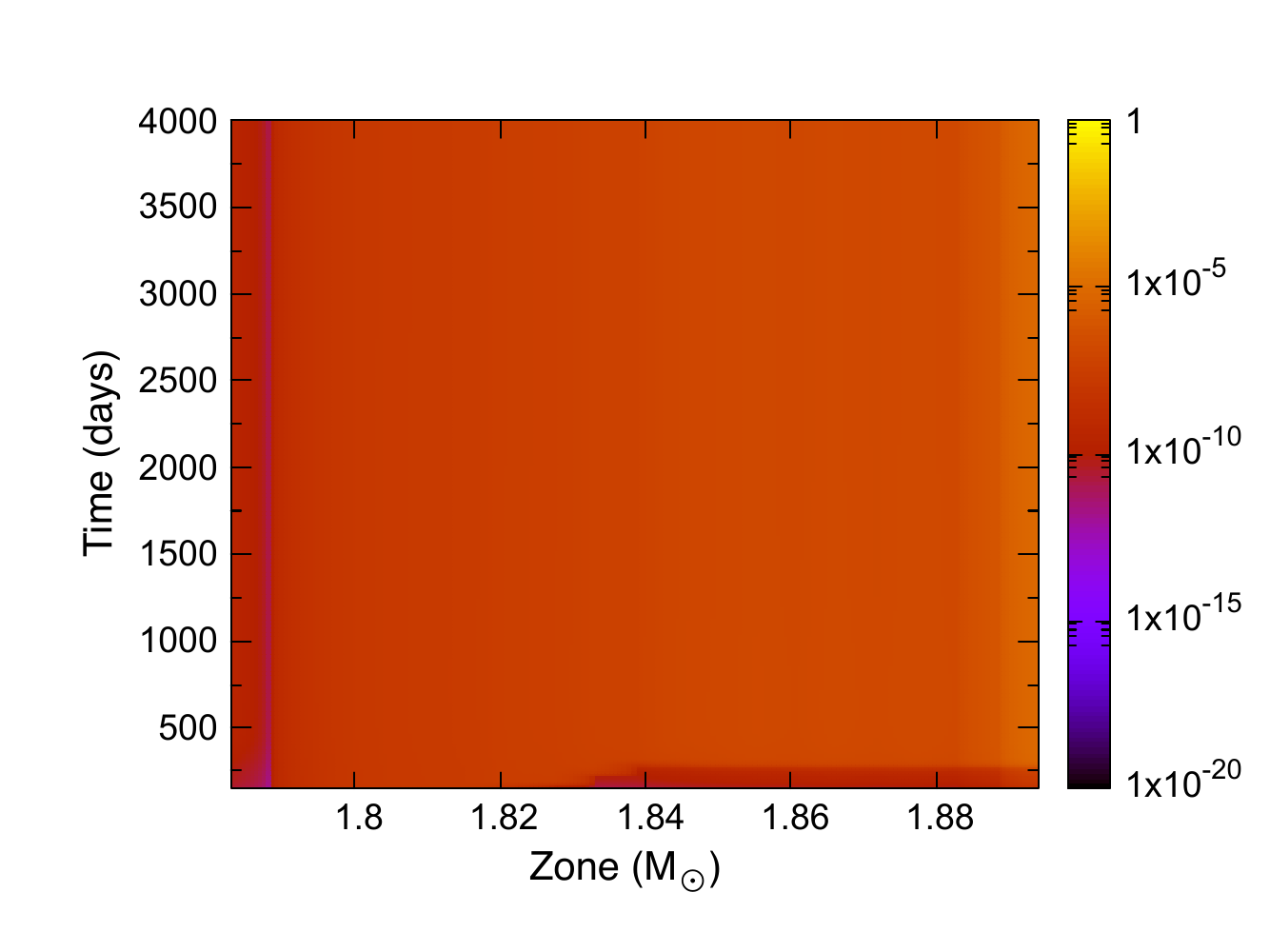}
%        \caption{ }
        \end{subfigure}
\hfill
\begin{subfigure}{0.495\textwidth}
        \includegraphics[width=1.\textwidth]{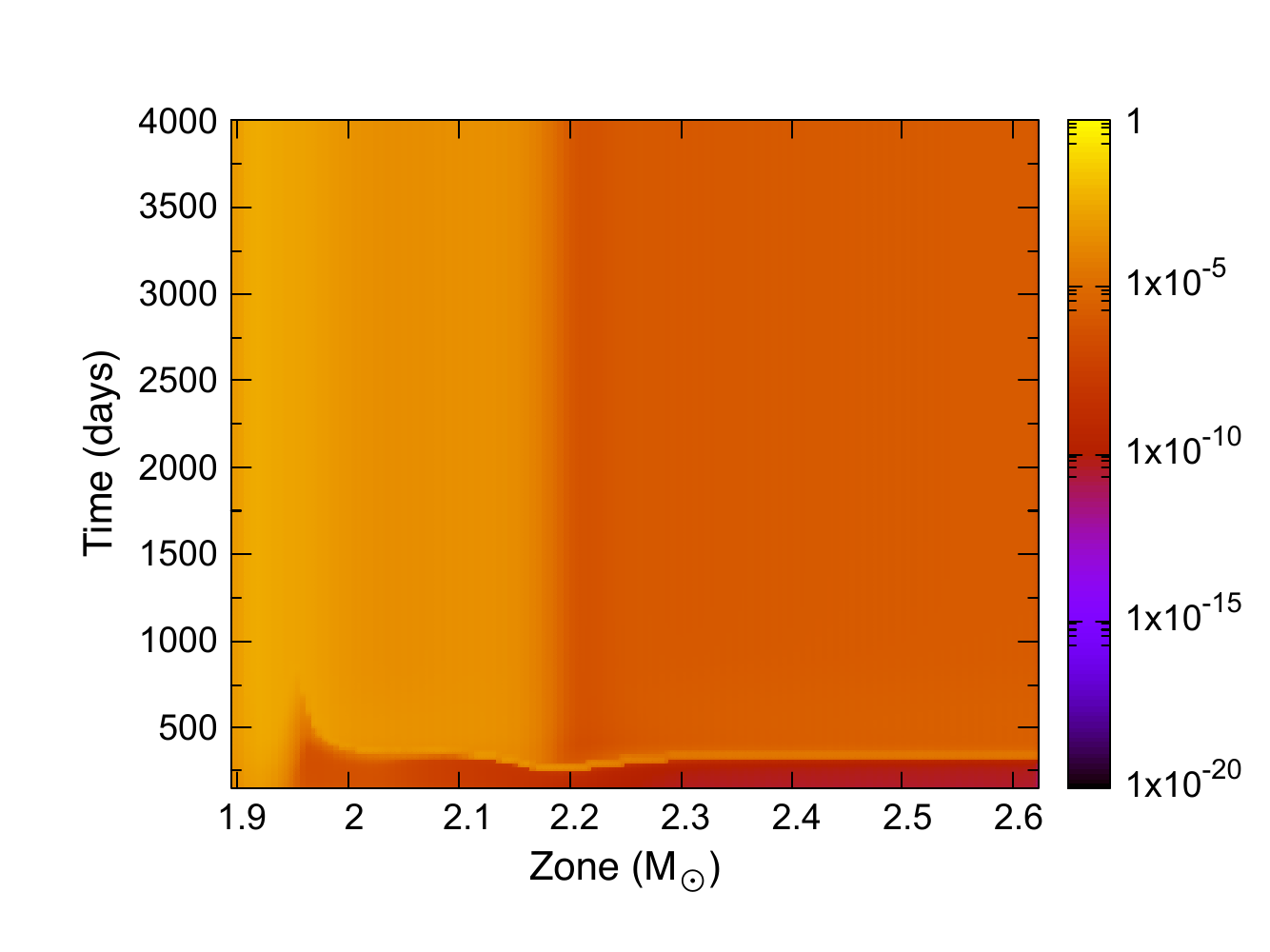}
%         \caption{ }
        \end{subfigure}
\hfill
\begin{subfigure}{0.495\textwidth}
        \includegraphics[width=1.\textwidth]{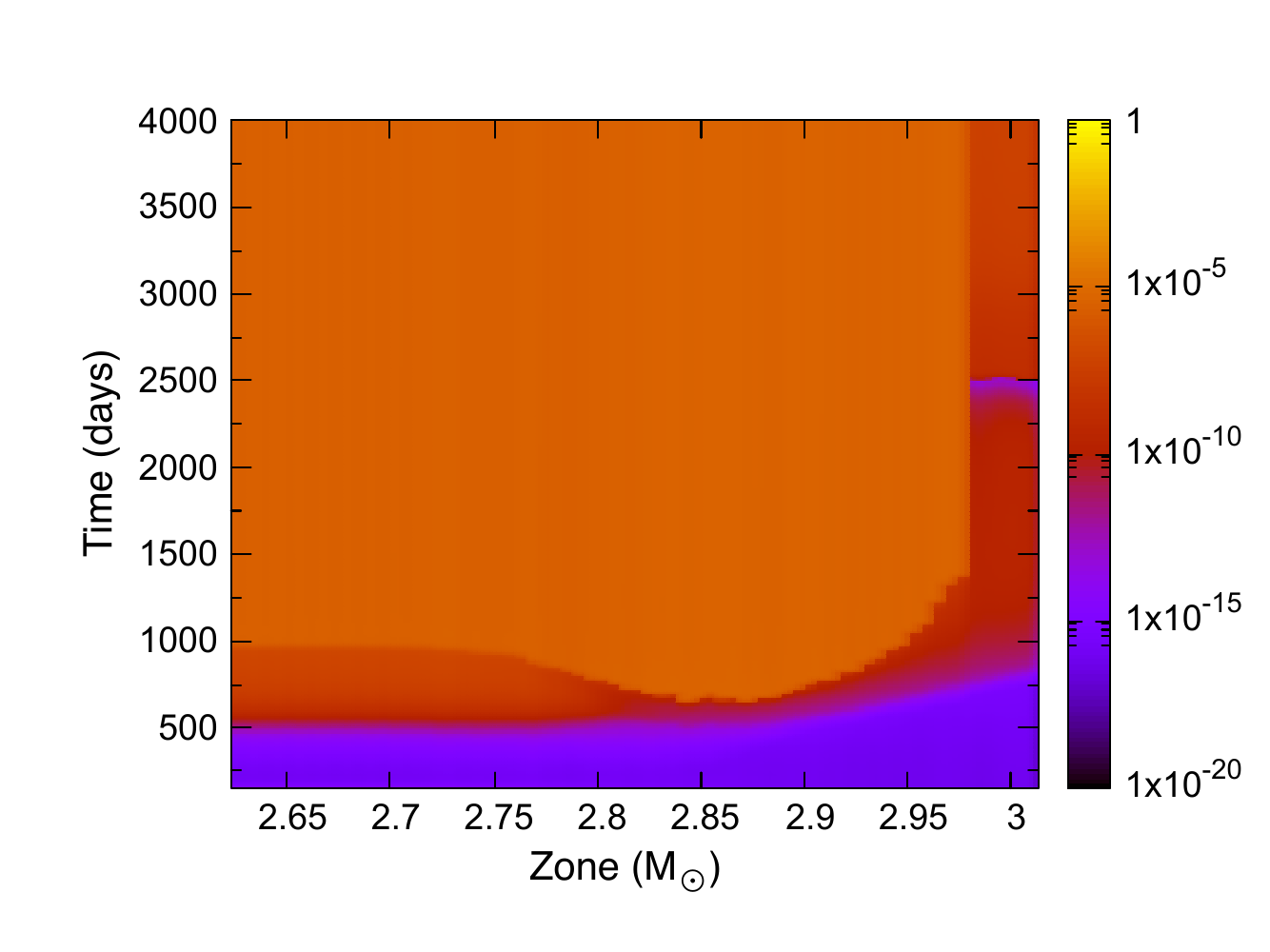}
%         \caption{ }
        \end{subfigure}
\hfill
\begin{subfigure}{0.495\textwidth}
        \includegraphics[width=1.\textwidth]{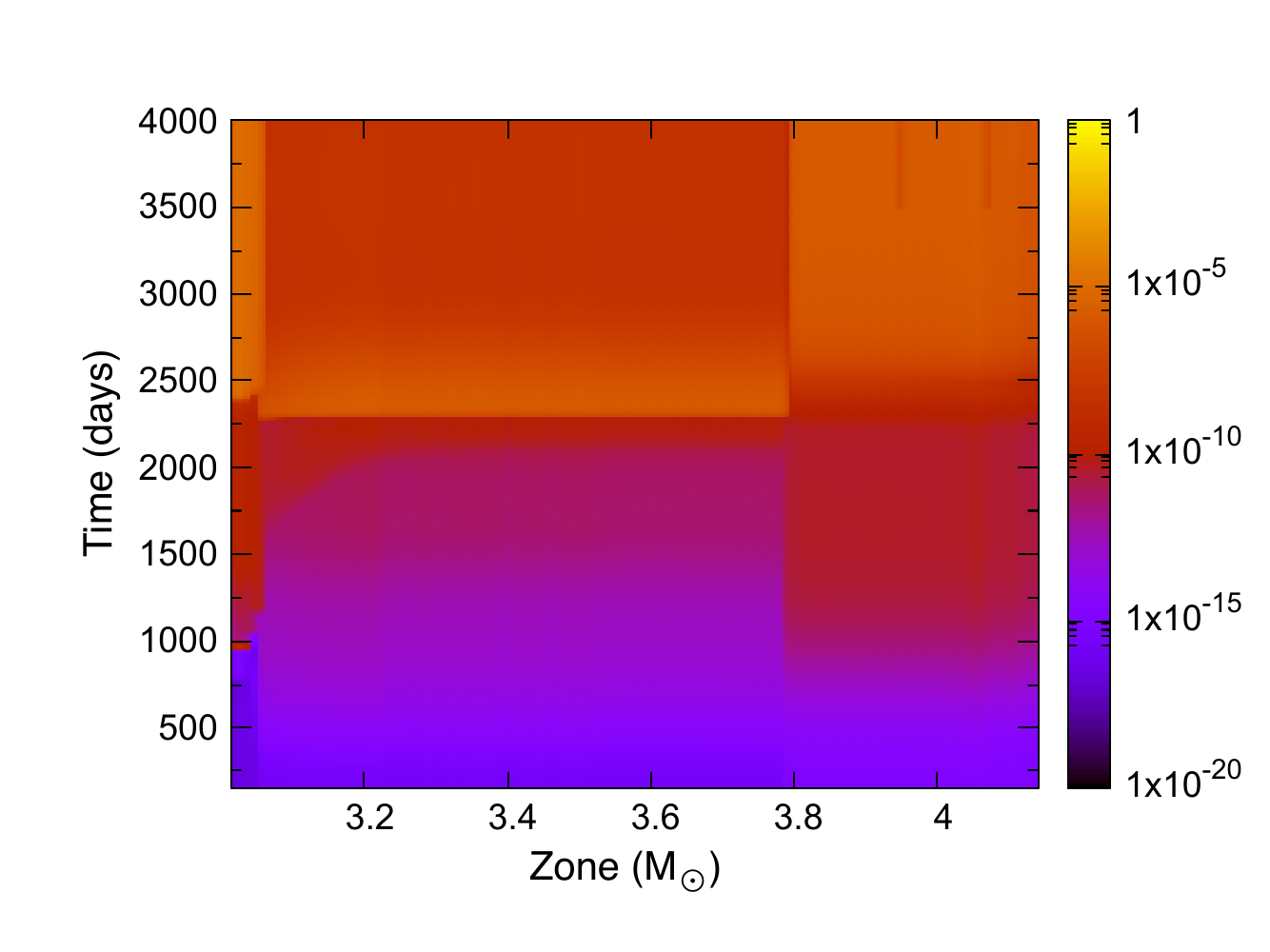}
%         \caption{ }
        \end{subfigure}
 \caption{SiO mass maps for the Standard Case. Top left: SiS/Ca region Top right: O/Si/Mg region; Bottom Left: O/C/Mg region; Bottom right: He/C/N region. }
       \label{fig19}
  \end{figure*}
% -----------------------------------------------------
The first overtone bands (${\rm \Delta v=2}$) of carbon monoxide CO located at $2.3 - 2.5$~\mic\ are observed as early as day 112 until day 349 in SN~1987A  \citep{spyro88}, from day 124 till day 205 in SN~2017eaw \citep{rho18}, and recently from day 199 till day 307 in SN 2023ixf \citep{park25}. Recent JWST data on SN 2023ixf by \citet{med25} show the excitation of both the first overtone bands and the fondamental bands (${\rm \Delta v=1}$) near $4.6$~\mic\ at day 252 while at day 373, the first overtone bands already fade owing to gas cooling while the fundamental bands are still excited. An early CO formation is also predicted by chemical models \citep{lepp90,cher08,cher09,sar13,slu18}. 

Our new model also confirms an early CO formation around day 200, as seen in Fig.~\ref{fig18}, where mass maps of CO for the entire ejecta are presented as a function of time and $z$ position. Firstly, we see that CO is produced in all ejecta regions although with very small efficiency in the Si/S/Ca region. While it forms quite homogeneously as early as day 200 in the two regions of the oxygen core, the O/C/Mg region is by far the most efficient of the two at forming CO, as already found by \citet{liu95} and \citet{cher09}. CO also forms in the outer carbon-rich region but not before day 750 and with an inhomogeneous distribution and a smaller efficiency in space and time compared to the previous region. An early CO synthesis is also seen in Fig.~\ref{fig20} where the total masses of CO and SiO are shown as a function of time and regions, along with masses derived from the above observations and NIR and ALMA data of SN1987A. Clearly $\sim 10^{-4}$~\Ms\ of CO forms at day 200, in accord with observations and grow to reach $\sim$ 0.2~\Ms\ at day 3000. This mass at late times agrees well with the CO line analysis of ALMA data by \citet{mat17}. 
\subsection{SiO}
\label{SiO}

The fundamental band of SiO between $7.5-9$~\mic\ was detected and the emission modelled in a few SNe: in SN1987A, from day 260 and 519 \citep{roch91,liu94}, in SN 2005af at day~214 \citep{kot06}, and in SN 2004et, between day 300 and 795 \citep{kot09}. Interestingly for this last object, the SiO emission decrease with post-explosion time was correlated to the onset of dust formation. The modelled SiO masses were comprised between $\sim 10^{-4}-10^{-3}$~\Ms. The molecule is also detected in SN 1987A with ALMA at much later times and the line analysis results in a SiO mass ranging between $4\times 10^{-5}$ and $2\times 10^{-3}$~\Ms\ \citep{mat17}. 

\begin{figure} 
  \centering
   \includegraphics[width=1.0\linewidth]{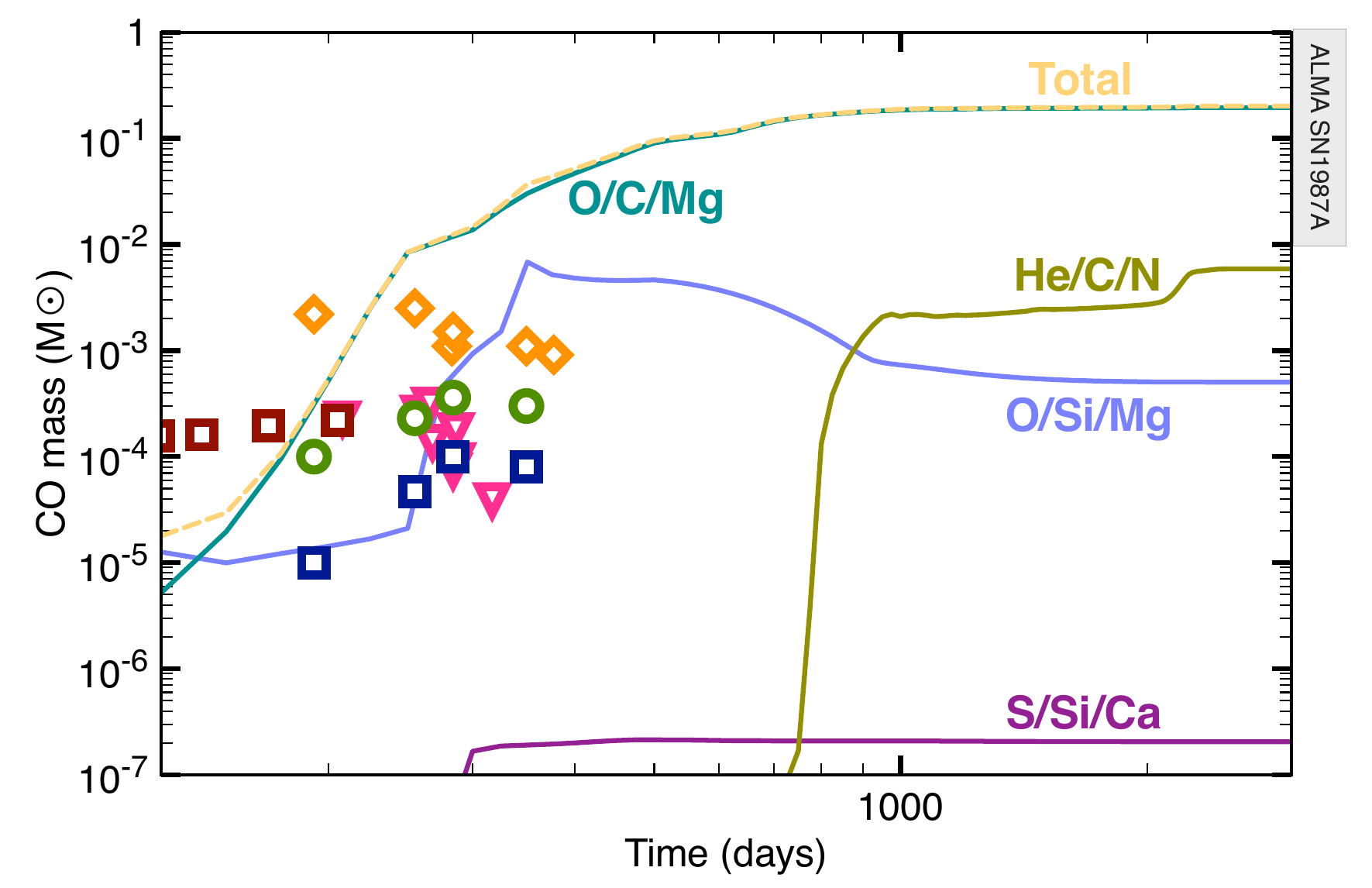}
  \includegraphics[width=1.0\linewidth]{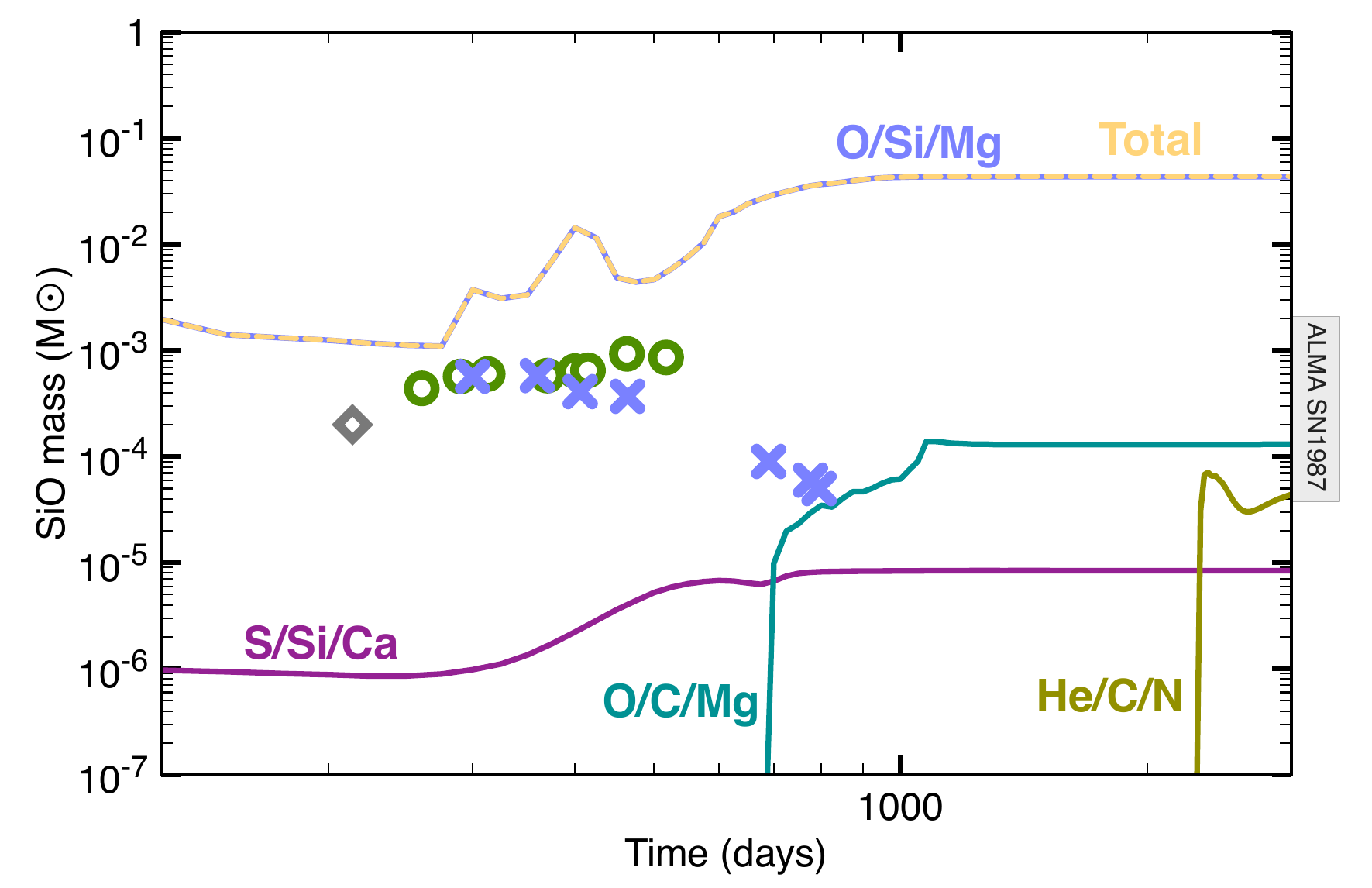}
\caption{Total CO and SiO masses and masses formed per region as a function of time for the Standard Case case and available observations for various Type II SNe. Top: CO - Brown empty square SN2017eaw  \citep{rho18}, Fuschia empty triangle SN2023ixf \citep{park25}, Dark blue empty square SN1987A \citep{spyro88}, Green empty circle and orange empty rhombus SN1987A, LTE and non-LTE case, respectively \citep{liu92}; Bottom:  SiO - Dark grey empty rhombus SN2005af \citep{kot06}, Green empty circle SN1987A - non LTE \citep{liu94}, Lavender cross SN2004et \citep{kot09}. Value ranges for both CO and SiO derived by \citet{mat17} with ALMA are also indicated with a grey filled rectangle.}
\label{fig20}
\end{figure}

We present mass maps of SiO for the entire ejecta in Fig.~\ref{fig19} as a function of time and $z$ position as well as the total masses of SiO as a function of time and regions in Fig.~\ref{fig20}. Inspection of Fig.~\ref{fig19} reveals that SiO forms in all ejecta regions but with rather low masses except for a maximum reached in the silicate/silica-forming region O/Si/Mg, between $1.9$ and $2.2$~\Ms, where the initial Si mass fraction is large. The SiO distribution also shows more inhomogeneities compared to that of CO and the molecule forms at later times, especially in region O/C/Mg and He/C/N. We see from Fig.~\ref{fig20} that the SiO mass increases with time and shows a reasonable agreement with astronomical data before day 500. It reaches a value of $\sim 4\times 10^{-2}$~\Ms\ at day 4000 (see also Table~\ref{finmas}), which is larger than the upper limit of SiO mass derived from ALMA data by more than a factor $\sim 20$. However, we see at least two reasons for not being alarmed by such a difference. Firstly, SN 1987A has a progenitor mass of $19$ \Ms\ with a different initial elemental composition. Secondly, this study only investigates the molecular component of a fictitious, spherically symmetric ejecta and not the solid component, which will form out of cluster coagulation and surface deposition. We believe surface deposition of SiO and SiO-related species onto silicate/silica grains may reduce the final SiO mass and this matter will be explored in a forthcoming paper. 

One positive aspect is that our new chemical model does not totally deplete the SiO mass as the former model did because we use a novel chemical description of the formation of silicate clusters where SiO dimerisation plays a minor role. Indeed, \citet{sar13} find the SiO mass drastically decreases from day 200 to 1500, where the mass is $\sim 10^{-6}$ and already much less than the lower SiO mass limit found from ALMA data. A similar low SiO mass is derived in a recent study of SN 2005af by \citet{sar25}, where the same chemical model as of \citet{sar13} is used and the SiO mass has already dropped to $\sim 10^{-5}$~\Ms\ at day 1000. We are thus encourage by the present findings and will derive final SiO masses when modelling dust synthesis. 

Finally, our results cannot explain the 3D SiO distribution in SN 1987A as reconstructed by \citet{abel17} whereby SiO extends to greater radial distances ($\equiv$ larger velocities) than CO in some directions, but where most of the CO emission presents a maximum extension larger than that of SiO. Here, we investigate a 1D physico-chemical model applied to a spherically-symmetric, stratified ejecta and apply a unique chemical model to the entire ejecta regions. However, our findings confirm most of the SiO mass forms inwards to that of CO, in agreement with the reconstruction, and shows SiO formation in the outer ejecta region, i.e. at higher velocities. Furthermore, both CO and SiO distributions are stratified but the latter shows more inhomogeneities resulting from chemistry, specifically in the oxygen core and the outer He/C/N region. Therefore, we propose chemistry may contribute to some extent the spread and inhomogeneous SiO distribution in SN 1987A, on top of dynamic instabilities, as suggested by \citet{abel17}. 
%_____________________________________________________________
%                              FIG 20 - CO and SiO masses compared to observations
%-------------------------------------------------------------
% -----------------------------------------------------------------
\subsection{Other molecules}
\label{mol}
A few other species are identified in the ejecta of SN 1987A, and include SO, SO$_2$, SiS, and CS. 

Sulphur monoxide, SO, is predicted to form in the oxygen core of SNe a few years after outburst by theoretical models \citep{cher09,sar13} and detected with ALMA in SN1987A decades after explosion \citep{mat17}. The derived mass from line analysis is $4\times 10^{-5}$~\Ms. The present total modelled mass is $1.4\times 10^{-4}$~\Ms\ and agrees well with observations. The molecule is essentially formed in the O/Si/Mg~region that produces silicates and silica, and in the O/C/Mg region, which forms most of CO. 

We also predict the formation of a large quantity of sulphur dioxide, SO$_2$, in the O/Si/Mg~region, with a mass of $1.3\times 10^{-2}$~\Ms. Lines of SO$_2$ are tentatively identified in the ALMA spectral coverage of SN~1987A. However, the SO$_2$ lines are weak or blended with other lines (e.g. HCO$^+$) and no mass is derived \citep{mat17}. Therefore, a strong disagreement exists between our large SO$_2$ mass formed in the first years after explosion and that derived from ALMA data, unless SO$_2$ is partially withdrawn from the ejected gas by some chemical or physical processes. In the context of the study of volcano gas plume interaction with solid ashes during volcanic eruptions or gas-aerosol interaction on planets and exoplanets, the reaction of hot/warm gaseous SO$_2$ on the surface of silicate glasses is studied and results in SO$_2$ depletion and formation of solid sulphates and silica in the case the solid is olivine \citep{king18}. SO$_2$ and silica/silicates having the same formation locus in the ejecta, we suggest depletion of SO$_2$ from the gas phase might occur through a similar process. 

As already discussed in \$~\ref{SiSreg} and \S~\ref{innertot}, silicon monosulphide, SiS, is the prominent species formed in the Si/S/Ca~region, while a residual mass is also produced in the O/Si/Mg~region. The total SiS mass at day 4000 is large with a value of $\sim 5\times 10^{-2}$~\Ms\ (see Table~\ref{finmas}). The molecule is tentatively identified at late times in SN 1987A with ALMA and a mass upper limit of $6\times 10^{-5}$~\Ms\ is estimated \citep{mat17}. This is much less than the value we derive a few years after explosion, even when considering differences in initial elemental compositions owing to the different progenitor masses (15~\Ms\ for our study and 19~\Ms\ for SN1987A). However, we do not consider the synthesis of the non-organic solid SiS$_2$ in our model. Solid silicon disulphide was proposed as a potential carrier of the 21~\mic\ band observed in proto-planetary nebular \citep{goe93}. Glassy SiS$_2$ shows a strong emission band around $\sim 20$~\mic\ and a minor peak at 16.5 \mic\ \citep{beg96}. If SiS$_2$ clusters were to form in significant amount, we would expect the final mass of SiS to decrease in the inner ejecta regions. The formation of solid SiS$_2$ will be investigated in a forthcoming paper. 

Finally, the first-overtone emission of carbon sulphide, CS, is tentatively identified around at 3.88~\mic\ in SN~1987A a few hundred days after outburst \citep{meik89,meik93} and at late times with ALMA \citep{mat17}. Existing models \citep{lepp90,cher09,sar13}  predict a mass between $10^{-5}$ and $10^{-4}$~\Ms, whereas an upper limit of $7\times 10^{-6}$~\Ms\ is derived from ALMA data. We find CS forms essentially in the He/C/N~regions with a mass of $5\times 10^{-4}$~\Ms\ at day 4000. Inspection of Figs.~\ref{fig14} and \ref{fig16} reveals CS does respond to gas density enhancement by shifting its formation onset from day 1000 for the Standard Case to day 500 for the high-density case. The final masses in both cases are rather similar. In any case, we see an early detection of CS cannot arise before a few hundred days after explosion and traces the density of the ejecta gas. According to our model, the cyano radical, CN, forms along with CS in the C-rich region, shows a similar trend in the onset of its formation and as such, can also be regarded as a gas density tracer through the detection of its fundamental band at 5~\mic\ in the near IR \citep{civ23}. 
 
Apart from SO, SO$_2$, SiS, and CS, a few molecules form in large amounts in the ejecta but are not yet observed in SN environments. They are O$_2$, CO$_2$, C$_3$ and CaS. All molecules are detected in space except for CaS. Sub-millimetre transitions of O$_2$ between $119$~GHz and $1121$~GHz were observed in molecular clouds with the Submillimeter Wave Astronomy Satellite \citep{mel00}, the Odin Satellite \citep{lar07}, and the Herschel Space Observatory \citep{gold11}. The CO$_2$ molecule is important as coolant through emission in the v2 bending mode at $15$~\mic\ in planetary atmosphere \citep{kute24}. The antisymmetric stretch v3 mode of C$_3 $ at $\sim 4.9$~\mic\ was observed in the carbon-rich AGB star IRC+10216 \citep{hin88} while the presence of the molecule was confirmed in the same object through observation of the v2 bending mode at $\sim 150$~\mic\ with the Infrared Space Telescope \citep{cer00}. As for CaS, rotational lines of the isovalent molecule CaO was recently detected in the ISM \citep{rey24} and the CaS molecule might be searched for through its rich vibrational-rotational spectrum \citep{taka89}. 

%Discuss DUST: CO emission at d 300 accompanied by dust emission to satisfy background flux - Rho chooses carbon but it should silicates which forms starting day 250. Probably silicates rather than carbon Medler/JWST observes CO  bands at day 252 and dust emission rising around 10 \mic\ with an excess in flux at 18 \mic. No SiO emission ?????? there is a strong IR excess that grows in strength relative to the SN flux at late times.
%In 2023ixf: jacobson 25 (0.5−1)×10−3  M⊙ of silicate dust in the cold dense shell and/or inner SN ejecta can effectively reproduce the global properties of the late-time (>300 days) UV-to-IR spectra of SN 2023ixf. utntil 620 days.

%CO+: Small masses at early times. While Spyro observed CO+ at 2.26 \mic\ in SN1987A, Park 2025: but we do not find CO+ emission at 2.26 μm in our spectra, nor was it present in SN 2017eaw (Rho et al. 2018a).

 \subsection{Dust clusters}
 \label{clus}
 %_____________________________________________________________
%                              FIG 21 - Dust clusters compared to observations
%-------------------------------------------------------------
\begin{figure} 
  \centering
   \includegraphics[width=1.0\linewidth]{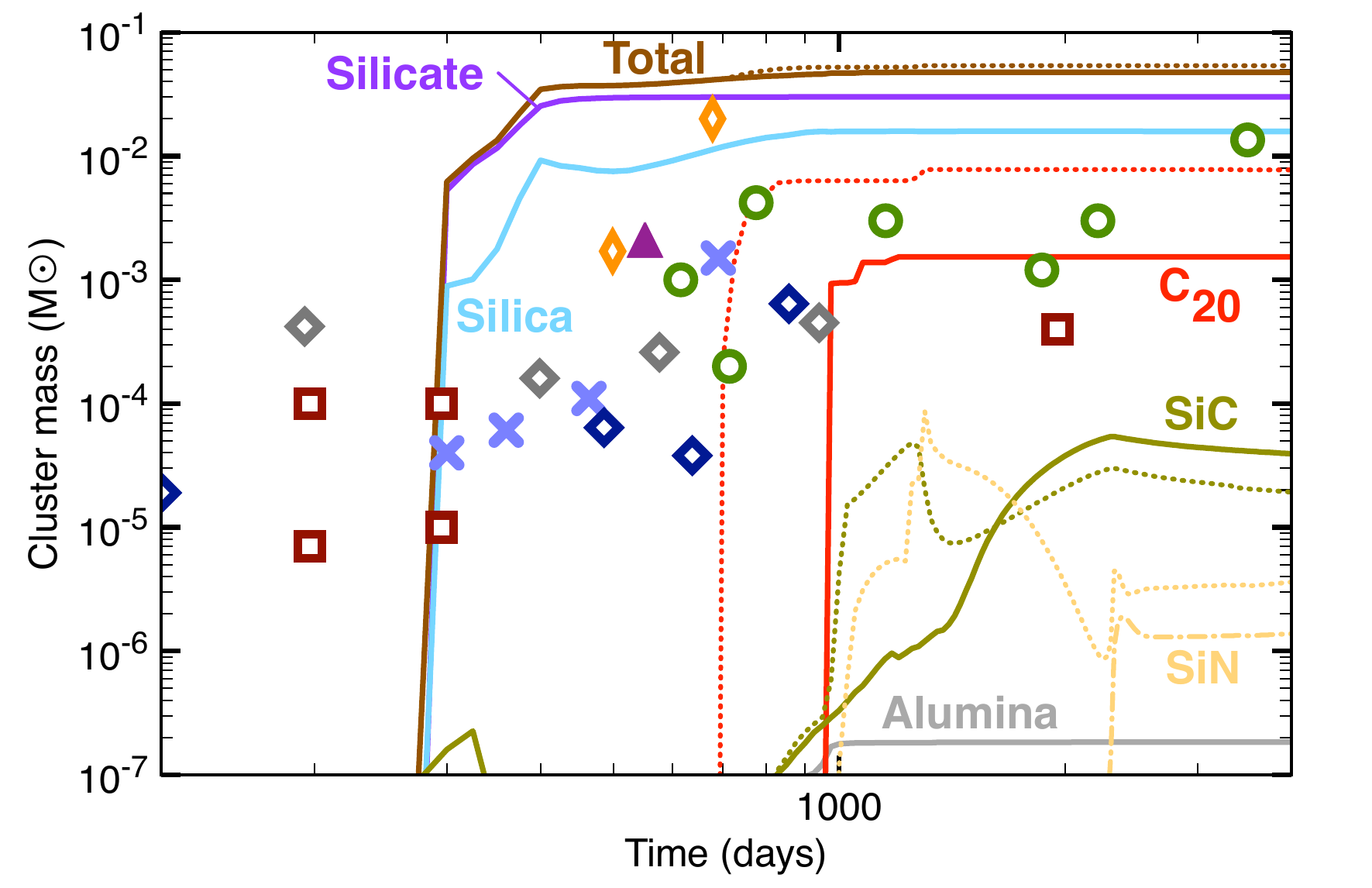}
  \includegraphics[width=1.0\linewidth]{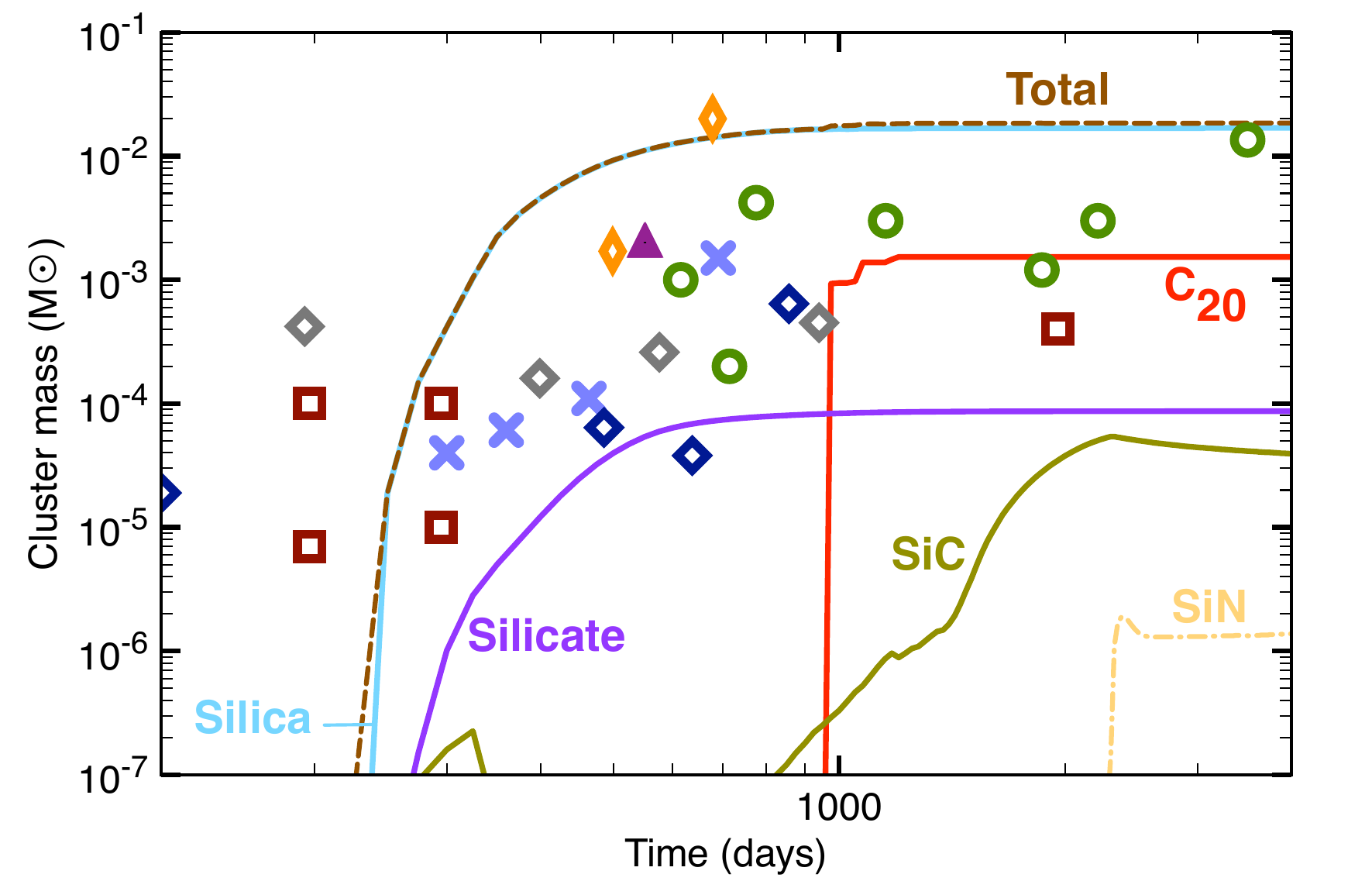}
\caption{Total dust cluster masses as a function of post-explosion time. Top: full lines are for the Standard Case and dotted lines are for the Standard Case with a high density He/C/N~region (see \S~\ref{HD} for details); Bottom: Standard Case with a low-temperature O/Si/Mg~region (see \S~\ref{LT} for details). Available observational data on dust are also represented for a few Type IIP SNe:  Brown empty square SN2017eaw, Lavender cross SN2004et, Grey empty rhombus SN2005af, Green empty circle SN1987A, Orange empty rhombus SN2003gd, Dark blue empty rhombus SN2011ja, Purple filled rhombus SN2006bc.}
\label{fig21}
\end{figure}
% -----------------------------------------------------------------
We present in Fig.~\ref{fig21} the masses of dust clusters derived for the Standard Case,  the Standard Case with a high-density He/C/N~ region and the Standard Case with a low-temperature O/Si/Mg~region. We also plot the dust masses derived from observations of a few Type~IIP SNe and SN~1987A. Since we deal only with the ejecta molecular phase in this paper and do not model the coagulation and growth of the dust clusters we consider, the mass values of Fig.~\ref{fig21} provide indications on the final dust mass but does not reflect the time evolution of dust grains, specifically the time at which coagulation of clusters begins to be efficient. 

The coagulation efficiency involves two opposite processes, the adhesion mechanisms due to interaction forces and the thermal rebounds effects directly linked to the translational kinetic energy of the particles, and thus the gas temperature. At high temperature, thermal rebounds prevails and coagulation efficiency greatly diminishes \citep{jo01,siri13}. It is therefore reasonable to consider coagulation initiates at the temperature regime of flame experiments. For the low-temperature case of region O/Si/Mg, the synthesis of silica and silicate clusters starts at $\sim$ day 250 where $T_{\rm gas} \sim 1600$~K, implying the clusters of Fig.~\ref{fig21} will readily coagulate while this may not the case for the Standard Case, where silica and silicate clusters start forming around day 300 and at $T_{\rm gas} \sim 3500$~K. In this case, coagulation will start at later times. 

Besides coagulation issues, we see the mass derived from observational data shown in Fig.~\ref{fig21} are consistent with modelling trends for cluster formation with a better agreement when lower temperatures are considered for the O/Si/Mg~region. Since this region produces large quantities of SiO, it would be of interest to consider SiO cooling when deriving temperature profiles for this zone. These trends agree also with recent observations of SN~2023ixf with the JWST reported by \citet{med25}. They find the detection of CO bands at day 252 is accompanied by a dust emission growing in strength relative to the SN flux around 10 \mic\ and an excess in flux at 18 \mic, indicative of silicate/silica-rich dust \citep{jae03}.

\section{Comparison with other modelling studies}
\label{dis}
Several studies partially \citep{slu18,lil20} or entirely \citep{sar22,sar222,sar25} use the chemical scheme developed by \citet{cher08,cher09} and \citet{sar13}. This scheme was a first attempt in the description of the formation of molecules and dust molecular clusters in a H-free environment, drawing on available data from combustion, aerosol and atmospheric chemistry. As already mentioned, further studies proved some proposed chemical routes were unlikely, e.g. for silicates \citep{brom16,kim22}. Regarding carbon clusters, the chemistry mainly relied on a unique study by \citet{clay99} at a time when the structures of small carbon clusters were not well identified and for which formation rates were artificially large. 

Although it is difficult to directly compare results from the present study with those of existing investigations since the chemistry and the physical models are different, we can nevertheless outline some differences and drawbacks in recent investigations. Regarding the study of \citet{sar13} who model a SNe with a 15~\Ms\ progenitor and a smaller \Ni\ mass ( \Ni\ $=0.75$~\Ms), we find the new chemical scheme is more efficient at forming molecules and dust clusters despite initial gas densities at day 100 lower by a factor $100$. Sarangi \& Cherchneff find at day 1500 a molecular mass of 29.4 \%\ and a dust mass of 1.6 \%\ the ejecta mass when we have $39.95$~\%\ and $2$~\%\ for molecule and cluster masses, respectively. We also find a different composition of dust clusters: the dust composition in \citet{sar15} is mainly made of carbon, alumina, and silicates in order of decreasing mass, whereas this new model has a dust cluster composition dominated by silicates and silica, with small amount of alumina and eventually carbon clusters that form in high-density clumps. As for SiC and SiN, we might expect small masses of SiC and Si$_3$N$_4$ clusters to form in these carbon-rich, dense clumps, with SiC clusters always more abundant than those of Si$_3$N$_4$ by a few orders of magnitude. The present SiC mass at day 4000 for the Standard Case agree with the SiC dust mass derived at day 2000 by \citet{sar15}. More generally, the SiC and SiN results are consistent with the finding of meteorite studies where rare Si$_3$N$_4$ and SiC grains are isolated with both isotopic anomalies pointing to a common Type II SN~origin \citep{nit95}. Finally, we do not form small clusters of pure Si, Mg and Fe in the inner Si/S/Ca~region anymore, because of the low dimerisation rates we use in line with the low rates for SiO dimerisation. Therefore, the new chemistry tends to optimise molecular and dust formation in the ejecta despite a larger \Ni\ content and lower gas densities. 

In his study of SN1987A, \citet{sar22} uses the chemical model of \citet{sar13} and the coagulation formalism of \citet{sar15} applied to a SN with a 19~\Ms\ progenitor. The study provides a dust composition made of silicate, alumina, carbon and silicon carbide, in order of decreasing masses, with the carbon dust synthesis occurring at \env~2500 days. As discussed in \S~\ref{carbon}, the scheme for carbon cluster growth relies on chemical reactions derived by \citet{clay99}. Such processes guarantee an "artificial" and efficient formation of carbon clusters over time, even at low gas temperature. At \env~2500 days, the gas densities are \env~ a few $1\times 10^{6}$~\cmcub and the gas temperature is \env~$100$~K according to \citet{sar22}. The rapid coagulation of carbon clusters into carbon dust at very late times seems extremely unlikely under these ejecta conditions since our results show no carbon clusters in the form of large rings that could act as coagulation seeds, do form (e.g. C$_{20}$) for low gas densities,  except for small carbons chains according to the low-temperature chemical pathways provided by \citet{loi14}. Therefore, the very late carbon dust formation in SN1987A presented by \citet{sar22} and the resulting infrared emission spectra need serious reconsideration. 

Finally, \citet{slu18} revisited the study by \citet{sar13} by investigating dust synthesis in a SN with a 20~\Ms\ progenitor and using a similar chemistry, an exhaustive description of dust growth including accretion and destruction processes, and a three-phase ejecta corresponding to low-density nickel bubbles, high-density shells of material swept up by the expanding bubbles, and an intermediate density region unaffected by the bubbles. Although we cannot directly compare results of both studies since the initial elemental compositions of the ejecta are different, we notice they derive temperature profiles quite different from those used by \citet{sar22} and in this study. Furthermore, our Standard Case has a gas density at day 100 similar to their intermediate density region, while our high-density case for the C-rich outer region is similar to their shell density. The dust mass they obtain for their ambiant phase is larger than ours by a factor of $10$, while the dust composition is made of magnesia, iron sulphide and pure silicon grains, in total contrast with our derived composition. Similarly, they find a dust mass larger than ours by a factor of $10$ in their shell phase, with again a totally different chemical composition including silicon, magnesia and forsterite dust. It is unclear whether this large discrepancy in dust masses can be attributed to considering accretion at the surface of dust grains since the authors do not compare results with and without accretion in their study. 

%We believe the present study represents a major step forward the understanding of chemical processes at play in SN ejecta and their role in the dust formation process.

%Observation of Al2O at 10 mic?

\section{Summary and conclusions}
\label{con}
In this paper we present a new exhaustive model for the chemistry taking place in the expanding ejecta of a non-interactive SN with progenitor mass of 15~\Ms. We focus on the formation of molecules and dust clusters until $\sim 11$ years post-explosion and propose new chemical routes to silicate and carbon dust synthesis. One unique and exhaustive chemical scheme is applied to all regions in the ejecta. By doing so, we remove the artificial chemical selection resulting from using a specific scheme for each region, as it is done in existing models. We test the impact of clumps and gas molecular cooling on the model and highlight the following outcomes: 

1)  Existing models are based on a silicate formation scheme involving the dimerisation of SiO as a first step, which artificially depletes the SiO content. Such a depletion is not seen in SN1987A. Furthermore, it was shown the dimerisation of SiO was extremely slow and inefficient for the ejecta conditions, thus invalidating these models. The new proposed scheme for silicate formation is efficient at forming dust clusters essentially in the inner oxygen core and represents a major improvement to existing models since SiO is just partially depleted in the dust formation process, in better agreement with observations,

2) The ejecta has a large molecular component corresponding to $\sim$ 40~\% of its mass while the dust represents $\sim 2$~\%. The most abundant species are, in order of decreasing masses, O$_2$, CO, SiS, SiO, CO$_2$, SO$_2$, CaS, N$_2$, and CS. However, these molecules form in different ejecta regions. For example, SiS is typical of the inner ejecta, C$_3$ and N$_2$ formation peaks in the outer ejecta, while SO$_2$ forms in the inner oxygen core and CO and CO$_2$ in the outer oxygen core. The CO and SiO molecules form all over the ejecta with a peak in the outer and inner oxygen core, respectively, The SiO distribution is patchier than that of CO, in agreement with the reconstruction of CO and SiO emission distribution as observed with ALMA in SN~1987A. 

3) We show carbon dust clusters need over-densities or clumps to form in the outer part of the ejecta, while low temperatures in the oxygen core are not conducive to the synthesis of silicates but favour the formation of silica. Cooling by SiO in the silicate-forming region may be investigated to better model the temperature dependance of this region. We find the molecules CS, CN, and C$_3$ are tracers of gas conditions in the outer ejecta zones, 
 
4) The exotic P- and F-bearing molecules form in small amounts in the oxygen core and the carbon-rich outer region, respectively. Only two species, PO and NF, have final masses of or above $1\times 10^{-8}$~\Ms\ at day 4000,

5) Finally, our results show that any dust directly formed in the ejecta do so a few hundred days after explosion since the chemistry requires high enough gas temperatures and densities to efficiently proceed. The final total dust cluster mass is $\sim 1.7 \times 10^{-2}$ for our Standard Case considering a low-temperature profile mimicking the impact of SiO cooling in the O/Si/Mg~region, and $\sim 5.3 \times 10^{-2}$ considering our Standard Case and a high-density He/C/N region. This value range puts a stringent limit to the total dust mass that forms within the ejecta when surface deposition processes are not considered.  

In a forthcoming paper, we will address the formation, growth, and time evolution of dust in SNe for various progenitor masses, and derive budgets for molecules and dust obtained with our new physico-chemical model of Type IIP SNe.  

%%%%%%%%%%.  ACKNOWLEDGEMENTS %%%%%%%%%%%%%%
\begin{acknowledgements}
  I.C and D.T. acknowledge the French National Research Agency (ANR) for financial support of the ODUST project ANR-22-CE31-0008. The project is provided with computing ressource HPC/AI/QUANTUM and storage resources by GENCI at CINES/IDRIS/TGCC through the grants 2023 AD010805116R2 and 2024 AD010805116R3 on the supercomputer JeanZay/JoliotCurie/Adastra's   
SKL/ROME/CSL/GENOA/V100/A100/H100/MI250x/MI300 partition. 
The authors thank Dr. Alix Gombert for DFT calculations on reactions of small carbon chains.
\end{acknowledgements}
%%%%%%%%%%%%%%%%%%%%%%%%%%%%%%%%%%%%%%%%%%%%%%%%%%%%%%%%%%%%%%

 \bibliographystyle{aa} % style aa.bst
 \bibliography{isa} % your references Yourfile.bib
 %
 %%%%%%%%%%%%%%%%%%%%%%%.   APPENDIX   %%%%%%%%%%%%%%%%%%%%%%%%%%%%
 \begin{appendix}
\FloatBarrier
 \section{Additional tables. }
 
% We list below all types of thermal and non-thermal chemical processes included in the model. 
 %
 % --- Table 1 - chemical processes -----
\begin{table*}[]
\caption{Thermal and non-thermal processes included in the chemical model for Type II-P SN ejecta (adapted from \citet{bis14}. }
\label{tabap1}   
\centering
\begin{tabular}{l rl l l l}
\hline \hline
\multicolumn{5}{c}{Reaction description} & Gas regime/location \\
\hline 
\multicolumn{6}{c}{THERMAL } \\
%THERMAL & & & & & \\
%\hline
%Unimolecular & AB &$\longrightarrow$& A + B & Thermal decomposition & Very high temperature\\
\hline
Bimolecular & AB+ C &$\longrightarrow$& BC + A & Neutral exchange & High temperature \\
& A + B &$\longrightarrow$&  AB + h$\nu$ & Radiative association & T independent \\
& AB + M &$\longrightarrow$& A + B + M & Collision dissociation & High density\\
& AB$^+$ + C &$\longrightarrow$&  BC$^+$ + A & Ion$-$Molecule & T independent \\
& AB$^+$ + C &$\longrightarrow$&  AB + C$^+$ & Charge exchange & T independent\\
& A $^+$ + e$^-$ &$\longrightarrow$& A + h$\nu$&Radiative recombination&  T independent \\
& AB $^+$ + e$^-$ &$\longrightarrow$& A + B& Dissociative recombination&  T independent \\
\hline
Termolecular & A + B + M &$\longrightarrow$&  AB + M & Three-body association & High density\\

\hline
\multicolumn{6}{c}{NON-THERMAL } \\
%NON-THERMAL & & & & & \\
\hline
& A + CE &$\longrightarrow$&  A$^+$ + e$^-$ + CE& Ionisation by Compton e$^-$  & entire SN ejecta \\
& AB + CE &$\longrightarrow$&  A + B + CE& Dissociation by Compton e$^-$  & entire SN ejecta \\
& AB + CE &$\longrightarrow$&  A$^+$ + B  + e$^-$+ CE& Ionisation by Compton e$^-$  & entire SN ejecta \\
& AB + CE &$\longrightarrow$&  A + B$^+$  + e$^-$+ CE& Ionisation by Compton e$^-$  & entire SN ejecta \\
\hline
\end{tabular}
\end{table*}
%
%----------- DFT energies -------
\begin{table*}[!h]
\caption{Absolute energies at $T=0$~K calculated at the DFT - B3LYP/6-31G(d) level and corrected for the zero-point energies (in Hartree). }                 % title of Table
\label{tabap2}    % is used to refer this table in the text
\centering                        % used for centering table
%\resizebox{\columnwidth}{!}{\begin{tabular}{c c c c c c}      % centered columns (4 columns)
\begin{tabular}{l c l c l c}  
\hline\hline               % inserts double horizontal lines
Species & E & Species & E & Species & E \\
\hline
SiO$_3$ & -515.06253 & MgO$_2$ & -350.38592 & Mg$_2$O &-475.40517 \\
Mg$_2$O$_2$ &-550.64069 & MgSiO \tablefootmark{a} & -564,79813 & MgSiO$_2$ & -640.10426 \\
MgSiO$_3$ & -715.34519 & Mg$_2$SiO$_2$ & -840.22978 & Mg$_2$SiO$_3$ &-915.52110 \\
Mg$_2$SiO$_4$\tablefootmark{a} & -990,77914 & & & & \\
\hline
\end{tabular}
\tablefoot{
\tablefoottext{a} {Calculation is from \citet{brom12}.}}
\end{table*}
%
%-----------------------------  TABLE: Final masses -------------------------------

\begin{table*}[]
\caption{Final masses of important molecules and dust clusters (in \Ms) at day 4000 for a low-temperature O/Si/Mg region, a high-density He/C/N region and the full ejecta including mass values for the low-temperature O/Si/Mg~region and the other ejecta regions as given in Table \ref{finmas}.}                 % title of Table
\label{tabap3}    % is used to refer this table in the text
\centering                        % used for centering table
\begin{tabular}{c c c c }      % centered columns (4 columns)
\hline\hline               % inserts double horizontal lines
Species/Region& O/Si/Mg (Low T)&  He/C/N (High D)& Total ejecta \tablefootmark{b} \\         % table heading
\hline                      % inserts single horizontal line
 \multicolumn{4}{c}{Detected molecules \tablefootmark{a}}\\
\hline 
  {\rm CO} &{\rm $1.66\times 10^{-3}$} & {\rm $4.09\times 10^{-3}$}& {\rm $2.03\times 10^{-1}$}  \\ 
   {\rm SiS} & {\rm $9.25\times 10^{-3}$} & {\rm $2.09\times 10^{-5}$}  & {\rm $5.01\times 10^{-2}$}  \\   
  {\rm SiO} & {\rm $5.46\times 10^{-2}$} & {\rm $1.07\times 10^{-4}$}&  {\rm $5.49\times 10^{-2}$}   \\ 
  {\rm SO$_2$} & {\rm $1.22\times 10^{-2}$} & $-$ & {\rm $1.23\times 10^{-2}$} \\ 
  {\rm CS} & {\rm $2.19\times 10^{-10}$} & {\rm $5.28\times 10^{-4}$}& {\rm $5.16\times 10^{-4}$}   \\ 
   {\rm SO} &  {\rm $6.45\times 10^{-4}$} &  {\rm $4.28\times 10^{-6}$}& {\rm $6.94\times 10^{-4}$}  \\     
\hline                                  %inserts single line
 \multicolumn{4}{c}{Potentially detectable molecules \tablefootmark{a}}\\
 \hline
  {\rm O$_2$} & {\rm $4.52\times 10^{-1}$} & {\rm $3.31\times 10^{-4}$}& {\rm $5.74\times 10^{-1}$} \\ 
  {\rm CO$_2$} & {\rm $1.08\times 10^{-2}$} &  {\rm $8.38\times 10^{-4}$}& {\rm $3.26\times 10^{-2}$} \\ 
  {\rm C$_{3}$} & $ - $& {\rm $1.22\times 10^{-3}$} & {\rm $2.45\times 10^{-2}$}  \\  
  {\rm CaS} &  {\rm $1.58\times 10^{-3}$} &  {\rm $4.00\times 10^{-7}$}& {\rm $4.84\times 10^{-3}$}  \\ 
   {\rm N$_2$} & $ - $  & {\rm $4.08\times 10^{-3}$}& {\rm $2.77\times 10^{-3}$}  \\   
 {\rm MgS} &  {\rm $7.52\times 10^{-5}$} & {\rm $4.46\times 10^{-6}$}& {\rm $8.76\times 10^{-5}$}   \\  
   {\rm CN} & $ - $  &  {\rm $1.70\times 10^{-3}$}& {\rm $4.23\times 10^{-4}$}  \\   
   {\rm Al$_2$O} & {\rm $2.24\times 10^{-5}$}&$-$ & {\rm $2.35\times 10^{-5}$} \\
 {\rm PO} &  {\rm $1.76\times 10^{-6}$} &  $ - $ & {\rm $1.87\times 10^{-6}$}   \\ 
  {\rm FeS} &  {\rm $2.33\times 10^{-8}$} & {\rm $1.12\times 10^{-7}$} & {\rm $3.87\times 10^{-5}$}   \\ 
  {\rm FeO} &  {\rm $1.32\times 10^{-6}$} &  {\rm $4.03\times 10^{-7}$} & {\rm $2.07\times 10^{-6}$}   \\ 
   {\rm SiO$_2$} &{\rm $3.50\times 10^{-8}$}& {\rm $6.67\times 10^{-8}$}& {\rm $7.24\times 10^{-8}$}  \\ 
     {\rm NF} &$ - $  & $-$& {\rm $9.92\times 10^{-9}$} \\  
      {\rm NO}  &$ - $ & {\rm $1.01\times 10^{-7}$}& {\rm $1.05\times 10^{-10}$} \\  
\hline
Total mass & {\rm $5.74\times 10^{-1}$} & {\rm $1.29\times 10^{-2}$} &  {\rm $9.33\times 10^{-1}$} \\
 \hline
  \multicolumn{4}{c}{Dust clusters \tablefootmark{a}}\\
 \hline
   {\rm Mg$_2$Si$_2$O$_6$} - enstatite& {\rm $8.76\times 10^{-5}$} &  $-$& {\rm $8.76\times 10^{-5}$}  \\  
   {\rm Mg$_4$Si$_2$O$_8$} - forsterite &{\rm $1.02\times 10^{-7}$} &$-$& {\rm $1.05\times 10^{-7}$}     \\  
   {\rm Si$_3$O$_5$} - quartz &  {\rm $1.48\times 10^{-2}$} & $-$& {\rm $1.48\times 10^{-2}$}  \\ 
    {\rm Si$_3$O$_6$} - quartz &  {\rm $2.04\times 10^{-3}$} & $-$ &{\rm $2.04\times 10^{-3}$}  \\  
    {\rm Al$_4$O$_6$} - alumina &  {\rm $4.01\times 10^{-12}$} & $-$ & {\rm $1.94\times 10^{-10}$} \\  
    {\rm C$_{20}$} - carbon & $ - $& {\rm $7.73\times 10^{-3}$} & {\rm $7.73\times 10^{-3}$}  \\  
     {\rm SiC} - silicon carbide&  {\rm $2.04\times 10^{-11}$} & {\rm $1.58\times 10^{-5}$} & {\rm $3.56\times 10^{-5}$} \\ 
     {\rm SiN} - silicon nitride&  {\rm $7.22\times 10^{-13}$}  & {\rm $3.61\times 10^{-6}$} & {\rm $1.38\times 10^{-6}$} \\ 
  \hline
  Total mass & {\rm $1.69\times 10^{-2}$} &{\rm $7.75\times 10^{-3}$} &$1.70\times 10^{-2}$  \\  
  \hline
\end{tabular}
\tablefoot{
\tablefoottext{a}{Very low masses less than $10^{-10}$ \Ms\ are labelled as $ - $. }
\tablefoottext{b}{Masses summed over the Si/S/Ca, O/C/Mg, and He/C/N regions of Table~\ref{finmas} and the low-temperature O/Si/Mg region as given by $1^{\rm st}$ column of this table}. 
 }
\end{table*}
% ------------------------------------------------------------
\section{Results on the O/Si/Mg region for the low-temperature case.}
%_____________________________________________________________
%                              FIG 8 - O/Si/Mg MOL+DUST mass LOW TEMP
%-------------------------------------------------------------
\begin{figure*}
   \centering
\begin{subfigure}{0.48\textwidth}
        \includegraphics[width=1.\textwidth]{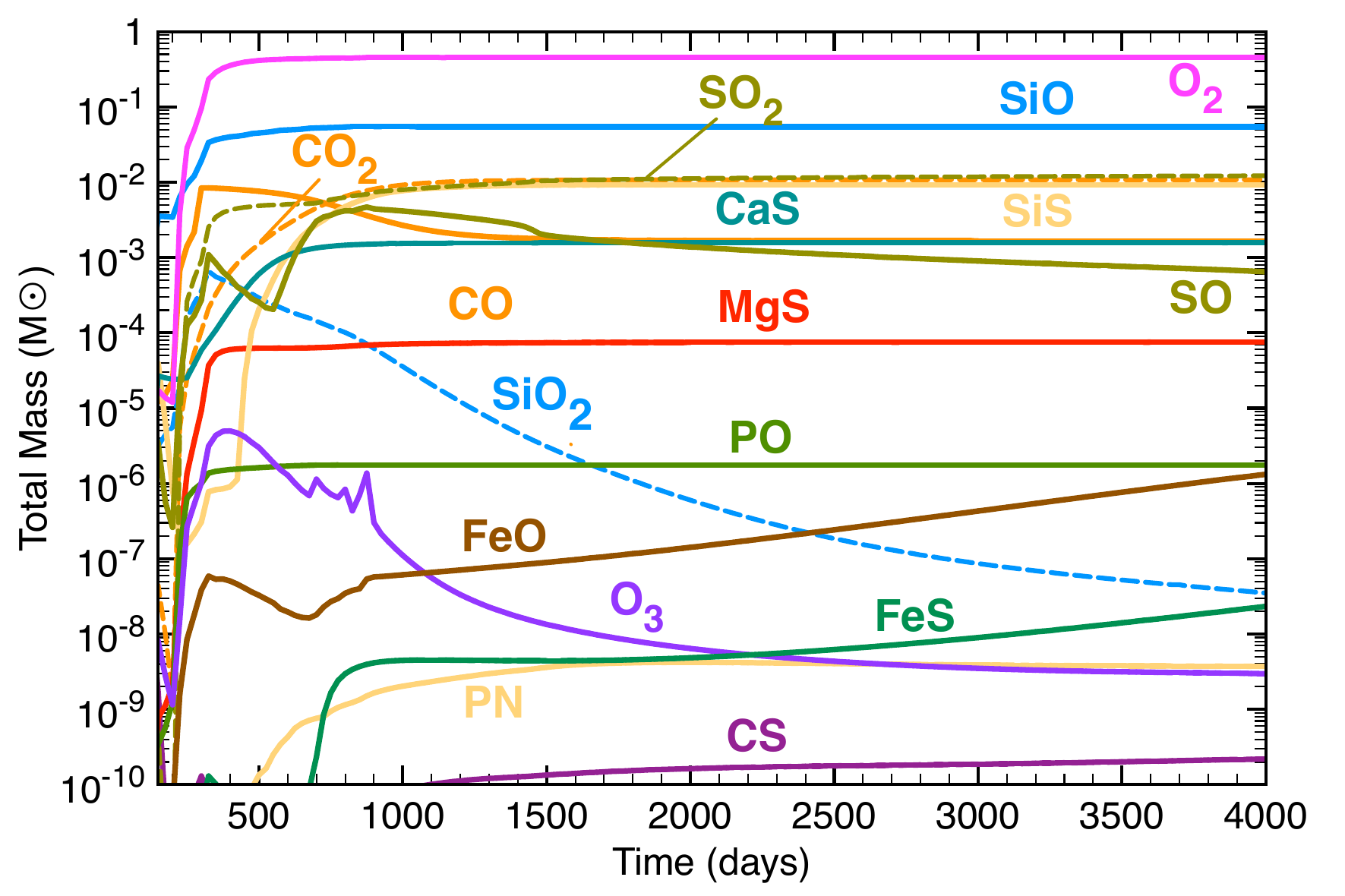}
%        \caption{ }
        \end{subfigure}
\hfill
\begin{subfigure}{0.48\textwidth}
        \includegraphics[width=1.\textwidth]{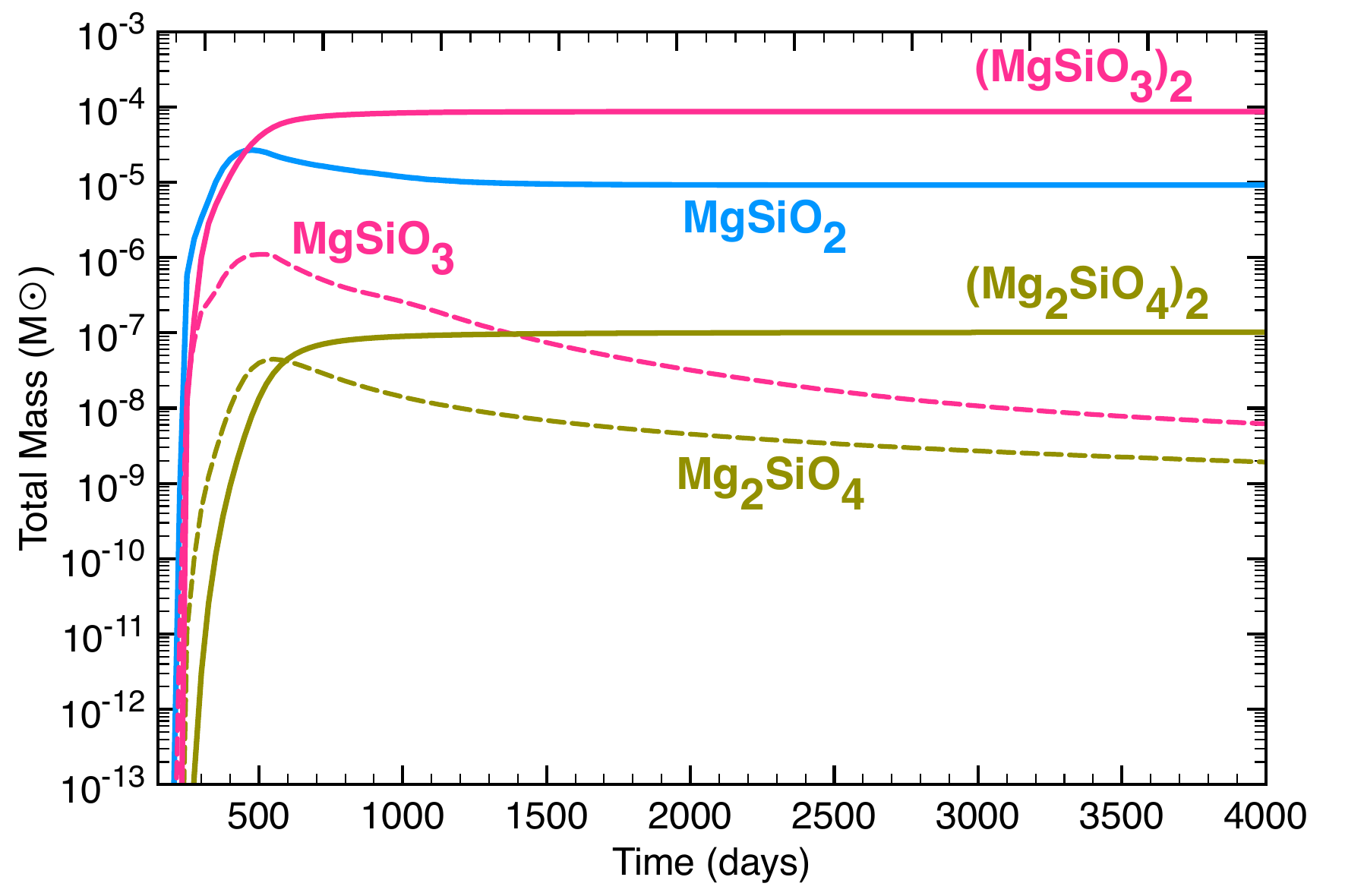}
%         \caption{ }
        \end{subfigure}
\hfill
\begin{subfigure}{0.48\textwidth}
        \includegraphics[width=1.\textwidth]{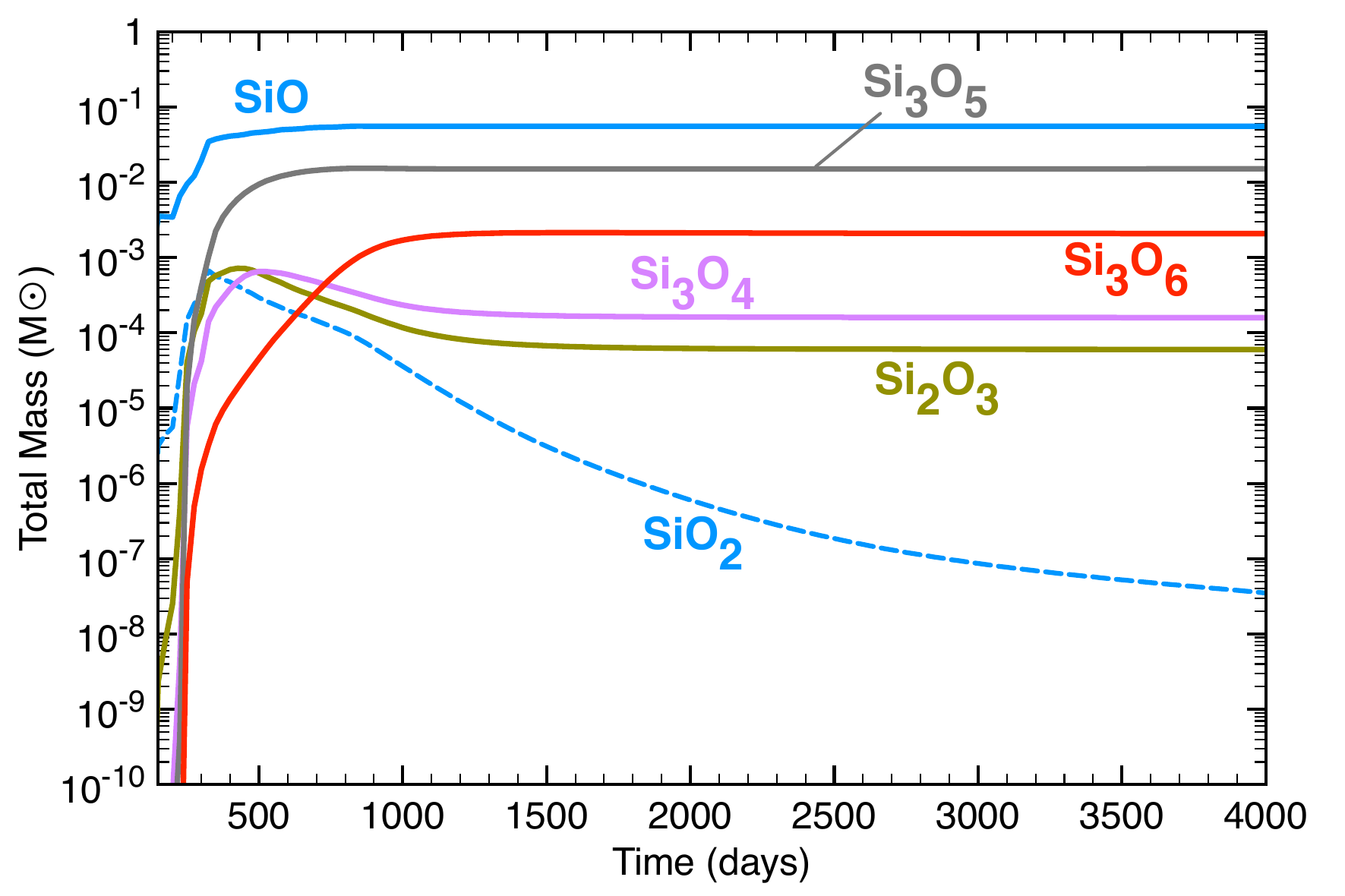}
%         \caption{ }
        \end{subfigure}
\hfill
\begin{subfigure}{0.48\textwidth}
        \includegraphics[width=1.\textwidth]{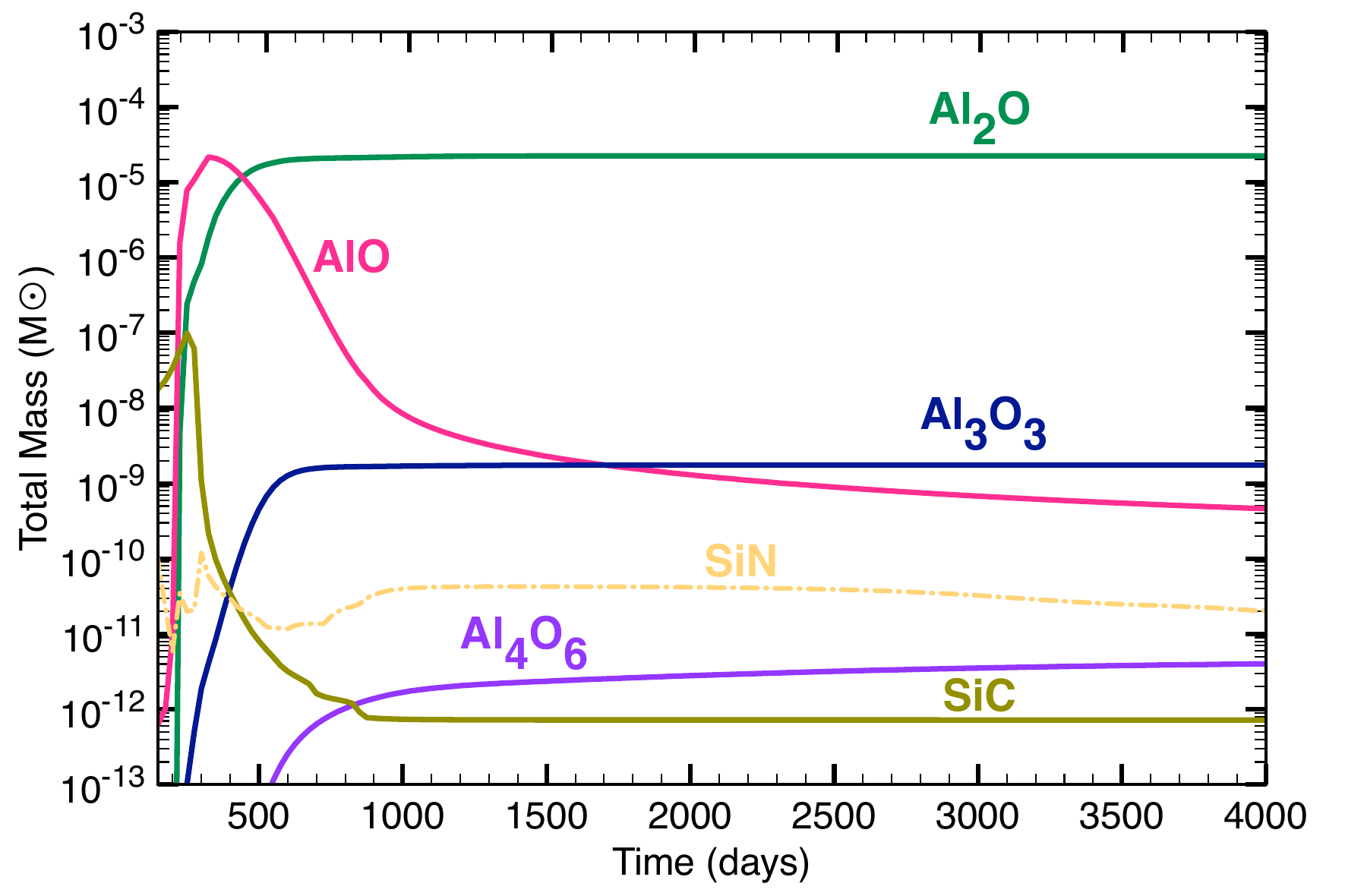}
 %        \caption{ }
        \end{subfigure}
 \caption{Total mass of molecules and dust clusters produced in the O/Si/Mg region as a function of post-explosion time for the low-temperature case discussed in \S~\ref{LT}. Top left: molecules; Top right: silicates; Bottom left: silica; Bottom right: alumina.}
       \label{figALT}
  \end{figure*}
% -----------------------------------------------------.

 \end{appendix}
\end{document}